\documentclass[12pt]{article}

\usepackage{axodraw}

\makeatletter
 
\def\paragraph{\@startsection{paragraph}{4}{\z@}{+2.00ex plus
 +1ex minus +.2ex}{1.5ex plus .2ex}{\it\normalsize}}
 
\def\section{\@startsection {section}{1}{\z@}{+3.0ex plus +1ex minus
  +.2ex}{2.3ex plus .2ex}{\normalsize\bf}}
\def\subsection{\@startsection{subsection}{2}{\z@}{+2.5ex plus +1ex
minus +.2ex}{1.5ex plus .2ex}{\normalsize\bf}}
\def\subsubsection{\@startsection{subsubsection}{3}{\z@}{+3.25ex plus
 +1ex minus +.2ex}{1.5ex plus .2ex}{\normalsize\bf}}

\@addtoreset{equation}{section}

\def\appendix{\par
 \setcounter{section}{0} 
 \setcounter{subsection}{0}
 \setcounter{equation}{0}
 \def\thesection{\Alph{section}}}
 
\expandafter\ifx\csname mathrm\endcsname\relax\def\mathrm#1{{\rm #1}}\fi
 
\makeatletter

\newcount\@tempcntc
\def\@citex[#1]#2{\if@filesw\immediate\write\@auxout{\string\citation{#2}}\fi
  \@tempcnta\z@\@tempcntb\m@ne\def\@citea{}\@cite{\@for\@citeb:=#2\do
    {\@ifundefined
       {b@\@citeb}{\@citeo\@tempcntb\m@ne\@citea
        \def\@citea{,\penalty\@m\ }{\bf ?}\@warning
       {Citation `\@citeb' on page \thepage \space undefined}}%
    {\setbox\z@\hbox{\global\@tempcntc0\csname
b@\@citeb\endcsname\relax}%
     \ifnum\@tempcntc=\z@ \@citeo\@tempcntb\m@ne
       \@citea\def\@citea{,\penalty\@m}
       \hbox{\csname b@\@citeb\endcsname}%
     \else
      \advance\@tempcntb\@ne
      \ifnum\@tempcntb=\@tempcntc
      \else\advance\@tempcntb\m@ne\@citeo
      \@tempcnta\@tempcntc\@tempcntb\@tempcntc\fi\fi}}\@citeo}{#1}}

\def\@citeo{\ifnum\@tempcnta>\@tempcntb\else\@citea
  \def\@citea{,\penalty\@m}%
  \ifnum\@tempcnta=\@tempcntb\the\@tempcnta\else
   {\advance\@tempcnta\@ne\ifnum\@tempcnta=\@tempcntb \else
\def\@citea{--}\fi
    \advance\@tempcnta\m@ne\the\@tempcnta\@citea\the\@tempcntb}\fi\fi}

\def\asymp#1%
{\mathrel{\raisebox{-.4em}{$\widetilde{\scriptstyle #1}$}}}
\def\Nlim#1{\mathrel{\raisebox{-.4em}
{$\stackrel{\disp\longrightarrow}{\scriptstyle#1}$}}}
\def\Nequal#1%
{\mathrel{\raisebox{-.5em}{$\stackrel{=}{\scriptstyle\rm#1}$}}}

\def\beq#1\eeq{\begin{equation}#1\end{equation}}
\def\beqar{\begin{eqnarray}}
\def\eeqar{\end{eqnarray}}
\def\barr#1{\begin{array}{#1}}
\def\earr{\end{array}}
\def\bfi{\begin{figure}}
\def\efi{\end{figure}}
\def\btab{\begin{table}}
\def\etab{\end{table}}
\def\bce{\begin{center}}
\def\ece{\end{center}}
\def\nn{\nonumber}
\def\disp{\displaystyle}

\def\De{\Delta}

\def\refeq#1{\mbox{(\ref{#1})}}

\def\reffi#1{\mbox{Fig.~\ref{#1}}}

\def\refta#1{\mbox{Table~\ref{#1}}}

\def\refse#1{\mbox{Section~\ref{#1}}}
\def\refses#1{\mbox{Sections~\ref{#1}}}
\def\refapp#1{\mbox{App.~\ref{#1}}}
\def\citere#1{\mbox{Ref.~\cite{#1}}}
\def\citeres#1{\mbox{Refs.~\cite{#1}}}
 
\newcommand{\TeV}{\unskip\,\mathrm{TeV}}
\newcommand{\GeV}{\unskip\,\mathrm{GeV}}
\newcommand{\MeV}{\unskip\,\mathrm{MeV}}
\newcommand{\eV}{\unskip\,\mathrm{eV}}
\newcommand{\mb}{\unskip\,\mathrm{mb}}
\newcommand{\pb}{\unskip\,\mathrm{pb}}

\newcommand{\ri}{{\mathrm{i}}}

\newcommand{\rd}{{\mathrm{d}}}

\newcommand{\M}{{\cal{M}}}

\def\mathswitchr#1{\relax\ifmmode{\mathrm{#1}}\else$\mathrm{#1}$\fi}

\newcommand{\PW}{\mathswitchr W}
\newcommand{\Pw}{\mathswitchr w}
\newcommand{\PZ}{\mathswitchr Z}

\newcommand{\Pe}{\mathswitchr e}

\newcommand{\Pt}{\mathswitchr t}

\newcommand{\Pep}{\mathswitchr {e^+}}
\newcommand{\Pem}{\mathswitchr {e^-}}

\def\mathswitch#1{\relax\ifmmode#1\else$#1$\fi}

\newcommand{\MW}{\mathswitch {M_\PW}}

\newcommand{\MZ}{\mathswitch {M_\PZ}}

\newcommand{\Me}{\mathswitch {m_\Pe}}

\newcommand{\Mt}{\mathswitch {m_\Pt}}
 
\newcommand{\sw}{\mathswitch {s_\Pw}}
\newcommand{\cw}{\mathswitch {c_\Pw}}

\def\Li{\mathop{\mathrm{Li}}\nolimits}

\newcommand{\QED}{{\mathrm{QED}}}

\newcommand{\virt}{\mathrm{virt}}
\newcommand{\soft}{\mathrm{soft}}

\newcommand{\rem}{\mathrm{rem}}

\newcommand{\cut}{\mathrm{cut}}

\newcommand{\z}{\setbox0\hbox{+}\hbox to \wd0{\hss0\hss}}

\def\slash#1{{\setbox0=\hbox{$#1$}
  \rlap{\ifdim\wd0>.7em\kern.22\wd0\else\kern.1\wd0\fi /}#1}}

\def\braket#1#2{\left\langle #1\vphantom{#2}
  \right. \kern-2.5pt\left| #2\vphantom{#1}\right\rangle }

\def\M{{\cal M}}
\def\O{{\cal O}}

\def\cut{\mathswitchr{cut}}

\def\sub{{\mathrm{sub}}}
\def\gsub{g^{(\sub)}}
\def\Gsub{G^{(\sub)}}
\def\cGsub{{\cal G}^{(\sub)}}

\def\gin{g^{\mathrm{(in)}}}
\def\gout{g^{\mathrm{(out)}}}
\def\goutin{g^{\mathrm{(out/in)}}}

\begin{document}

\thispagestyle{empty}
\def\thefootnote{\fnsymbol{footnote}}
\setcounter{footnote}{1}
\null
\strut\hfill BI-TP 99/09 \\
\strut\hfill hep-ph/9904440
\vskip 0cm
\vfill
\begin{center}
{\Large \bf 
\boldmath{A General Approach To \\[.5em]
Photon Radiation Off Fermions}%
\par} \vskip 2.5em
{\large
{\sc Stefan Dittmaier}\\[1ex]
{\normalsize \it Theoretische Physik, Universit\"at Bielefeld,
D-33615 Bielefeld, Germany}
}
\par \vskip 1em
\end{center} \par
\vskip 3cm 
{\bf Abstract:} \par
Soft or collinear photon emission potentially poses numerical problems
in the phase-space integration of radiative processes. In this paper, a
general subtraction formalism is presented that removes such
singularities from the integrand of the numerical integration and adds
back the analytically integrated contributions that have been
subtracted. The method is a generalization of the {\it dipole formalism}
of Catani and Seymour, which was formulated for NLO QCD processes with
massless unpolarized particles. The presented formalism allows for 
arbitrary mass
and helicity configurations in processes with charged fermions and any
other neutral particles. Particular attention is paid to the limit of
small fermion masses, in which collinear singularities cause potentially
large corrections. The actual application and the efficiency of the
formalism are demonstrated by the discussion of photonic corrections to
the processes $\gamma\gamma\to\Pt\bar\Pt(\gamma)$, 
$\Pem\gamma\to\Pem\gamma(\gamma)$, and
$\mu^+\mu^-\to \nu_\Pe\bar\nu_\Pe(\gamma)$.
\par
\vfill
\noindent 
April 1999 \par
\vskip 1cm 
\null
\setcounter{page}{0}
\clearpage
\def\thefootnote{\arabic{footnote}}
\setcounter{footnote}{0}

\section{Introduction}
\label{se:intro}

Precision experiments with $\Pe^\pm$ beams, such as at LEP, at the SLC, 
or at future linear colliders, allow for an investigation of electroweak 
processes with a typical accuracy of some per cent down to some fractions 
of a per cent. An adequate description of such reactions---and a theoretical 
understanding of them that goes beyond a qualitative level---forces us to 
control higher-order corrections in perturbative predictions. 
An important source of such radiative corrections is due to the virtual 
exchange and the real emission of photons, or of gluons if quarks are 
involved. Although photonic corrections are formally of $\O(\alpha)$
relative to the lowest order, leading to the naive expectation of 
$\sim$1\% as the typical size, the actual effects very often amount to 
$\sim$10\% or more. Apart from large kinematical effects caused by real 
photon radiation in particular processes, this enhancement mainly 
originates from collinear photon emission off highly relativistic
particles, such as $\Pe^\pm$ at the GeV scale, and from the
corresponding virtual photon exchange. For initial-state radiation off
electrons, this kind of correction is proportional to $\alpha\ln(\Me/Q)$,
where $Q\gg\Me$ is a typical energy scale of the process. The remaining
$\O(\alpha)$ corrections amount to one to a few per cent and have to be
included in precision calculations as well. For per-cent accuracy even
the leading $\O(\alpha^2)$ corrections, or higher, can be relevant.

In this paper we focus on the calculation of the full $\O(\alpha)$
correction that is induced by real photon radiation. Such calculations will 
be performed for practically all realistic observables numerically, owing
to the complexity of the squared amplitudes of the most interesting 
processes and the necessity of phase-space cuts. Usually the integration 
over the multidimensional phase space is performed by Monte Carlo
integration. Thus, a linear increase in accuracy is
roughly accompanied by a quadratic increase in the CPU time needed for the
evaluation. In this context, the singularities of a squared amplitude
cause problems. For example, the integrand of a bremsstrahlung process
blows up if the photon energy becomes small, leading to the well-known
logarithmic IR singularity in the phase-space integral. Following a 
frequently applied standard procedure, known as phase-space slicing, 
one introduces a small cutoff energy $\Delta E$ and integrates over the 
photon energy only down to $\Delta E$ numerically. The soft-photon part,
$E_\gamma<\Delta E$,
is known to factorize from the Born cross section, and the corresponding
correction factor, which contains the IR singularity, can be calculated 
analytically. Since the results obtained this way are correct up to
$\O(\Delta E/Q)$, precise predictions require rather small values of
$\Delta E$. For $\Delta E\to 0$ the numerical integration result grows
like $\alpha\ln(\Delta E/Q)$. Consequently, more and more CPU time is
wasted in the precise calculation of this known singular term that
cancels in the final result anyhow. Therefore, procedures that avoid 
such singular numerical integrations are desirable.

Similar problems arise from collinear photon emission off a charged
particle with mass $m\ll Q$. Integrating over small emission angles
$\theta$ results in mass-singular corrections proportional to 
$\alpha\ln(m/Q)$. Applying phase-space slicing, the collinearity region
is excluded by a small cutoff angle $\Delta\theta$ so that the
singularity appears as a $\alpha\ln(\Delta\theta)$ contribution to the
numerical integration result. The missing contribution from the region
$\theta<\Delta\theta$ is related to the lowest-order cross section and
can be obtained without singular numerical integration, similar to the
IR case. Concerning the precision of the integration procedure,
$\Delta\theta$ plays a similar role as $\Delta E$ above, and a procedure
that avoids the singular integration is preferable.

Singular numerical integrations are absent in so-called subtraction
methods. The idea of such methods is to subtract and
to add a simple auxiliary function from the singular integrand.
This auxiliary function has to be chosen in such a way that it cancels all
singularities of the original integrand so that the phase-space
integration of the difference can be performed numerically, even over the
singular regions of the original integrand.
In this difference the original matrix element can be
evaluated without regulators for IR or collinear singularities, i.e.\ it
is possible to apply powerful spinor techniques 
(see e.g.\ \citere{ca81,be82,di99} and references therein)
that have been developed for four space-time dimensions.
The auxiliary function has to be simple enough so that it can
be integrated over the singular regions analytically, when the
subtracted contribution is added again. This part contains the singular
contributions and requires regulators.
In general, the statistical uncertainty of the finally obtained correction 
is smaller than the one of the corresponding result of phase-space slicing,
because the absolute value of the numerical integral is usually much
smaller for the subtraction method, owing to the absence of
singular contributions. Unfortunately, the above
requirements set highly non-trivial conditions on the subtraction
functions, rendering the construction of a general subtraction procedure
difficult. Although various subtraction formalisms have been described
for NLO corrections in massless QCD \cite{el81,ca96},
to the best of our knowledge, up to now no general
subtraction method has been presented that is able to deal with
massive particles in any given process. For the special case of
heavy-quark correlations in hadron--hadron collisions, a subtraction
procedure has been described in \citere{ma92}.

In the following we work out a rather general subtraction method for the
treatment of photon radiation for any given process involving massive or
massless, polarized or unpolarized fermions and any kind of neutral bosons. 
The inclusion of
charged bosons is completely straightforward. Our method follows the
guideline provided by the {\it dipole formalism}, which has been presented
by Catani and Seymour \cite{ca96} for QCD with massless, unpolarized 
partons. Since
the colour flow in QCD processes is more involved than the charge flow
in electroweak processes, our presentation is simpler than the one in
\citere{ca96} in this respect. However, the generalization of the
dipole formalism to arbitrary masses turns out to be highly non-trivial.
Even in the limit of small fermion masses, which is of particular
interest, there is an important difference between our approach and the
subtraction procedures of \citeres{el81,ca96} for massless QCD partons. 
We consistently regularize IR and collinear singularities with finite
masses, as it is commonly applied to photon radiation in electroweak
processes, whereas the above-mentioned QCD studies are carried out in
dimensional regularization.

Although we treat only photon radiation explicitly, one should realize
that the presented results can also be used for gluon radiation in
processes that involve only massive quarks as QCD partons; in this case
the colour flow has to be handled as described in \citere{ca96}. 
Our work also represents a first step towards the generalization of the 
dipole formalism in QCD to include massive partons. 

The paper is organized as follows: in the next section we review the
general structure of IR and collinear singularities, and describe the
strategy of the subtraction procedure. In \refse{se:subfunc0} we
anticipate our results on the subtraction function and its
integrated counterpart for the special case of light fermions, in order 
to illustrate the structure of the formalism. The general results for
arbitrary fermion masses are given in \refse{se:subfunc}, where the
details of the derivation are described, too.
Section~\ref{se:appl} contains the numerical examples, including
discussions of the photonic ${\cal O}(\alpha)$ corrections to the processes
$\gamma\gamma\to\Pt\bar\Pt(\gamma)$, 
$\Pem\gamma\to\Pem\gamma(\gamma)$, and
$\mu^+\mu^-\to\nu_\Pe\bar\nu_\Pe(\gamma)$.
In \refse{se:disc} we discuss salient features of subtraction formalisms
and of the dipole approach. The discussion, in particular, includes comments on 
the implementation of phase-space cuts, some practical advice, and
remarks on the partial generalization to QCD.
Our summary is presented in \refse{se:sum}. In the appendix
we provide important special cases, further details of the calculation,
and the virtual photonic corrections to $\mu^+\mu^-\to\nu_\Pe\bar\nu_\Pe$.

\section{General strategy}

\subsection{Preliminary remarks and conventions}

We consider photon emission in processes that involve arbitrary fermions
and any massive neutral bosons. The initial state may
also contain photons. The presented method remains applicable to reactions 
with more than one photon in the final state if only a single photon
can become soft or collinear with a light fermion in phase space. Note
that situations with more than one photon being soft or collinear
correspond to corrections of ${\cal O}(\alpha^2)$, or higher, relative to 
the lowest-order process without photon emission. In other words, the
method to be described covers all kinds of real-photonic
${\cal O}(\alpha)$ corrections
to processes involving charged fermions and any neutral particles.

The relative charge and the mass of a fermion $f$ are 
denoted by $Q_f$ and $m_f$, the momentum and the helicity of $f$ 
are assigned to $p_f$ and $\kappa_f$, respectively. 
Instead of the general indices $f,f'$ for any fermions, we use the
indices $a,b$ only for initial-state fermions and $i,j$ only for final-state
fermions.
Moreover, we define the sign factors $\sigma_f=\pm 1$ for the charge
flow related to the fermion $f$; specifically, we set $\sigma_f=+1$ for
incoming fermions and outgoing anti-fermions, and $\sigma_f=-1$ for
outgoing fermions and incoming anti-fermions.
Consequently, charge conservation of the whole reaction implies
\beq
\sum_f Q_f \sigma_f = 0.
\label{eq:Qcons}
\eeq

Since IR and collinear divergences are regularized by particle masses,
we consistently work within four space-time dimensions. The invariant
phase-space measure is abbreviated by
\beq
\rd\phi(k_1,\dots,k_n;K) = 
\left[ \prod_{l=1}^{n} \frac{\rd^4 k_l}{(2\pi)^3} 
\theta(k_l^0) \delta(k_l^2-m_l^2) \right]
(2\pi)^4 \, \delta^{(4)}\left( K-\sum_{l=1}^{n} k_l \right).
\eeq
In the following, $\M_1$ is the transition matrix element of the
considered process that involves an outgoing photon with momentum $k$.
The matrix element of the corresponding process without photon emission
is denoted by $\M_0$. For brevity, we explicitly write down only those
momenta and helicities as arguments of $\M_1$ and $\M_0$ that are 
important in the considered equation. 
The collections of all momenta of the corresponding 
reactions are abbreviated by $\Phi_1$ and $\Phi_0$, and the respective
phase-space measures by $\rd\Phi_1$ and $\rd\Phi_0$.

\subsection{IR and collinear singularities}
\label{se:sing}

If the momentum $k$ of the radiated photon becomes soft ($k\to 0$), 
the squared matrix element $\sum_{\lambda_\gamma}|\M_1|^2$, summed
over all photon polarizations $\lambda_\gamma$, becomes IR-singular and 
asymptotically proportional to $|\M_0|^2$ in the well-known form
(see e.g.\ \citere{de93a})
\beq
\sum_{\lambda_\gamma} |\M_1|^2 \asymp{k\to 0} 
-\sum_{f,f'} Q_f \sigma_f Q_{f'} \sigma_{f'} e^2
\frac{p_f p_{f'}}{(p_f k)(p_{f'}k)} |\M_0|^2,
\label{eq:IRlimit}
\eeq
where the sums on the r.h.s.\ run over all charged fermions of the
reaction, and 
$e$ is the positron charge. Eq.~\refeq{eq:IRlimit} is valid for all
polarization configurations separately.
The phase-space integral of \refeq{eq:IRlimit} 
over the soft-photon region is logarithmically divergent. We choose an
infinitesimal photon mass $m_\gamma$ as regulator, yielding singular
contributions proportional to $\alpha\ln(m_\gamma)$ to the real 
${\cal O}(\alpha)$ corrections. According to the Bloch--Nordsieck
theorem \cite{bl37}, these singular contributions cancel against
IR-singular counterparts in the virtual corrections.

\begin{sloppypar}
Another type of singularity occurs in the
limit of a vanishing fermion mass, $m_f\to 0$, if the region of collinear 
photon emission from $f$ is included in the phase-space integration. 
The squared amplitude $|\M_1|^2$ develops poles in $(p_f k)\to 0$, leading
to logarithmic singularities in the phase-space integral.
The asymptotic form of $\sum_{\lambda_\gamma} |\M_1|^2$ in the
collinearity regions is related to the squared amplitude $|\M_0|^2$ and
well known \cite{be82,ba73}. Distinguishing between photon
emission from outgoing and incoming fermions, we have
\beqar
\sum_{\lambda_\gamma} |\M_1(p_i;\kappa_i)|^2 
& \asymp{p_i k\to 0} & 
Q_i^2 e^2 \gout_{i,\tau}(p_i,k) |\M_0(p_i+k;\tau\kappa_i)|^2,
\nn\\*[.5em]
\sum_{\lambda_\gamma} |\M_1(p_a;\kappa_a)|^2 
& \asymp{p_a k\to 0} &
Q_a^2 e^2 \gin_{a,\tau}(p_a,k) |\M_0(x_a p_a;\tau\kappa_a)|^2,
\label{eq:collimit}
\eeqar
where the signs $\tau=\pm$ account for a possible spin flip of the
considered fermion. Whenever $\tau$ appears more than once in products,
we assume summation over $\tau=\pm$.
Note that we take helicity eigenstates as polarization basis throughout.
The functions $\goutin_{f,\tau}$ are given by
\beqar
\gout_{i,+}(p_i,k) &=& \frac{1}{p_i k}
\biggl[ P_{ff}(z_i)-\frac{m_i^2}{p_i k} \biggr]
-\gout_{i,-}(p_i,k),
\nn\\[.5em]
\gout_{i,-}(p_i,k) &=& \frac{m_i^2}{2(p_i k)^2} \frac{(1-z_i)^2}{z_i}, 
\nn\\[.5em]
\gin_{a,+}(p_a,k)  &=& \frac{1}{x_a(p_a k)}
\biggl[ P_{ff}(x_a)-\frac{x_a m_a^2}{p_a k} \biggr]
-\gin_{a,-}(p_a,k),
\nn\\[.5em]
\gin_{a,-}(p_a,k)  &=& \frac{m_a^2}{2(p_a k)^2} \frac{(1-x_a)^2}{x_a},
\label{eq:ginout}
\eeqar
where $P_{ff}(y)$ is the usual splitting function,
\beq
P_{ff}(y) = \frac{1+y^2}{1-y}.
\eeq
If one is only interested in unpolarized fermions, the summation 
of $\goutin_{f,\tau}|\M_0(\tau\kappa_f)|^2$ 
over $\tau$ in \refeq{eq:collimit} 
reduces to $(\goutin_{f,+}+\goutin_{f,-})|\M_0(\kappa_f)|^2$. In the
following, all formulas are written in a form that facilitates this
replacement. 
The variables $z_i$ and $x_a$ are the fractions of the fermion 
energies that are kept by the fermions after photon emission,
\beq
z_i = \frac{p_i^0}{p_i^0+k^0}, \qquad
x_a = \frac{p_a^0-k^0}{p_a^0}.
\label{eq:zx}
\eeq
While final-state radiation does not change any momentum other than
$p_i$ in the hard 
scattering process, initial-state radiation scales the fermion momentum
$p_a$ down to $x_a p_a \sim p_a-k$, thereby reducing the centre-of-mass
(CM) energy of the hard scattering process.
Integrating the squared amplitudes over a collinearity region yields
contributions proportional to $\alpha\ln(m_f)$ to the real
${\cal O}(\alpha)$ corrections%
\footnote{There are also corrections of the forms
$\alpha\ln(m_f)\ln(m_\gamma)$ and $\alpha\ln^2(m_f)$, which originate
from soft photons. These corrections, however, always cancel against
virtual corrections.}.
According to the 
Kinoshita--Lee--Nauenberg theorem \cite{ki62}, 
the mass-singular corrections $\alpha\ln(m_i)$, which originate from
final-state radiation, cancel against their counterparts in the
virtual corrections in the total cross section. Mass singularities from
final-state fermions can only survive in specific distributions, such as
distributions of invariant-masses that are built of fermion momenta
only, i.e.\ without taking into account photon recombination. For
fermions in the initial state the sum of real and virtual corrections
remains mass-singular, and the $\alpha\ln(m_a)$ terms are a potential 
source of large corrections.
\end{sloppypar}

The asymptotic relations \refeq{eq:IRlimit} and \refeq{eq:collimit}
both relate the full squared matrix element $|\M_1(\Phi_1)|^2$ for 
the radiative process to $|\M_0(\Phi_0)|^2$, which corresponds to 
the process without photon emission. Note that the phase spaces on which 
these functions are defined are different. 
In order to guarantee that both sides of \refeq{eq:IRlimit} and
\refeq{eq:collimit} are defined on the phase space spanned by the
momenta $\Phi_1$, one has to specify an appropriate mapping from 
$\Phi_1$ to $\Phi_0$ that respects all mass-shell relations.
The definition of such mappings is of central importance in the
construction of a subtraction function $|\M_\sub|^2$ that is 
parametrized by $\Phi_1$ and has the same asymptotic behaviour as
$\sum_{\lambda_\gamma} |\M_1|^2$ in the singular limits.

\subsection{The dipole subtraction formalism}

Our final aim is to perform the phase-space integral of 
$\sum_{\lambda_\gamma} |\M_1|^2$, which involves IR and collinear
singularities, without carrying out singular numerical integrations. The
basic idea in a subtraction method is to subtract and to add the integral
of an appropriate subtraction function $|\M_\sub|^2$,
\beq
\int\rd\Phi_1\, \sum_{\lambda_\gamma} |\M_1|^2 = 
\int\rd\Phi_1\, \Biggl(\sum_{\lambda_\gamma}|\M_1|^2-|\M_\sub|^2\Biggr)
+ \int\rd\Phi_1\, |\M_\sub|^2,
\label{eq:intM1}
\eeq
where $|\M_\sub|^2$ possesses the same asymptotic behaviour as
$\sum_{\lambda_\gamma} |\M_1|^2$ in the singular limits. Specifically,
we demand
\beq
|\M_\sub|^2 \asymp{\phantom{k\to 0}} \sum_{\lambda_\gamma} |\M_1|^2 
\qquad \mbox{for} \quad k\to 0 
\quad \mbox{or} \quad p_i k\to 0 
\quad \mbox{or} \quad p_a k\to 0,
\label{eq:m2subasymp}
\eeq
where $i$ and $a$ label all outgoing and incoming light fermions. Owing
to \refeq{eq:m2subasymp}, the phase-space integration of the difference
$(\sum_{\lambda_\gamma}|\M_1|^2-|\M_\sub|^2)$ in \refeq{eq:intM1} is
non-singular, i.e.\ it can be performed numerically without regulators. 
The singular contributions of the original integral 
$\int\rd\Phi_1\sum_{\lambda_\gamma}|\M_1|^2$ are completely contained in
$\int\rd\Phi_1|\M_\sub|^2$. If $|\M_\sub|^2$ is chosen appropriately the
singular integrations can be carried out analytically. To this end, the
phase-space integral $\int\rd\Phi_1$ is factorized into a part 
$\int\rd\tilde\Phi_0$ connected to the non-radiative process and a part
$\int [\rd k]$ connected to the photon phase space,
\beq
\int\rd\Phi_1 = \int\rd\tilde\Phi_0 \otimes \int [\rd k].
\eeq
The sign ``$\otimes$'' indicates that this factorization is not an
ordinary product, but may contain also summations and convolutions.
Since $\sum_{\lambda_\gamma}|\M_1(\Phi_1)|^2$ is related to 
$|\M_0(\Phi_0)|^2$ of the non-radiative process in the singular limits,
the subtraction function $|\M_\sub(\Phi_1)|^2$ can be defined in such a
way that it depends on the momenta of $\tilde\Phi_0$ only via 
$|\M_0(\tilde\Phi_0)|^2$. The integration variables of $\int [\rd k]$
occur only in the remaining terms of the subtraction function. Since
those terms are process-independent, the singular integration of
$|\M_\sub(\Phi_1)|^2$ over $[\rd k]$ can be performed analytically once 
and for all. Finally, the integral
$\int\rd\Phi_1\sum_{\lambda_\gamma}|\M_1|^2$ takes the schematic form
\beq
\int\rd\Phi_1\, \sum_{\lambda_\gamma} |\M_1|^2 = 
\int\rd\Phi_1\, \Biggl(\sum_{\lambda_\gamma}|\M_1|^2-|\M_\sub|^2\Biggr)
+ \int\rd\tilde\Phi_0\,\otimes\left(\int [\rd k] \, |\M_\sub|^2\right).
\eeq
The integrations over $\rd\Phi_1$ and $\rd\tilde\Phi_0$ on the r.h.s.\
are free of singularities, and thus are well-suited for numerical
evaluations. Since the singularities in $\int [\rd k] |\M_\sub|^2$ are
controlled analytically, they can be easily combined with their
counterparts in the virtual corrections.

We have seen that the subtraction function $|\M_\sub|^2$ has to obey two
non-trivial conditions.
It must possess the asymptotic behaviour given in \refeq{eq:m2subasymp},
and it must still be simple enough so that it can be
integrated over the singular regions analytically. Note that all the 
collinearity regions of phase space overlap and have the IR part 
($k\to 0$) in common. Therefore, the naive sum of all collinear
singularities, which are proportional to $Q_f^2$, leads to an
overcounting of the IR singularity, and thus cannot be used in the
subtraction function. In the following we show how
this overcounting is avoided and how the subtraction function is
constructed within the dipole formalism. In contrast to
\citere{ca96}, where this formalism is described for massless,
unpolarized partons in QCD, we have to take care of fermion masses
and polarizations.

The subtraction function $|\M_\sub|^2$ is constructed from auxiliary
functions $\gsub_{ff',\tau}$, which are labelled 
by pairs of different fermions $f\neq f'$:
\beq
|\M_\sub(\Phi_1)|^2 = 
-\sum_{f\neq f'} Q_f \sigma_f Q_{f'} \sigma_{f'} e^2
\gsub_{ff',\tau}(p_f,p_{f'},k) 
\left|\M_0\left(\tilde\Phi_{0,ff'};\tau\kappa_f\right)\right|^2.
\label{eq:m2sub}
\eeq 
Since only the kinematics of fermion $f$ gives rise to singular
contributions in the subtraction function, $f$ is called {\it emitter}, 
whereas $f'$ is called {\it spectator}. 
The summation over $\tau$ accounts for the helicity flip of the emitter $f$.
The auxiliary functions $\gsub_{ff',\tau}$ have to 
possess an appropriate asymptotic behaviour. In the IR limit one 
globally demands 
\beqar
\gsub_{ff',+}(p_f,p_{f'},k) & \asymp{k\to 0} &
\frac{1}{p_f k} \biggl[ 
\frac{2(p_f p_{f'})}{p_f k+p_{f'} k}
-\frac{m_f^2}{p_f k} \biggr],
\nn\\[.5em]
\gsub_{ff',-}(p_f,p_{f'},k) & \asymp{k\to 0} & \O(1).
\label{eq:IRcond}
\eeqar
In the collinear limits one demands separate conditions for 
final- and initial-state fermions $f=i,a$:
\beqar
\gsub_{if',\tau}(p_i,p_{f'},k) & \asymp{p_i k\to 0} & \gout_{i,\tau}(p_i,k),
\nn\\[.5em]
\gsub_{af',\tau}(p_a,p_{f'},k) & \asymp{p_a k\to 0} &  \gin_{a,\tau}(p_a,k),
\label{eq:colcond}
\eeqar
where $f'$ can be outgoing or incoming. 
The analytical form of $\gsub_{ff',\tau}$ is, of course, 
not uniquely determined by the asymptotic conditions. 
A convenient choice for these auxiliary functions, which are graphically
represented by the effective diagrams of \reffi{fig:effdiags},
is given in the next sections. 
\bfi
\centerline{
\begin{picture}(360,230)(0,0)
\put(0,120){
  \begin{picture}(160,120)(0,0)
  \Line(80,50)(140, 90)
  \Line(80,50)(120, 20)
  \LongArrow( 90,30)(105, 18)
  \LongArrow(115,85)(130, 95)
  \LongArrow(115,55)(130, 45)
  \Photon(110,70)(140,50){2}{4}
  \Vertex(110,70){2.5}
  \GCirc(80,50){10}{1}
  \put(145, 88){$i$}
  \put(145, 46){$\gamma$}
  \put(125, 15){$j$}
  \put( 10,110){\underline{$\gsub_{ij}$}:}
  \put( 90, 14){$p_j$}
  \put(115, 99){$p_i$}
  \put(116, 40){$k$}
  \end{picture} }
\put(180,120){
  \begin{picture}(160,120)(0,0)
  \Line(80,50)(140, 90)
  \Line(80,50)( 40, 20)
  \LongArrow( 52,18)( 67, 30)
  \LongArrow(115,85)(130, 95)
  \LongArrow(115,55)(130, 45)
  \Photon(110,70)(140,50){2}{4}
  \Vertex(110,70){2.5}
  \GCirc(80,50){10}{1}
  \put(145, 88){$i$}
  \put(145, 46){$\gamma$}
  \put( 28, 18){$a$}
  \put( 10,110){\underline{$\gsub_{ia}$}:}
  \put( 65, 18){$p_a$}
  \put(115, 99){$p_i$}
  \put(116, 40){$k$}
  \end{picture} }
\put(0,-10){
  \begin{picture}(160,120)(0,0)
  \Line(80,50)( 20, 90)
  \Line(80,50)(120, 20)
  \LongArrow( 90,30)(105, 18)
  \LongArrow( 65,70)( 80, 80)
  \LongArrow( 20,80)( 35, 70)
  \Photon( 50,70)( 80,90){2}{4}
  \Vertex( 50,70){2.5}
  \GCirc(80,50){10}{1}
  \put( 10, 88){$a$}
  \put( 87, 90){$\gamma$}
  \put(125, 15){$i$}
  \put( 10,110){\underline{$\gsub_{ai}$}:}
  \put( 90, 14){$p_i$}
  \put( 20, 63){$p_a$}
  \put( 76, 65){$k$}
  \end{picture} }
\put(180,-10){
  \begin{picture}(160,120)(0,0)
  \Line(80,50)( 20, 90)
  \Line(80,50)( 40, 20)
  \LongArrow( 52,18)( 67, 30)
  \LongArrow( 65,70)( 80, 80)
  \LongArrow( 20,80)( 35, 70)
  \Photon( 50,70)( 80,90){2}{4}
  \Vertex( 50,70){2.5}
  \GCirc(80,50){10}{1}
  \put( 10, 88){$a$}
  \put( 87, 90){$\gamma$}
  \put( 28, 18){$b$}
  \put( 10,110){\underline{$\gsub_{ab}$}:}
  \put( 65, 18){$p_b$}
  \put( 20, 63){$p_a$}
  \put( 76, 65){$k$}
  \end{picture} }
\end{picture} } 
\caption{Effective diagrams for the different emitter/spectator cases.}
\label{fig:effdiags}
\efi

Finally, we have to specify the conditions on the momenta to be inserted
in $|\M_0|^2$ in \refeq{eq:m2sub}. As explained above, it is necessary 
to define a
mapping $\tilde\Phi_{0,ff'}$ from the momenta of $\Phi_1$ to the ones 
of $\Phi_0$ that respects all mass-shell relations. The symbol 
$\tilde\Phi_{0,ff'}$ indicates that different mappings are used for 
different pairs $ff'$. 
Denoting the momenta of $f$ and $f'$ in $\tilde\Phi_{0,ff'}$ by 
$\tilde p_f$ and $\tilde p_{f'}$, and writing $k_n$ and $\tilde k_n$ for
the remaining momenta in $\Phi_1$ and $\tilde\Phi_{0,ff'}$,
respectively, we require
\beq
\tilde p_f \Nlim{k\to 0} p_f, \qquad
\tilde p_{f'} \Nlim{k\to 0} p_{f'}, \qquad
\tilde k_n \Nlim{k\to 0} k_n 
\label{eq:IRmom}
\eeq
for the IR limit and
\beq
\tilde p_i \Nlim{p_i k\to 0} p_i+k, \qquad
\tilde p_a \Nlim{p_a k\to 0} x_a p_a, \qquad
\tilde p_{f'} \Nlim{p_f k\to 0} p_{f'}, \qquad
\tilde k_n \Nlim{p_f k\to 0} k_n 
\label{eq:colmom}
\eeq
for the collinear limits.
The mass-shell conditions $\tilde p_f^2=p_f^2=m_f^2$,
$\tilde p_{f'}^2=p_{f'}^2=m_{f'}^2$, and $\tilde k_n^2 = k_n^2 = m_n^2$
have to be fulfilled for arbitrary photon momentum $k$.

Using \refeq{eq:IRcond} and \refeq{eq:colcond}, it is rather easy to 
check that the subtraction function \refeq{eq:m2sub} possesses the
asymptotic behaviour required in \refeq{eq:m2subasymp}.
For the IR limit, the asymptotic relation is verified upon inserting 
\refeq{eq:IRcond}
into \refeq{eq:m2sub} and rearranging the terms in $\sum_{f\neq f'}$:
\beqar
\sum_{f\neq f'} Q_f \sigma_f Q_{f'} \sigma_{f'} 
\frac{1}{p_f k} \biggl[ 
\frac{2(p_f p_{f'})}{p_f k+p_{f'} k}
-\frac{m_f^2}{p_f k} \biggr]
&=& \sum_{f\neq f'} Q_f \sigma_f Q_{f'} \sigma_{f'} 
\frac{p_{f'} p_f}{(p_{f'} k)(p_f k)}
+\sum_f Q_f^2 \frac{m_f^2}{(p_f k)^2}
\nn\\*[.5em]
&=& \sum_{f,f'} Q_f \sigma_f Q_{f'} \sigma_{f'} 
\frac{p_{f'} p_f}{(p_{f'} k)(p_f k)}.
\eeqar
Note that charge conservation \refeq{eq:Qcons} was used in the form
\beq
\sum_{f' (f'\neq f)} Q_{f'} \sigma_{f'}=-Q_f\sigma_f
\label{eq:Qcons2}
\eeq
in the term proportional to $m_f^2$. The arguments of
$|\M_0|^2$ in $|\M_\sub|^2$ behave in the desired way owing to
\refeq{eq:IRmom}. The asymptotic relation \refeq{eq:m2subasymp} for the
collinear limits follows after inserting the conditions 
\refeq{eq:colcond} into \refeq{eq:m2sub} and using again charge
conservation \refeq{eq:Qcons2}. The correct behaviour of the momenta of 
$|\M_0|^2$ in $|\M_\sub|^2$ is guaranteed by \refeq{eq:colmom}.

\section{Subtraction functions and integrated 
counterparts---special case of light fermions}
\label{se:subfunc0}

Before we turn to the treatment of the general case of massive fermions
in the next section,
we first describe the dipole subtraction formalism for light fermions,
i.e.\ we neglect fermion masses in this section whenever possible.%
\footnote{The dipole subtraction formalism for this important special
case has been worked out independently by M.~Roth \cite{ro99}. Comparing
both approaches, we find full consistency.}
In this way, the structure of the formalism becomes clear without being
obscured by all kind of complications that are related to particle
masses. Moreover, this section provides a condensed instruction to the
formalism for light fermions, since details of the method that are only 
relevant for its derivation are also postponed to the next section.

\subsection{Final-state emitter and final-state spectator}
\label{se:ij0}

We define the auxiliary functions $\gsub_{ij,\tau}$, which correspond 
to a final-state emitter $i$ and a final-state spectator $j$, by
\beqar
\gsub_{ij,+}(p_i,p_j,k) &=&
\frac{1}{(p_i k)(1-y_{ij})} \biggl[
\frac{2}{1-z_{ij}(1-y_{ij})}-1-z_{ij} \biggr],
\nn\\[.5em]
\gsub_{ij,-}(p_i,p_j,k) &=& 0,
\label{eq:gij0}
\eeqar
where the variables $y_{ij}$ and $z_{ij}$ are given by
\beq
y_{ij} = \frac{p_i k}{p_i p_j + p_i k + p_j k}, \qquad
z_{ij} = \frac{p_i p_j}{p_i p_j + p_j k}.
\label{eq:yijzij}
\eeq
Since we assume $m_{i,j}\to 0$, the explicit mass terms in
$\gout_{i,\tau}$ and in $\gsub_{ij,\tau}$ are negligible in the
difference $(|\M_\sub|^2-\sum_{\lambda_\gamma} |\M_1|^2)$. In
particular, this implies that $\gsub_{ij,-}$ vanishes. Note, however,
that those mass terms are relevant in the integration of $\gsub_{ij,\tau}$
over the photonic part of phase space (see next section).
It is straightforward to check that the functions $\gsub_{ij,\tau}$ of
\refeq{eq:gij0} obey the asymptotic
conditions \refeq{eq:IRcond} and \refeq{eq:colcond} in the IR and
collinear limits, in which we get
\beq
y_{ij} \Nlim{k\to 0} 0, \qquad 
z_{ij} \Nlim{k\to 0} 1, \qquad
y_{ij} \Nlim{p_i k\to 0} 0, \qquad 
z_{ij} \Nlim{p_i k\to 0} z_i,
\label{eq:yzijlim}
\eeq
with $z_i$ from \refeq{eq:zx}. 
For the evaluation of $\left|\M_0\left(\tilde\Phi_{0,ij}\right)\right|^2$ 
we have to define the mapping $\tilde\Phi_{0,ij}$ from $\Phi_1$ to
$\Phi_0$. Of course, it is desirable to leave as many momenta unchanged
as possible. Therefore, we redefine only the momenta of $f$ and $f'$,
and leave all other momenta $k_n$ unaffected, $\tilde k_n=k_n$.
The momenta $\tilde p_i$ and $\tilde p_j$ are chosen as
\beq
\tilde p_i^\mu = p_i^\mu + k^\mu - \frac{y_{ij}}{1-y_{ij}} p_j^\mu, 
\qquad
\tilde p_j^\mu = \frac{1}{1-y_{ij}} p_j^\mu.
\label{eq:tmomij0}
\eeq
The on-shell relations $\tilde p_j^2=\tilde p_i^2=0$ and the validity of 
the required asymptotic behaviour \refeq{eq:IRmom} and \refeq{eq:colmom} 
can be checked easily. Moreover, momentum conservation,
\beq
P_{ij}=p_i+p_j+k=\tilde p_i+\tilde p_j,
\eeq
is trivially fulfilled, and $P_{ij}^2\ge 0$ holds for all phase-space points.
The above definitions comprise all ingredients for the evaluation of the
$ij$ contribution to the difference 
$(\sum_{\lambda_\gamma} |\M_1|^2-|\M_\sub|^2)$, which is integrated over
the full phase space numerically.
We recall that this integration can be performed with vanishing photon
and fermion masses.

The construction of the above contribution to the subtraction function
entirely follows the pattern of \citere{ca96} for massless QCD partons. 
The same applies to the other emitter/spectator cases.
Note that we have added a factor of
$1/(1-y_{ij})$ in $\gsub_{ij,+}$ that is not included in the approach
of \citere{ca96}. This factor, which is introduced for 
convenience (see massive case in \refse{se:ij}), 
affects only non-singular contributions. 

The differences between \citere{ca96} and our approach for light
fermions become apparent in the analytical integration
of the subtraction function over the photonic part of phase space, where
both IR and collinear regulators have to be taken into account. The
photonic part of phase space is defined by extracting the phase-space
measure $\rd\tilde\Phi_{0,ij}$, which is spanned by the momenta 
$\tilde p_i$, $\tilde p_j$, and $k_n$, from the full phase-space
measure $\rd\Phi_1$, which is spanned by the momenta $p_i$, $p_j$, $k$, 
and $k_n$. The details of this splitting and of the integration over the
photonic part can be found in \refse{se:ij} and in \refapp{app:PS}.
Denoting the integral of 
$\gsub_{ij,\tau}$ over the photonic phase space by $\Gsub_{ij,\tau}$,
and including an appropriate normalization factor,  
the $ij$ contribution $|\M_{\sub,ij}(\Phi_1)|^2$
to the phase-space integral of the subtraction function reads
\beq
\int\rd\Phi_1\,|\M_{\sub,ij}(\Phi_1)|^2
= -\frac{\alpha}{2\pi} Q_i\sigma_i Q_j\sigma_j  \, 
\int\rd\tilde\Phi_{0,ij}\,
\Gsub_{ij,\tau}(P_{ij}^2) |\M_0(\tilde p_i,\tilde p_j;\tau\kappa_i)|^2,
\label{eq:intij}
\eeq
where $\alpha=e^2/(4\pi)$ is the fine-structure constant. 
The functions $\Gsub_{ij,\tau}$ are explicitly given by
\beq
\Gsub_{ij,+}(P_{ij}^2) = {\cal L}(P_{ij}^2,m_i^2) - \frac{\pi^2}{3} + 1,
\qquad
\Gsub_{ij,-}(P_{ij}^2) = \frac{1}{2},
\label{eq:Gij0}
\eeq
where the singular terms are contained in the function
\beq
{\cal L}(P^2,m^2) =
\ln\biggl(\frac{m^2}{P^2}\biggr)
\ln\biggl(\frac{m_\gamma^2}{P^2}\biggr)
+ \ln\biggl(\frac{m_\gamma^2}{P^2}\biggr)
- \frac{1}{2}\ln^2\biggl(\frac{m^2}{P^2}\biggr)
+ \frac{1}{2}\ln\biggl(\frac{m^2}{P^2}\biggr).
\label{eq:L}
\eeq
As required, only the emitter mass $m_i$ gives rise to logarithmic
singularities, whereas the spectator mass $m_j$ can be set to zero
exactly. The spin-flip part $\Gsub_{ij,-}$ is non-vanishing 
and entirely induced by photons that are emitted collinearly to the
emitter $i$. Since all singular terms are 
factorized into $\Gsub_{ij,+}$, the
masses $m_\gamma$, $m_i$, and $m_j$ can be set to zero everywhere in
\refeq{eq:intij} apart from $\Gsub_{ij,+}$.

\subsection{Final-state emitter and initial-state spectator, and vice
versa}
\label{se:iaai0}

Emitter/spectator pairs from the final/initial state and vice versa,
i.e.\ the cases $ia$ and $ai$, always occur in pairs for a given
process. Since the kinematics is identical in both cases, we treat them
in one go. The corresponding auxiliary functions $\gsub_{ff',\tau}$ are
given by
\beqar
\gsub_{ia,+}(p_i,p_a,k) &=&
\frac{1}{(p_i k)x_{ia}} \biggl[ \frac{2}{2-x_{ia}-z_{ia}}-1-z_{ia} \biggr], 
\nn\\[.5em]
\gsub_{ai,+}(p_a,p_i,k) &=&
\frac{1}{(p_a k)x_{ia}} \biggl[ \frac{2}{2-x_{ia}-z_{ia}}-1-x_{ia} \biggr], 
\nn\\[.5em]
\gsub_{ia,-}(p_i,p_a,k) &=& 
\gsub_{ai,-}(p_a,p_i,k) = 0,
\eeqar
with the variables
\beq
x_{ia} = \frac{p_a p_i + p_a k - p_i k}{p_a p_i + p_a k}, \qquad
z_{ia} = \frac{p_a p_i}{p_a p_i + p_a k}.
\label{eq:xiazia}
\eeq
The desired asymptotic behaviour
\refeq{eq:IRcond} and \refeq{eq:colcond} in the singular limits, which imply
\beqar
&& x_{ia} \Nlim{k\to 0} 1, \qquad 
z_{ia} \Nlim{k\to 0} 1, \qquad
x_{ia} \Nlim{p_i k\to 0} 1, \qquad 
z_{ia} \Nlim{p_i k\to 0} z_i, 
\nn\\[.5em] &&
x_{ia} \Nlim{p_a k\to 0} x_a, \qquad 
z_{ia} \Nlim{p_a k\to 0} 1,
\label{eq:xzialim}
\eeqar
can be verified easily.
The modified momenta $\tilde p_i$ and $\tilde p_a$ of the sets
$\tilde\Phi_{0,ia}=\tilde\Phi_{0,ai}$ are chosen as
\beqar
\tilde p_i^\mu = p_i^\mu+k^\mu-(1-x_{ia})p_a^\mu, \qquad
\tilde p_a^\mu = x_{ia} p_a^\mu,
\label{eq:tpitpa0}
\eeqar
and the remaining momenta $\tilde k_n$ coincide with the corresponding
momenta $k_n$ of $\Phi_1$.
The on-shell relations $\tilde p_i^2=\tilde p_a^2=0$, 
the required asymptotic behaviour \refeq{eq:IRmom} and \refeq{eq:colmom},
as well as momentum conservation,
\beq
P_{ia}=p_i+k-p_a=\tilde p_i-\tilde p_a,
\eeq
can be checked easily. For massless fermions $i$ and $a$ we always have
$P_{ia}^2\le 0$.
This completes the definitions of all quantities
for the evaluation of the $ia$ and $ai$ parts of the difference
$(\sum_{\lambda_\gamma} |\M_1|^2-|\M_\sub|^2)$, which can be performed
numerically with vanishing photon and fermion masses.

The analytical integration of the $ia$ and $ai$ parts of the subtraction 
function is more involved than in the $ij$ case, since the modified
momenta $\tilde p_i$ and $\tilde p_a$ correspond to a new initial state.
In the following, we consider a scattering reaction with
the two incoming light-like momenta $p_a$ and $p_b$. 
Owing to $\tilde p_a= x_{ia} p_a$, the CM frames of $p_a+p_b$ and 
$\tilde p_a+p_b$ are related by a boost along the beam axis. The strength 
of this boost is determined by $x_{ia}$, which is the ratio of the
corresponding squared CM energies $s$ and $\tilde s$,
\beq
\tilde s = 2(\tilde p_a p_b) = 2x_{ia} (p_a p_b) = x_{ia} s.
\eeq
The photonic part of the phase space, which results from the
extraction of the phase-space measure $\rd\tilde\Phi_{0,ia}$ from
the full measure $\rd\Phi_1$, involves an integration over $x_{ia}$.
This integration over $x_{ia}$ cannot be carried out analytically, since
the complete phase space spanned by the new momenta $\tilde\Phi_{0,ia}$
implicitly depends on $x_{ia}$ via the CM energy $\sqrt{\tilde s}$.
Thus, the integral over the photonic part of phase space is written in
terms of a convolution over $x_{ia}$.
Including an appropriate normalization, the contributions to
the phase-space integral of the subtraction function read
\beqar
\lefteqn{\int\rd\Phi_1\,|\M_{\sub,ff'}(\Phi_1)|^2
= -\frac{\alpha}{2\pi} Q_a\sigma_a Q_i\sigma_i } &&
\nn\\*
&& {} \times \int_0^1\rd x\, 
\int\rd\tilde\Phi_{0,ia}(x)\, 
\frac{1}{x} \, 
\cGsub_{ff',\tau}(P_{ia}^2,x)\, 
\left|\M_0\Big(x p_a+P_{ia},x p_a;\tau\kappa_f\Big)\right|^2,
\label{eq:intiaai01}
\eeqar
with $ff'=ia,ai$ for the two different cases. In this convolution $x$
plays the role of $x_{ia}$, and the argument of the phase-space measure
$\rd\tilde\Phi_{0,ia}(x)$ indicates that each value of $x$ determines a
different phase space. The momenta to be inserted in $|\M_0|^2$ are
$\tilde p_i=x p_a+P_{ia}$ and $\tilde p_a=x p_a$, where $p_a$ is fixed,
and $P_{ia}$ varies with the phase-space point in
$\tilde\Phi_{0,ia}(x)$. 

Since the distributions
$\cGsub_{ff',+}(P_{ia}^2,x)$ become IR-singular at the point $x\to
1$, the convolution is not yet suited for a numerical evaluation. A
possible way out is provided by the application of the $[\dots]_+$
prescription to this distribution,
\beq
\int_0^1\rd x\, [f(x)]_+ g(x) = \int_0^1\rd x\, f(x) [g(x)-g(1)].
\eeq
Using this trick, the IR singularities in the endpoint contributions
\beq
\Gsub_{ff',\tau}(P_{ia}^2) = \int_0^1\rd x\,
\cGsub_{ff',\tau}(P_{ia}^2,x)
\eeq
are separated from $\cGsub_{ff',\tau}(P_{ia}^2,x)$, and the convolution
reads
\beqar
\lefteqn{\int\rd\Phi_1\,|\M_{\sub,ff'}(\Phi_1)|^2
= -\frac{\alpha}{2\pi} Q_a\sigma_a Q_i\sigma_i } &&
\nn\\*
&& {} \times \Biggl\{
\int_0^1\rd x\, 
\Biggl[
\int\rd\tilde\Phi_{0,ia}(P_{ia}^2,x)\, \frac{1}{x} \, 
\cGsub_{ff',\tau}(P_{ia}^2,x)\, 
\left|\M_0\Big(x p_a+P_{ia},x p_a;\tau\kappa_f\Big)\right|^2
\nn\\
&& \phantom{ {}\times \Biggl\{ \int_0^1\rd x\, \Biggl[ } {}
-\int\rd\tilde\Phi_{0,ia}(P_{ia}^2,1)\, 
\cGsub_{ff',\tau}(P_{ia}^2,x)\, 
\left|\M_0\Big(p_a+P_{ia},p_a;\tau\kappa_f\Big)\right|^2 \Biggr]
\nn\\
&& \phantom{ {}\times \Biggl\{ }{}
+ \int\rd\tilde\Phi_{0,ia}(P_{ia}^2,1) \, \Gsub_{ff',\tau}(P_{ia}^2) 
\left|\M_0\Big(p_a+P_{ia},p_a;\tau\kappa_f\Big)\right|^2
\Biggr\}.
\label{eq:intiaai02}
\eeqar
Note that we have included $P_{ia}^2$ as additional argument in the
phase-space measure $\rd\tilde\Phi_{0,ia}(P_{ia}^2,x)$, in order to
signalize that we have kept $P_{ia}^2$ fixed during the integration over
$x$ in the calculation of the endpoint contributions
$\Gsub_{ff',\tau}(P_{ia}^2)$. 
In \refeq{eq:intiaai02} all singular
contributions are factorized into $\cGsub_{ff',\tau}$ or $\Gsub_{ff',\tau}$
so that this equation is well-suited for numerical evaluations, and the
photon and fermion masses can be set to zero everywhere apart from 
$\cGsub_{ff',\tau}$ and $\Gsub_{ff',\tau}$.

Finally, we give the explicit form of the distributions
$\cGsub_{ff',\tau}(P_{ia}^2,x)$,
\beqar
\cGsub_{ia,+}(P_{ia}^2,x)  &=&
\frac{1}{1-x}\left[2\ln\biggl(\frac{2-x}{1-x}\biggr)-\frac{3}{2}\right],
\nn\\[.5em]
\cGsub_{ai,+}(P_{ia}^2,x) &=&
P_{ff}(x)\left[\ln\biggl(\frac{|P_{ia}^2|}{m_a^2 x}\biggr)-1\right]
-\frac{2}{1-x}\ln(2-x)+(1+x)\ln(1-x),
\nn\\[.5em]
\cGsub_{ia,-}(P_{ia}^2,x)  &=& 0,
\nn\\[.5em]
\cGsub_{ai,-}(P_{ia}^2,x) &=& 1-x,
\eeqar
and the corresponding endpoint parts $\Gsub_{ff',\tau}(P_{ia}^2)$,
\beqar
\Gsub_{ia,+}(P_{ia}^2) &=&
{\cal L}(|P_{ia}^2|,m_i^2)
- \frac{\pi^2}{2} + 1,
\nn\\[.5em]
\Gsub_{ai,+}(P_{ia}^2) &=&
{\cal L}(|P_{ia}^2|,m_a^2)
+ \frac{\pi^2}{6} - \frac{3}{2},
\nn\\[.5em]
\Gsub_{ia,-}(P_{ia}^2) &=& \Gsub_{ai,-}(P_{ia}^2) = \frac{1}{2},
\eeqar
where ${\cal L}$ is the function defined in \refeq{eq:L}, which contains
the logarithmic singularities.
Only the emitter masses lead to mass singularities, as it
should be. From the general discussion of mass singularities in 
\refse{se:sing}, it is also clear that the singularity of final-state
emitter appears only in the endpoint contribution $\Gsub_{ia,+}$. 
For an initial-state emitter also the distribution $\cGsub_{ai,+}$ contains
a mass-singular part, which is proportional to the splitting
function $P_{ff}(x)$. The spin-flip contributions are regular, 
and the one for a final-state emitter is completely contained in the
endpoint part.

\subsection{Initial-state emitter and initial-state spectator}
\label{se:ab0}

For an emitter $a$ and a spectator $b$ from the initial state we
introduce the variables
\beq
x_{ab} = \frac{p_a p_b-p_a k-p_b k}{p_a p_b}, \qquad
y_{ab} = \frac{p_a k}{p_a p_b}
\label{eq:xabyab}
\eeq
and define the auxiliary functions $\gsub_{ab,\tau}$ by
\beqar
\gsub_{ab,+}(p_a,p_b,k) &=&
\frac{1}{(p_a k)x_{ab}} \biggl[ \frac{2}{1-x_{ab}}-1-x_{ab} \biggr], 
\nn\\[.5em]
\gsub_{ab,-}(p_a,p_b,k) &=& 0.
\eeqar
They posses the required asymptotic behaviour in the singular limits,
which are characterized by
\beq
x_{ab} \Nlim{k\to 0} 1, \qquad 
y_{ab} \Nlim{k\to 0} 0, \qquad
x_{ab} \Nlim{p_a k\to 0} x_a, \qquad 
y_{ab} \Nlim{p_a k\to 0} 0.
\label{eq:xyablim}
\eeq
Following the guideline of \citere{ca96}, the construction of the
modified momenta $\tilde\Phi_{0,ab}$, which are used to evaluate
$\left|\M_0\left(\tilde\Phi_{0,ab}\right)\right|^2$ in \refeq{eq:m2sub},
differs from the previous cases. Instead of changing only the emitter
and spectator momenta, we now keep the spectator momentum $p_b$ fixed
and change all outgoing momenta $k_j$ other than $k$. Note that $k_j$ 
also includes the momenta of neutral outgoing particles, i.e.\ we have
\beq
P_{ab} = p_a + p_b - k = \sum_j k_j.
\eeq
The new momenta 
\beq
\tilde p_a^\mu = x_{ab} p_a^\mu, \qquad
\tilde P_{ab}^\mu = x_{ab} p_a^\mu + p_b^\mu
\label{eq:tpatP0}
\eeq
are chosen in such a way that $\tilde p_a^2=0$ and $\tilde P_{ab}^2=P_{ab}^2$.
In the IR limit they obviously tend to $p_a$ and $P_{ab}$, respectively;
in the collinear limit $\tilde p_a^\mu$ approaches $x_a p_a$ with $x_a$
from \refeq{eq:zx}. The individual momenta $k_j$ are modified by a
Lorentz transformation as follows,
\beq
\tilde k_j^\mu = \Lambda^\mu_{\phantom{\mu}\nu} k_j^\nu
\label{eq:tkj}
\eeq
with
\beq
\Lambda^\mu_{\phantom{\mu}\nu} = g^\mu_{\phantom{\mu}\nu}
-\frac{(P_{ab}+\tilde P_{ab})^\mu(P_{ab}+\tilde P_{ab})_\nu}
{P_{ab}^2+P_{ab}\tilde P_{ab}}
+\frac{2\tilde P_{ab}^\mu P_{ab,\nu}}{P_{ab}^2},
\label{eq:LT}
\eeq
so that the mass-shell relations $\tilde k_j^2=k_j^2$ are retained. The
necessary condition 
$\Lambda^\mu_{\phantom{\mu}\nu}\Lambda_\mu^{\phantom{\mu}\rho}=
g_\nu^{\phantom{\mu}\rho}$
and the relation $\sum_j \tilde k_j=\tilde P_{ab}$
are easily checked by direct calculation using
$\tilde P_{ab}^2=P_{ab}^2$.
The above relations comprise the necessary input for the construction of
the $ab$ part $|\M_{\sub,ab}|^2$ of the subtraction function.

Concerning the analytical integration of the $ab$ contribution to the 
subtraction function over the photonic part of phase space, the
situation is similar to the previous section. The incoming momenta $p_a$
and $p_b$ are modified in an analogous way, namely $p_a$ is scaled down
to $\tilde p_a$ by the variable $x_{ab}$, while $p_b$ is kept fixed. 
The CM frames of $p_a+p_b$ and $\tilde p_a+p_b$ are related by a boost
along the beam axis, and the squared CM energies $s$ and $\tilde s$ are
related by
\beq
\tilde s = 2(\tilde p_a p_b) = 2x_{ab} (p_a p_b) = x_{ab} s.
\eeq
Note also that $P_{ab}^2 =\tilde s=x_{ab} s$, owing to definition 
\refeq{eq:xabyab}. The separation of the photonic part of phase space 
again leads to a convolution over $x=x_{ab}$ for the integrated $ab$
contribution $|\M_{\sub,ab}|^2$ to the subtraction function,
\beqar
\lefteqn{\int\rd\Phi_1\,|\M_{\sub,ab}(\Phi_1)|^2
= -\frac{\alpha}{2\pi} Q_a\sigma_a Q_b\sigma_b } &&
\nn\\*
&& {} \times \int_0^1\rd x\, 
\frac{1}{x} \, \cGsub_{ab,\tau}(s,x)\, 
\int\rd\tilde\Phi_{0,ab}(x)\, 
\left|\M_0\Big(x p_a,\tilde k_n(x);\tau\kappa_a\Big)\right|^2.
\label{eq:intab01}
\eeqar
The argument of the modified momenta $\tilde k_n(x)$ indicates that the new
phase space implicitly depends on $x$. For the numerical evaluation of
the convolution, it is appropriate to separate the IR-singular endpoint 
part
\beq
\Gsub_{ab,\tau}(s) = \int_0^1\rd x\, \cGsub_{ab,\tau}(s,x)
\label{eq:Gab}
\eeq
from the distribution $\cGsub_{ab,\tau}$ with the help of the
$[\dots]_+$ prescription.
The numerically accessible form of the convolution is
\beqar
\lefteqn{\int\rd\Phi_1\,|\M_{\sub,ab}(\Phi_1)|^2
= -\frac{\alpha}{2\pi} Q_a\sigma_a Q_b\sigma_b } &&
\nn\\*
&& {} \times \Biggl\{
\int_0^1\rd x\, 
\cGsub_{ab,\tau}(s,x)\, 
\Biggl[ \frac{1}{x} \, \int\rd\tilde\Phi_{0,ab}(s,x)\, 
\left|\M_0\Big(x p_a,\tilde k_n(x);\tau\kappa_a\Big)\right|^2
\nn\\
&& \phantom{ {}\times \Biggl\{ \int_0^1\rd x\, 
	     \cGsub_{ab,\tau}(s,x)\, \Biggl[ } {}
-\int\rd\tilde\Phi_{0,ab}(s,1)\, 
\left|\M_0\Big(p_a,\tilde k_n(1);\tau\kappa_a\Big)\right|^2 \Biggr]
\nn\\
&& \phantom{ {}\times \Biggl\{ }{}
+ \Gsub_{ab,\tau}(s) 
\int\rd\tilde\Phi_{0,ab}(s,1) \, 
\left|\M_0\Big(p_a,\tilde k_n(1);\tau\kappa_a\Big)\right|^2
\Biggr\},
\label{eq:intab02}
\eeqar
where the distributions are given by
\beq
\cGsub_{ab,+}(s,x) =
P_{ff}(x)\left[\ln\biggl(\frac{s}{m_a^2}\biggr)-1\right], \qquad
\cGsub_{ab,-}(s,x) = 1-x,
\eeq
and the endpoint parts read
\beq
\Gsub_{ab,+}(s) = {\cal L}(s,m_a^2) -\frac{\pi^2}{3} + \frac{3}{2}, \qquad
\Gsub_{ab,-}(s) = \frac{1}{2}.
\eeq
Note that the original squared CM energy $s$ is kept fixed in the
integration over $x$ that defines the endpoint contributions in 
\refeq{eq:Gab}. This is also indicated in the phase-space measure 
$\rd\tilde\Phi_{0,ab}(s,x)$, which is to be parametrized for fixed $s$ and
$x$. The mass singularities of the emitter $a$ are completely factorized
into $\cGsub_{ab,+}$ and $\Gsub_{ab,+}$ so that the convolution over
$x$ can be carried out numerically for vanishing photon and fermion
masses. Of course, the spectator mass is set to zero everywhere.

\section{Subtraction functions and integrated 
counterparts---general case}
\label{se:subfunc}

In this section we turn to the case of arbitrary finite fermion masses.
Here we include also details of the derivation, which have been omitted
in the previous section for brevity. The anticipated results for light
fermions can be obtained from the general ones of this section by
carefully expanding the corresponding formulas for small fermion masses.

Moreover, it is phenomenologically important to consider the case of
light fermions only in the initial state, which is of particular interest 
for $\Pep\Pem$ collisions at high energies, as observed at LEP or the SLC. 
In this case, the dipole formalism is also considerably simpler than for
general fermion masses. The results of the corresponding expansion are
listed in \refapp{app:ma0}.

\subsection{Final-state emitter and final-state spectator}
\label{se:ij}

In order to construct the contribution to $|\M_\sub|^2$ corresponding to
an emitter $i$ and a spectator $j$ from the final state, we have to define 
the auxiliary functions $\gsub_{ij,\tau}(p_i,p_j,k)$ and the embedding of 
the momenta $\tilde p_i$ and $\tilde p_j$ into the phase space spanned by 
$p_i$, $p_j$, and $k$. This embedding has to respect the mass-shell
conditions $\tilde p_i^2=p_i^2=m_i^2$, $\tilde p_j^2=p_j^2=m_j^2$, and
$k^2=m_\gamma^2$, where the photon mass $m_\gamma$ is taken to be
infinitesimal in the final result. In the following we make use of the
variables $y_{ij}$ and $z_{ij}$ of \refeq{eq:yijzij} and introduce the
abbreviations
\beq
P_{ij} = p_i+p_j+k, \qquad \bar P_{ij}^2 = P_{ij}^2-m_i^2-m_j^2-m_\gamma^2,
\qquad \lambda_{ij}=\lambda(P_{ij}^2,m_i^2,m_j^2),
\eeq
where 
\beq
\lambda(x,y,z) = x^2+y^2+z^2-2xy-2xz-2yz.
\eeq
In the physical phase space we always have $P_{ij}^2$, $\bar P_{ij}^2$,
$\lambda_{ij}>0$. For later convenience, we define the auxiliary functions
\beqar
R_{ij}(y) &=& 
\frac{\sqrt{(2m_j^2+\bar P_{ij}^2-\bar P_{ij}^2 y)^2-4P_{ij}^2 m_j^2}}
{\sqrt{\lambda_{ij}}}, 
\nn\\[.5em]
r_{ij}(y) &=& 
1-\frac{2m_j^2(2m_i^2+\bar P_{ij}^2)}{\lambda_{ij}}\frac{y}{1-y}.
\eeqar
Their actual form is convention except for their behaviour near $y=0$,
where they are regular with $R_{ij}(0)=r_{ij}(0)=1$ for $m_\gamma=0$.

Using the above quantities,
we define the functions $\gsub_{ij,\tau}$ as follows:
\beqar
\gsub_{ij,+}(p_i,p_j,k) &=&
\frac{1}{(p_i k)R_{ij}(y_{ij})} \biggl[
\frac{2}{1-z_{ij}(1-y_{ij})}-1-z_{ij}
-\frac{m_i^2}{p_i k} \biggr]
-\gsub_{ij,-}(p_i,p_j,k),
\nn\\[.5em]
\gsub_{ij,-}(p_i,p_j,k) &=&
\frac{m_i^2}{2(p_i k)^2} \, 
\frac{(1-z_{ij})^2}{z_{ij}} \, 
\frac{r_{ij}(y_{ij})}{R_{ij}(y_{ij})}.
\eeqar
It is straightforward to check that these functions obey the asymptotic
conditions \refeq{eq:IRcond} and \refeq{eq:colcond} in the IR and
collinear limits, in which $y_{ij}$ and $z_{ij}$ behave as given in
\refeq{eq:yzijlim}. Note that the limits \refeq{eq:yzijlim}
implicitly assume $m_\gamma\to 0$; the collinear limit additionally
requires $m_i\to 0$. 
In the mapping $\tilde\Phi_{0,ij}$ from $\Phi_1$ to $\Phi_0$ we leave
all momenta $k_n$ other than $p_i$, $p_j$, and $k$ unaffected, as in the
case of light fermions.
The momenta $\tilde p_i$ and $\tilde p_j$ are chosen as
\beqar
\tilde p_j^\mu &=& 
\frac{\sqrt{\lambda_{ij}}}
{\sqrt{\lambda\left((p_i+k)^2,P_{ij}^2,m_j^2\right)}}
\biggl( p_j^\mu-\frac{P_{ij} p_j}{P_{ij}^2}P_{ij}^\mu \biggr)
+\frac{P_{ij}^2+m_j^2-m_i^2}{2P_{ij}^2} P_{ij}^\mu,
\nn\\[.5em]
\tilde p_i^\mu &=& P_{ij}^\mu-\tilde p_j^\mu.
\label{eq:tmomij}
\eeqar
The on-shell relations $\tilde p_j^2=m_j^2$ and $\tilde p_i^2=m_i^2$
directly follow by expanding the squared momenta, and momentum
conservation $P_{ij}=p_i+p_j+k=\tilde p_i+\tilde p_j$ is fulfilled by
definition. Moreover, the validity of the required asymptotic behaviour
\refeq{eq:IRmom} and \refeq{eq:colmom} is obvious.

The relations given above include all ingredients that are necessary to
calculate the $ij$ contribution to the subtraction function 
$|\M_\sub(\Phi_1)|^2$. The phase-space integral of the difference 
$(\sum_{\lambda_\gamma}|\M_1|^2-|\M_\sub|^2)$
is non-singular in the IR limit, and also in the collinear limit, which
occurs for $m_i\to 0$. Therefore, this integral can be
evaluated with $m_\gamma=0$ everywhere, and with $m_f=0$ for light fermions.
However, we need the dependence on $m_\gamma$ and light fermion masses
$m_f$ in the integral of $|\M_\sub|^2$ over the photon phase space,
which is calculated next.

To this end, we separate the phase space spanned by the momenta 
$\tilde p_i$ and $\tilde p_j$ from the one that is spanned by $p_i$, 
$p_j$, and $k$ as follows:
\beq
\int\rd\phi(p_i,p_j,k;P_{ij}) = 
\int\rd\phi(\tilde p_i,\tilde p_j;P_{ij}) 
\int [\rd k(P_{ij}^2,y_{ij},z_{ij})].
\label{eq:PSijsplit}
\eeq
Our aim is to perform the integration of the $ij$ part of the
subtraction function over the remaining phase-space
variables contained in the measure $[\rd k(P_{ij}^2,y_{ij},z_{ij})]$. 
The explicit form of this measure, which is derived in \refapp{app:PS},
is given by
\beq
\int [\rd k(P_{ij}^2,y_{ij},z_{ij})] =
\frac{1}{4(2\pi)^3} \frac{\bar P_{ij}^4}{\sqrt{\lambda_{ij}}}
\int_0^{2\pi}\rd\varphi_{ij} 
\int_{y_1}^{y_2}\rd y_{ij} \, (1-y_{ij}) 
\int_{z_1(y_{ij})}^{z_2(y_{ij})}\rd z_{ij}.
\label{eq:dkij}
\eeq
The angle $\varphi_{ij}$ is the azimuthal angle of ${\bf p}_i$ with
respect to the ${\bf p}_j$ axis in the CM frame of $P_{ij}$. 
The integration boundary for the variables $y_{ij}$ and $z_{ij}$ is
given by
\beqar
y_1 &=& \frac{2m_i m_\gamma}{\bar P_{ij}^2}, \qquad
y_2 = 1-\frac{2m_j\left(\sqrt{P_{ij}^2}-m_j\right)}{\bar P_{ij}^2},
\nn\\[.5em]
z_{1,2}(y_{ij}) &=& 
\frac{ (2m_i^2+\bar P_{ij}^2 y_{ij})(1-y_{ij}) \mp
\sqrt{y_{ij}^2-y_1^2}\,\sqrt{\lambda_{ij}}\, R_{ij}(y_{ij}) }
{2(1-y_{ij})(m_i^2+m_\gamma^2+\bar P_{ij}^2 y_{ij})}.
\label{eq:yijzijbound}
\eeqar
Since the integrand $\gsub_{ij,\tau}(p_i,p_j,k) 
\left|\M_0\left(\tilde\Phi_{0,ij};\tau\kappa_i\right)\right|^2$ 
of the phase-space integral does not depend on the angle 
$\varphi_{ij}$, the integral over $\varphi_{ij}$ simply
yields a trivial factor of $2\pi$. Moreover,
$\left|\M_0\left(\tilde\Phi_{0,ij}\right)\right|^2$ 
is independent of
$y_{ij}$ and $z_{ij}$ so that the integrations over $y_{ij}$ and
$z_{ij}$ only concern the auxiliary functions $\gsub_{ij,\tau}$,
and we define
\beq
\Gsub_{ij,\tau}(P_{ij}^2) 
= \frac{\bar P_{ij}^4}{2\sqrt{\lambda_{ij}}}
\int_{y_1}^{y_2}\rd y_{ij} \, (1-y_{ij}) 
\int_{z_1(y_{ij})}^{z_2(y_{ij})}\rd z_{ij} \,
\gsub_{ij,\tau}(p_i,p_j,k).
\label{eq:Gijdef}
\eeq
While the integration over $z_{ij}$ is very simple, the one over
$y_{ij}$ is non-trivial, but can be performed analytically. Details of
the calculation can be found in \refapp{app:int}. We obtain
\beqar
\Gsub_{ij,+}(P_{ij}^2) &=&
\ln\left(\frac{m_\gamma^2 a_3^3}{m_i^2}\right)
-2\ln(1-a_3^2) + \frac{a_3^2}{2} + \frac{3}{2}
+\frac{\bar P_{ij}^2}{\sqrt{\lambda_{ij}}} 
\left[
\ln(a_1)\ln\left(\frac{m_\gamma^2 m_j^2}{\lambda_{ij}a_2}\right)
\right.
\nn\\ && {}
\left.
+2\Li_2(a_1)+4\Li_2\left(-\sqrt{\frac{a_2}{a_1}}\right)
-4\Li_2(-\sqrt{a_1 a_2})
+\frac{1}{2}\ln^2(a_1)-\frac{\pi^2}{3}
\right]
\nn\\ && {}
-\Gsub_{ij,-}(P_{ij}^2),
\nn\\[.5em]
\Gsub_{ij,-}(P_{ij}^2) &=&
\frac{\bar P_{ij}^4}{\lambda_{ij}} \left\{
\frac{2m_i^2}{\bar P_{ij}^2}\ln(a_3)
+\frac{4m_i^2 m_j^2}{\bar P_{ij}^4}
\ln\left(\frac{1+a_3^2}{2a_3}\right)
+\frac{m_i^2 m_j^2}{(\bar P_{ij}^2+m_i^2)^2} 
\ln\left(\frac{2m_i m_j a_3}{\bar P_{ij}^2} \right)
\right.
\nn\\ && {}
+\frac{m_i m_j(\bar P_{ij}^2+2m_i^2)}{\bar P_{ij}^4} \left[
4{\mathrm{arctan}}(a_3)-\pi\right]
\nn\\ && {}
\left.
+(1-a_3^2)
\left[\frac{1}{2}+\frac{m_i^2 P_{ij}^2}{\bar P_{ij}^2(\bar P_{ij}^2+m_i^2)}
+\frac{2m_i^3m_j}{a_3\bar P_{ij}^4} \right]
\right\},
\label{eq:Gij}
\eeqar
with the shorthands
\beq
a_1 = \frac{\bar P_{ij}^2+2m_i^2-\sqrt{\lambda_{ij}}}
{\bar P_{ij}^2+2m_i^2+\sqrt{\lambda_{ij}}}, \qquad
a_2 = \frac{\bar P_{ij}^2-\sqrt{\lambda_{ij}}}
{\bar P_{ij}^2+\sqrt{\lambda_{ij}}}, \qquad
a_3 = \frac{m_i}{\sqrt{P_{ij}^2}-m_j},
\label{eq:a123}
\eeq
and the usual dilogarithm $\Li_2(x)=-\int_0^1\rd t\ln(1-xt)/t$.
The contribution of the $ij$ part $|\M_{\sub,ij}(\Phi_1)|^2$ to the
phase-space integral of the subtraction function is formally the same as
given in \refeq{eq:intij} for light fermions, where we have to
insert the functions $\Gsub_{ij,\tau}$ of \refeq{eq:Gij} for finite 
fermion masses. Expanding these functions for $m_{i,j}\to 0$, we obtain
the results of \refeq{eq:Gij0}.

\subsection{Final-state emitter and initial-state spectator}
\label{se:ia}

For a final-state emitter $i$ and an initial-state spectator $a$ we keep
the definitions \refeq{eq:xiazia} of the variables $x_{ia}$, $z_{ia}$ 
and introduce the abbreviations
\beq
P_{ia} = p_i + k - p_a, \qquad
\bar P_{ia}^2 = P_{ia}^2-m_a^2-m_i^2-m_\gamma^2, \qquad
\lambda_{ia} = \lambda(P_{ia}^2,m_a^2,m_i^2).
\label{eq:Pia}
\eeq
In the following, we only consider fixed momenta $P_{ia}$ that obey
\beq
P_{ia}^2 < (m_a-m_i)^2,
\label{eq:Pia2limit}
\eeq
because other values of $P_{ia}^2$ do not admit the limits $k\to 0$ or
$p_i k\to 0$, as can be checked easily. In particular, 
\refeq{eq:Pia2limit} ensures that $\bar P_{ia}^2<0$.
Moreover, we introduce an auxiliary parameter $x_0$ with 
$0\le x_0<1$ that specifies the lower limit on $x_{ia}$ for which the 
subtraction function will be applied. 
We are forced to deviate from the simple choice $x_0=0$, because
$x_{ia}\to 0$ is not allowed for all configurations of $P_{ia}^2$,
$m_a$, and $m_i$. More precisely, one has to require
\beq
x_0 > \hat x = \frac{-\bar P_{ia}^2}{2m_a\left(m_a-\sqrt{P_{ia}^2}\right)}
\qquad
\mbox{if} \quad 0 < \sqrt{P_{ia}^2} < m_a-m_i,
\label{eq:x0lim}
\eeq
which is only possible for $P_{ia}^2>0$ and $m_a>m_i$.
Otherwise we can take any value for $x_0$ with $0\le x_0<1$. For
instance, it is possible to set $x_0=0$ for vanishing fermion masses,
as done in \refse{se:iaai0}.
For $P_{ia}^2>(m_a-m_i)^2$ or $x_{ia}<x_0$ the
subtraction functions $\gsub_{ia,\tau}$ are set to zero consistently. 
Note that both IR and collinear singularities appear at
$x_{ia}\to 1$, i.e.\ applying the subtraction function for
$x_0<x_{ia}<1$ correctly cancels these singularities.
The final results on observables must not depend on $x_0$, which will
not be further specified in the following. Checking the $x_0$
independence of observables is a non-trivial check on the complete 
subtraction prodecure.

It is convenient to introduce the auxiliary functions
\beqar
R_{ia}(x) &=& \frac{\sqrt{(\bar P_{ia}^2+2m_a^2 x)^2-4m_a^2 P_{ia}^2 x^2}}
{\sqrt{\lambda_{ia}}},
\nn\\[.5em]
r_{ia}(x) &=& 1+\frac{\bar P_{ia}^2(\bar P_{ia}^2+2m_a^2)}{\lambda_{ia}}
\frac{1-x}{x},
\label{eq:Ria}
\eeqar
with $R_{ia}(1)=r_{ia}(1)=1$ for vanishing photon mass $m_\gamma$.
The subtraction functions for finite fermion masses read
\beqar
\gsub_{ia,+}(p_i,p_a,k) &=&
\frac{1}{(p_i k)x_{ia}} \biggl[ \frac{2}{2-x_{ia}-z_{ia}}-1-z_{ia} 
-\frac{m_i^2}{p_i k} \biggr]
-\gsub_{ia,-}(p_i,p_a,k),
\nn\\[.5em]
\gsub_{ia,-}(p_i,p_a,k) &=&
\frac{m_i^2}{2(p_i k)^2}\, 
\frac{(1-z_{ia})^2}{z_{ia}}\, 
\frac{r_{ia}(x_{ia})}{x_{ia}}.
\eeqar
They possess the desired asymptotic behaviour 
\refeq{eq:IRcond} and \refeq{eq:colcond} in the singular limits. The
behaviour of $x_{ia}$ and $z_{ia}$ for $k\to 0$ or $p_i k\to 0$ is given
in \refeq{eq:xzialim}.
The auxiliary momenta $\tilde\Phi_{0,ia}$, which are needed to
evaluate $\left|\M_0\left(\tilde\Phi_{0,ia}\right)\right|^2$, are
constructed in a similar way as for light fermions.
The emitter and spectator momenta are given by
\beqar
\tilde p_i^\mu &=& 
\frac{\sqrt{\lambda_{ia}}}
{\sqrt{\lambda\Big((p_i+k)^2,P_{ia}^2,m_a^2\Big)}}
\left(p_a^\mu-\frac{P_{ia}p_a}{P_{ia}^2}P_{ia}^\mu\right)
+\frac{P_{ia}^2-m_a^2+m_i^2}{2P_{ia}^2}P_{ia}^\mu,
\nn\\[.5em]
\tilde p_a^\mu &=& \tilde p_i^\mu - P_{ia}^\mu,
\label{eq:tpitpa}
\eeqar
and the remaining momenta $\tilde k_n$ coincide with the corresponding
momenta $k_n$ of $\Phi_1$. The on-shell relations 
$\tilde p_i^2=p_i^2=m_i^2$ and  $\tilde p_a^2=p_a^2=m_a^2$
can be verified easily. Momentum
conservation $P_{ia} = p_i + k - p_a = \tilde p_i - \tilde p_a$ and the
validity of the required asymptotics \refeq{eq:IRmom} and \refeq{eq:colmom} 
are obvious.

Using the above relations, the $ia$ part of $|\M_\sub(\Phi_1)|^2$ can 
be evaluated. Concerning the role 
of the masses $m_\gamma$ and $m_f$ in the difference 
$(\sum_{\lambda_\gamma}|\M_1|^2-|\M_\sub|^2)$,
the remarks of the previous section apply as well.

As already explained in \refse{se:iaai0}, the photonic part of phase
space, which is obtained by separating $\rd\tilde\Phi_{0,ia}$ from
$\rd\Phi_1$, involves an integration over the variable $x_{ia}$. 
This integration over $x_{ia}$ turns into a convolution over $x=x_{ia}$
in the integration of the subtraction function, where $x$ determines the
CM energy of the reduced phase space spanned by the momenta
$\tilde\Phi_{0,ia}$. For the explicit splitting of phase space, we
decompose the momentum $P_{ia}=p_b-K_{ia}$ into the second incoming momentum 
$p_b$ and the total momentum $K_{ia}$ of all outgoing particles other than $i$
and the inspected photon.
The phase-space separation is defined by
\beqar
\lefteqn{
\hspace*{-1em} 
\int\rd\phi(p_i,k,K_{ia};p_a+p_b) \theta(x_{ia}-x_0) } &&
\nn\\[.5em] 
&=& \int_{x_0}^{x_1}\rd x
\int\rd\phi\Big(\tilde p_i(x),K_{ia};\tilde p_a(x)+p_b\Big) 
\int [\rd k(P_{ia}^2,x,z_{ia})],
\label{eq:PSiasplit}
\eeqar
with $x_1$ given below. The $x$-dependent momenta 
\beqar
\tilde p_a^\mu(x) &=& \frac{1}{R_{ia}(x)}
\left(x p_a^\mu+\frac{\bar P_{ia}^2+2m_a^2 x}{2P_{ia}^2}P_{ia}^\mu\right)
-\frac{P_{ia}^2+m_a^2-m_i^2}{2P_{ia}^2}P_{ia}^\mu,
\nn\\[.5em]
\tilde p_i^\mu(x) &=& \tilde p_a^\mu(x)+P_{ia}^\mu
\label{eq:tmomiax}
\eeqar
result from $\tilde p_a$ and $\tilde p_i$ upon substituting
$(p_a P_{ia})\to-m_a^2-\bar P_{ia}^2/(2x)$, which eliminates $(p_a P_{ia})$
in favour of $x$ and $P_{ia}^2$. Note that the substitution also concerns
$(p_i+k)^2=(p_a+P_{ia})^2$ in \refeq{eq:tpitpa}. 
If we now try to reconstruct 
$\tilde p_a^\mu(x)$ from $x$ and $p_a$, as it was possible for $m_a=0$
in \refse{se:iaai0}, we find that the knowledge of $x$ and $p_a$ is not
sufficient for $m_a\neq 0$.
This complication is due to the fact that the boost relating the CM frames 
of $p_a+p_b$ and $\tilde p_a+p_b$ does not simply go along the beam axis 
for finite $m_a$. We will come back to this problem and to its solution
at the end of this section and proceed by performing the integral of the
contribution to the subtraction function over $[\rd k(P_{ia}^2,x,z_{ia})]$.
In \refapp{app:PS} the explicit form of this measure is derived. The
result is
\beq
\int [\rd k(P_{ia}^2,x,z_{ia})] =
\frac{1}{4(2\pi)^3}
\frac{\bar P_{ia}^4}{\sqrt{\lambda_{ia}} R_{ia}(x)} \,
\frac{\rho_{ia}(\tilde s)}{x^2} \,
\int_{z_1(x)}^{z_2(x)}\rd z_{ia} \int_0^{2\pi}\rd\varphi_\gamma,
\label{eq:dkia}
\eeq
where $\varphi_\gamma$ is the azimuthal angle of the photon in the CM
frame of $p_i+k$. The function $\rho_{ia}(\tilde s)$ reads
\beq
\rho_{ia}(\tilde s) = 
\sqrt{\frac{\lambda(\tilde s,m_a^2,m_b^2)}{\lambda(s,m_a^2,m_b^2)}},
\eeq
where $s$ and $\tilde s$ denote the squared CM energies of $p_a+p_b$ and
$\tilde p_a+p_b$, respectively.
The integration boundary for $z_{ia}$ is given by
\beq
z_{1,2}(x) = 
\frac{\bar P_{ia}^2[\bar P_{ia}^2-x(\bar P_{ia}^2+2m_i^2)] \mp
\sqrt{\bar P_{ia}^4(1-x)^2-4m_i^2 m_\gamma^2 x^2}\,
\sqrt{\lambda_{ia}}\, R_{ia}(x) }
{2\bar P_{ia}^2[\bar P_{ia}^2-x(P_{ia}^2-m_a^2)]}.
\label{eq:z12}
\eeq
From this relation we read off that the maximal value of $x$ is given by
\beq
x_1 = \frac{\bar P_{ia}^2}{\bar P_{ia}^2-2m_i m_\gamma} = 
1-\frac{2m_i m_\gamma}{|\bar P_{ia}^2|} + \O(m_\gamma^2).
\eeq
\begin{sloppypar}
\noindent
While the integration of $\gsub_{ia,\tau}(p_i,p_a,k)
\left|\M_0\left(\tilde\Phi_{0,ia};\tau\kappa_i\right)\right|^2$ 
over $\varphi_\gamma$ yields a trivial factor of $2\pi$, the integration
over $z_{ia}$ depends on the actual form of $\gsub_{ia,\tau}$. Defining
\beq
\cGsub_{ff',\tau}(P_{ia}^2,x_{ia}) 
= \frac{\bar P_{ia}^4}{2\sqrt{\lambda_{ia}} R_{ia}(x_{ia})}
\int_{z_1(x_{ia})}^{z_2(x_{ia})}\rd z_{ia} \, \gsub_{ff',\tau}(p_f,p_{f'},k)
\label{eq:cGsubiaaidef}
\eeq
for $ff'=ia$, we obtain
\beqar
\cGsub_{ia,+}(P_{ia}^2,x) &=&
-\frac{\bar P_{ia}^2}{\sqrt{\lambda_{ia}}} \,
\frac{1}{R_{ia}(x)(1-x)} \, \left\{
2\ln\left[\frac{2-x-z_1(x)}{2-x-z_2(x)}\right]
\right.
\nn\\*
&& \left. \hspace*{1em} {}
+[z_1(x)-z_2(x)] \left[
1+\frac{z_1(x)+z_2(x)}{2}
-\frac{2m_i^2 x}{\bar P_{ia}^2(1-x)} \right] \right\}
-\cGsub_{ia,-}(P_{ia}^2,x),
\nn\\[.5em]
\cGsub_{ia,-}(P_{ia}^2,x) &=&
\frac{m_i^2}{\sqrt{\lambda_{ia}}}
\frac{x r_{ia}(x)}{(1-x)^2 R_{ia}(x)} \,
\left\{ \ln\left[\frac{z_2(x)}{z_1(x)}\right]
\right.
\nn\\*
&& \left. \hspace*{1em} {}
+[z_1(x)-z_2(x)]\left[2-\frac{z_1(x)+z_2(x)}{2}\right] \right\}
\label{eq:cGsubia}
\eeqar
after a simple integration. The function $\cGsub_{ia,+}$ is singular in
the limit $x\to 1$ for $m_\gamma=0$. Since our aim is to perform the
convolution over $x$ numerically with $m_\gamma=0$, we separate the
singularity at $x\to 1$ by introducing the $[\dots]_+$ prescription.
Considering that our lower integration limit $x_0$ can be different from
zero, we write
\beq
\cGsub_{ff',\tau}(P_{ia}^2,x) = 
\left[ \cGsub_{ff',\tau}(P_{ia}^2,x) \theta(x-x_0) \right]_+
+\Gsub_{ff',\tau}(P_{ia}^2,x_0)\delta(1-x)
\label{eq:Gplus}
\eeq
with
\beq
\Gsub_{ff',\tau}(P_{ia}^2,x_0) = \int_{x_0}^{x_1}\rd x\,
\cGsub_{ff',\tau}(P_{ia}^2,x),
\label{eq:Gsubiaaidef}
\eeq
where $ff'=ia$.
In writing \refeq{eq:Gplus}, we have already used that the photon mass
$m_\gamma$ can be set to zero, and thus $x_1$ set to 1, everywhere apart
from $\Gsub_{ia,+}$, which contains the IR singularity.
The integration over the distribution 
$\left[ \cGsub_{ia,\tau}(P_{ia}^2,x) \theta(x-x_0) \right]_+$
can be performed with $m_\gamma=0$ and, if desired, with $m_f=0$ for the
fermion masses. The endpoint contributions
$\Gsub_{ia,\tau}(P_{ia}^2,x_0)$ are obtained after
performing the non-trivial integration over $x$. Details of this
integration can be found in \refapp{app:int}.
The results are
\beqar
\Gsub_{ia,+}(P_{ia}^2,x_0) &=&
2\ln\left(\frac{m_\gamma m_i}{|\bar P_{ia}^2|(1-x_0)}\right)
+\frac{\bar P_{ia}^4}{2(\bar P_{ia}^2+m_i^2)^2}
\ln\left[\frac{m_i^2 x_0-\bar P_{ia}^2(1-x_0)}{m_i^2}\right]
\nn\\ &&{}
-\frac{\bar P_{ia}^4(1-x_0)}{2(\bar P_{ia}^2+m_i^2)
[\bar P_{ia}^2(1-x_0)-m_i^2 x_0]} + 2
\nn\\ &&{}
+\frac{\bar P_{ia}^2}{\sqrt{\lambda_{ia}}} \left\{
2\ln(b_1)\ln\left[\frac{b_0\sqrt{\lambda_{ia}(m_a^2+m_i^2-\bar P_{ia}^2)}}
{m_\gamma m_a^2}\right]
-\frac{1}{2}\ln^2(b_1)+\frac{\pi^2}{3}
\right.
\nn\\ && \hspace*{4em} \left. {}
+2\sum_{k=1}^5 (-1)^k \Li_2(b_k) 
    \vphantom{\left[\frac{\sqrt{\bar\lambda_{ia}}}{m_a^2}\right]}
\right\}
-\Gsub_{ia,-}(P_{ia}^2,x_0),
\nn\\[.5em]
\Gsub_{ia,-}(P_{ia}^2,x_0) &=&
\frac{2m_a m_i(\bar P_{ia}^2+2m_i^2)}{\lambda_{ia}}
{\mathrm{arctan}}\left[\frac{m_a}{m_i}(1-x_0)\right]
-\frac{m_i^2}{\sqrt{\lambda_{ia}}} \frac{R_{ia}(x_0)}{1-x_0} \ln(b_6)
\nn\\ &&{}
+\frac{m_i^2 \bar P_{ia}^2}{\lambda_{ia}} \left\{
\ln(x_0)
-\left[1+\frac{m_a^2 \bar P_{ia}^2}{(\bar P_{ia}^2+m_i^2)^2}\right]
\ln\left[x_0-\frac{\bar P_{ia}^2}{m_i^2}(1-x_0)\right]
\right.
\nn\\ && \hspace*{4.5em} {}
+\frac{2m_a^2}{\bar P_{ia}^2}\ln\left[1+\frac{m_a^2}{m_i^2}(1-x_0)^2\right]
-\frac{x_0[\bar P_{ia}^2-2m_a^2(1-x_0)]}{2[\bar P_{ia}^2(1-x_0)-m_i^2 x_0]}
\nn\\ && \hspace*{4.5em} \left. {}
-\frac{4m_a^2}{\bar P_{ia}^2}
+\frac{m_a^2(1-x_0)}{\bar P_{ia}^2+m_i^2}
\right\}
+\frac{\bar P_{ia}^4}{2\lambda_{ia}},
\label{eq:Gsubia}
\eeqar
with the abbreviations
\beqar
b_0 &=& \frac{-4m_a^2 \bar P_{ia}^2(1-x_0)}
{\lambda_{ia}[1+R_{ia}(x_0)]^2+4m_a^2(\bar P_{ia}^2+m_i^2)(1-x_0)^2},
\qquad
b_1 = \frac{2m_i^2-\bar P_{ia}^2-\sqrt{\lambda_{ia}}}
{2m_i^2-\bar P_{ia}^2+\sqrt{\lambda_{ia}}},
\nn\\[.5em]
b_{2,3} &=& \frac{2m_i^2+\bar P_{ia}^2\pm\sqrt{\lambda_{ia}}}
{-\bar P_{ia}^2\pm\sqrt{\lambda_{ia}}} \, b_0,
\qquad
b_{4,5} = \frac{2(m_i^2-m_a^2\mp\sqrt{\lambda_{ia}})}
{2m_a^2-\bar P_{ia}^2\mp\sqrt{\lambda_{ia}}} \, b_0,
\nn\\[.5em]
b_6 &=& 
\frac{2m_i^2 x_0-\bar P_{ia}^2(1-x_0)+\sqrt{\lambda_{ia}}R_{ia}(x_0)(1-x_0)}
{2m_i^2 x_0-\bar P_{ia}^2(1-x_0)-\sqrt{\lambda_{ia}}R_{ia}(x_0)(1-x_0)}.
\label{eq:bi}
\eeqar
We have checked that all arguments of the logarithms and dilogarithms in 
\refeq{eq:Gsubia} lie on the first Riemann sheet of the corresponding 
function for the allowed regions of the various parameters.
Although the spin-flip contribution $\cGsub_{ia,-}(P_{ia}^2,x)$ is
not IR-singular at $x\to 1$, we have nevertheless introduced the
$[\dots]_+$ distribution. This is advantageous for a large momentum
transfer, since in the limit $m_{a,i}\to 0$ the complete contribution of 
$\cGsub_{ia,-}(P_{ia}^2,x)$ is contained in the endpoint part 
$\Gsub_{ia,-}(P_{ia}^2,x_0)$.
If the limit $x_0$ can be set to zero, many terms in the endpoint
contributions $\Gsub_{ia,\tau}(P_{ia}^2,x_0)$ simplify.
In order to facilitate the application of our results, we list the
corresponding results
$\Gsub_{ia,\tau}(P_{ia}^2)=\Gsub_{ia,\tau}(P_{ia}^2,0)$ in
\refapp{app:x0} explicitly.
\end{sloppypar}

The final result for the $ia$ contribution to the phase-space integral 
of the subtraction function reads
\beqar
\lefteqn{\int\rd\Phi_1\,|\M_{\sub,ff'}(\Phi_1)|^2
= -\frac{\alpha}{2\pi} Q_a\sigma_a Q_i\sigma_i } &&
\nn\\*
&& {} \times \Biggl\{
\int_{x_0}^1\rd x\, 
\Biggl[
\int\rd\tilde\Phi_{0,ia}(K_{ia}^2,P_{ia}^2,x)\, 
\frac{\rho_{ia}(\tilde s)}{x^2} \, 
\cGsub_{ff',\tau}(P_{ia}^2,x)\, 
\left|\M_0\Big(\tilde p_a(x),\tilde p_i(x);\tau\kappa_f\Big)\right|^2
\nn\\
&& \phantom{ {}\times \Biggl\{ \int_0^1\rd x\, \Biggl[ } {}
-\int\rd\tilde\Phi_{0,ia}(K_{ia}^2,P_{ia}^2,1)\, 
\cGsub_{ff',\tau}(P_{ia}^2,x)\, 
\left|\M_0\Big(\tilde p_a(1),\tilde p_i(1);\tau\kappa_f\Big)\right|^2 \Biggr]
\nn\\
&& \phantom{ {}\times \Biggl\{ }{}
+ \int\rd\tilde\Phi_{0,ia}(K_{ia}^2,P_{ia}^2,1) \, 
\Gsub_{ff',\tau}(P_{ia}^2,x_0) 
\left|\M_0\Big(\tilde p_a(1),\tilde p_i(1);\tau\kappa_f\Big)\right|^2
\Biggr\},
\label{eq:intia}
\eeqar
with $ff'=ia$.
As already mentioned above, the phase-space integration over the
$x$-dependent momenta has to be performed carefully. First, one has to
determine the squared CM energy $\tilde s=(\tilde p_a(x)+p_b)^2$ of the
new initial state. The needed scalar product $p_a(x)p_b$ is obtained
upon contracting the first equation of \refeq{eq:tmomiax}
with $p_{b,\mu}$. The product $P_{ia}p_b$, which appears on the r.h.s.,
can be replaced by $P_{ia}p_b = (m_b^2+P_{ia}^2-K_{ia}^2)/2$,
according to the definition of the outgoing momentum $K_{ia}$. 
In summary, we obtain $\tilde s$ as a function of $x$, $P_{ia}^2$, and
$K_{ia}^2$,
\beqar
\tilde s &=& m_a^2+m_b^2 + \frac{1}{R_{ia}(x)}
\left[x (s-m_a^2-m_b^2)
+\frac{(\bar P_{ia}^2+2m_a^2 x)(m_b^2+P_{ia}^2-K_{ia}^2)}{2P_{ia}^2}
\right]
\nn\\[.5em] 
&& {}
-\frac{(P_{ia}^2+m_a^2-m_i^2)(m_b^2+P_{ia}^2-K_{ia}^2)}{2P_{ia}^2}.
\label{eq:tildesiaai}
\eeqar
Thus, it is necessary first to fix $x$, $P_{ia}^2$, and $K_{ia}^2$ in
the phase-space integration over 
$\rd\tilde\Phi_{0,ia}(K_{ia}^2,P_{ia}^2,x)$, before the other
phase-space-variables can be parametrized, which is indicated by the
arguments of $\rd\tilde\Phi_{0,ia}$. 

\subsection{Initial-state emitter and final-state spectator}
\label{se:ai}

The case of an initial-state emitter $a$ and a final-state spectator $i$
is kinematically identical with the previous one, where the roles played by
$a$ and $i$ are interchanged. Therefore, we can make use of the
variables and auxiliary quantities $x_{ia}$, $z_{ia}$, $P_{ia}$, etc.\ of
the previous section and adopt the same restrictions on $P_{ia}^2$ and
$x_{ia}$. 
For finite fermion masses the auxiliary functions $\gsub_{ai,\tau}$ read
\beqar
\gsub_{ai,+}(p_a,p_i,k) &=&
\frac{1}{(p_a k)x_{ia}} \biggl[ \frac{2}{2-x_{ia}-z_{ia}}
-R_{ia}(x_{ia})(1+x_{ia})
-\frac{x_{ia} m_a^2}{p_a k} \biggr]
-\gsub_{ai,-}(p_a,p_i,k),
\nn\\[.5em]
\gsub_{ai,-}(p_a,p_i,k) &=&
\frac{m_a^2}{2(p_a k)^2} \, \frac{(1-x_{ia})^2}{x_{ia}}. 
\label{eq:gai}
\eeqar
In the IR and collinear limits the functions \refeq{eq:gai}
behave as required in \refeq{eq:IRcond} and \refeq{eq:colcond}.
The behaviour of $x_{ia}$ and $z_{ia}$ in those limits is given in
\refeq{eq:xzialim}.
The auxiliary momenta $\tilde p_a$ and $\tilde p_i$ are constructed as
specified in \refeq{eq:tpitpa}, completing the 
construction prescription for the subtraction
contribution $|\M_{\sub,ai}|^2$.

The separation of the photon phase space also proceeds along the same
lines as in the previous section, leading to the same kind of
convolution over $x$.
In contrast to the previous case, in which the singularities appeared
for $x\to 1$, the collinear singularity ($p_a k\to 0$) is not restricted to a
single point in $x$, as can be seen in \refeq{eq:xzialim}.
Therefore, it is necessary to choose the lower limit $x_0$ for $x$ small 
enough so that the complete range of $x=x_{ia}$ is 
covered for small $m_a$; otherwise collinear singularities in 
$(\sum_{\lambda_\gamma}|\M_1|^2-|\M_\sub|^2)$
remain uncancelled. Since negative values of $x_{ia}$ (if they occur at all) 
can never lead to collinear singularities, our initial
restriction $x_0\ge 0$ is consistent.
If effects of $\O(m_a)$ are consistently neglected, one can simply take
$x_0=0$, as already done in \refse{se:subfunc0}.

The analytical integration of the subtraction function over the photonic 
phase space is performed as in the previous section. Hence, we define
$\cGsub_{ai,\tau}$ according to \refeq{eq:cGsubiaaidef} with $ff'=ai$
and carry out the simple integration over $z_{ia}$, yielding
\beqar
\cGsub_{ai,+}(P_{ia}^2,x) &=&
-\frac{\bar P_{ia}^2}{\sqrt{\lambda_{ia}}} \,
\frac{1}{R_{ia}(x)} \, \left\{
\frac{2}{1-x}\ln\left(\frac{[1-z_1(x)][2-x-z_2(x)]}
{[1-z_2(x)][2-x-z_1(x)]}\right)
\right.
\nn\\*
&& \left. \hspace*{1em} {}
+R_{ia}(x)(1+x)\ln\left[\frac{1-z_2(x)}{1-z_1(x)}\right]
+\frac{2m_a^2 x^2}{\bar P_{ia}^2}
\left[\frac{1}{1-z_2(x)}-\frac{1}{1-z_1(x)}\right] \right\}
\nn\\ && {}
-\cGsub_{ai,-}(P_{ia}^2,x),
\nn\\[.5em]
\cGsub_{ai,-}(P_{ia}^2,x) &=& 1-x.
\label{eq:cGsubai}
\eeqar
For $m_\gamma=0$, the function $\cGsub_{ai,+}(P_{ia}^2,x)$ becomes
singular at $x\to 1$. This singularity is split off by introducing
the $[\dots]_+$ distribution as specified in \refeq{eq:Gplus} and 
\refeq{eq:Gsubiaaidef} with $ff'=ai$, thereby defining the endpoint 
contributions $\Gsub_{ai,\tau}(P_{ia}^2,x_0)$.
Actually, this splitting is not needed for
$\cGsub_{ai,-}(P_{ia}^2,x)$, which is a simple regular function; we
proceed this way in order to keep the generic description of the
polarized and unpolarized cases.
The integration over $x$ can be performed analytically, yielding
\beqar
\lefteqn{ \hspace*{-1em} 
\Gsub_{ai,+}(P_{ia}^2,x_0) =
2\ln\left[\frac{m_\gamma m_i}{|\bar P_{ia}^2|(1-x_0)}\right]
+\frac{\bar P_{ia}^4[1-R_{ia}(x_0)]}{4m_a^2(\bar P_{ia}^2+m_i^2)}
+2(1-x_0) } &&
\nn\\ \hspace*{1em} && {}
+\frac{\bar P_{ia}^4(3\bar P_{ia}^2+2m_i^2)}
{2\sqrt{\lambda_{ia}}(\bar P_{ia}^2+m_i^2)^2} \left\{ 
\frac{1}{\gamma}\ln\left[
\frac{\bar P_{ia}^2 \gamma^2+2m_a^2+\gamma\sqrt{\lambda_{ia}}}
{\bar P_{ia}^2 \gamma^2+2m_a^2 x_0+\gamma\sqrt{\lambda_{ia}}R_{ia}(x_0)} 
\right] \right.
\nn\\ && \hspace*{4em} \left. {}
+\ln\left[x_0-\frac{\bar P_{ia}^2}{m_i^2}(1-x_0)\right]
+\ln\left[\frac{\bar P_{ia}^2(1-2\gamma^2)-2m_a^2+\sqrt{\lambda_{ia}}}
{\bar P_{ia}^2(1-2\gamma^2)-2m_a^2 x_0+\sqrt{\lambda_{ia}}R_{ia}(x_0)}\right]
\right\}
\nn\\ && {}
+\frac{\bar P_{ia}^2}{\sqrt{\lambda_{ia}}} \left\{
2\ln\left(\frac{m_\gamma m_a}{b_0\sqrt{\lambda_{ia}}}\right)
\ln\left(\frac{c_1}{c_0}\right)
-\ln\left(\frac{m_a^2+m_i^2-\bar P_{ia}^2}{m_a^2}\right)\ln(c_1)
+\frac{1}{2}\ln(c_0 c_1)\ln\left(\frac{c_1}{c_0}\right)
\right.
\nn\\ && \hspace*{4em}\left. {}
+\frac{x_0}{2}(2+x_0)
\ln\left(\frac{\bar P_{ia}^2+\sqrt{\lambda_{ia}}R_{ia}(x_0)}
{\bar P_{ia}^2-\sqrt{\lambda_{ia}}R_{ia}(x_0)}\right)
-\frac{1}{2}\ln(c_0)
-2\sum_{k=0}^5 (-1)^k \Li_2(c_k)
\right\}
\nn\\ && {}
-\Gsub_{ai,-}(P_{ia}^2,x_0),
\nn\\[.5em]
\lefteqn{ \hspace*{-1em} \Gsub_{ai,-}(P_{ia}^2,x_0) = 
\frac{1}{2}(1-x_0)^2, } &&
\eeqar
with the shorthands
\beqar
c_0 &=& \frac{\bar P_{ia}^2+\sqrt{\lambda_{ia}}}
{\bar P_{ia}^2-\sqrt{\lambda_{ia}}}, \qquad
c_1 = b_1, \qquad
c_{2,3} = -\frac{2m_a^2+\bar P_{ia}^2 \mp \sqrt{\lambda_{ia}}}{2m_a^2}b_0,
\qquad c_{4,5} = b_{4,5}, 
\nn\\[.5em]
\gamma &=& \frac{m_a}{\sqrt{-\bar P_{ia}^2-m_i^2+\ri\epsilon}}.
\eeqar
The variables $b_{0,1,4,5}$ are defined in \refeq{eq:bi}. Although the
variable $\gamma$ can become imaginary for some values of $P_{ia}^2$,
the result for $\Gsub_{ia,+}$ is always real and does not depend
on the sign of the infinitesimal imaginary part $\ri\epsilon$.
The endpoint contributions 
$\Gsub_{ai,\tau}(P_{ia}^2)=\Gsub_{ai,\tau}(P_{ia}^2,0)$ 
for the simpler value $x_0=0$ are explicitly 
listed in \refapp{app:x0}.

\begin{sloppypar}
The result for the $ai$ contribution to the integrated the subtraction 
function takes the same form as in the $ia$ case, but now we have to
identify $ff'=ai$ in \refeq{eq:intia}.
Concerning the phase-space integration over 
$\rd\tilde\Phi_{0,ia}(K_{ia}^2,P_{ia}^2,x)$,
the remarks made at the end of the previous section apply as well.
The IR singularity is contained in the endpoint part 
$\Gsub_{ai,+}(P_{ia}^2,x_0)$. However, the collinear singularity also 
appears in the function $\cGsub_{ai,+}(P_{ia}^2,x_0)$. Since all singular 
terms are factorized, the convolution over $x$ itself can be carried out with 
$m_\gamma=0$, and with $m_f=0$ for light fermions. 
\end{sloppypar}

\subsection{Initial-state emitter and initial-state spectator}

For an emitter $a$ and a spectator $b$ we keep the definition 
\refeq{eq:xabyab} of the variables $x_{ab}$ and $y_{ab}$. Moreover, we
introduce the abbreviations
\beq
P_{ab} = p_a+p_b-k, \qquad
s = (p_a+p_b)^2, \qquad \bar s = s-m_a^2-m_b^2, \qquad
\lambda_{ab}=\lambda(s,m_a^2,m_b^2)
\eeq
and the auxiliary function
\beq
R_{ab}(x) = \frac{\sqrt{(\bar s x+m_\gamma^2)^2-4m_a^2 m_b^2}}
{\sqrt{\lambda_{ab}}}.
\eeq
The function is regular at $x\to 1$ with $R_{ab}(1)=1$ for $m_\gamma=0$.
For the contribution to the subtraction function we define
\beqar
\gsub_{ab,+}(p_a,p_b,k) &=&
\frac{1}{(p_a k)x_{ab}} \biggl[ \frac{2}{1-x_{ab}}-1-x_{ab}
-\frac{x_{ab} m_a^2}{p_a k} \biggr]
-\gsub_{ab,-}(p_a,p_b,k),
\nn\\[.5em]
\gsub_{ab,-}(p_a,p_b,k) &=&
\frac{m_a^2}{2(p_a k)^2} \, \frac{(1-x_{ab})^2}{x_{ab}}. 
\label{eq:gab}
\eeqar
These functions posses the required asymptotic behaviour in the 
singular limits, which are characterized by \refeq{eq:xyablim}.
The functions $\gsub_{ab,\tau}$ are set to zero for $x_{ab}<x_0$, where
the kinematical lower bound
\beq
x_0 \ge \hat x = \frac{2m_a m_b-m_\gamma^2}{\bar s}
\eeq
has to be respected.
For $x_{ab}$ smaller than $\hat x$, collinear photon emission cannot
occur, and the following construction of new momenta would break down.
Note that
$x_0$ can be set to zero in either limit of $m_a\to 0$ or $m_b\to 0$, in
consistency with our treatment of light fermions above.

As already explained in the case of light fermions in \refse{se:ab0},
the spectator momentum $p_b$ is kept fixed, but the emitter momentum
$p_a$ and the momenta $k_j$ of all outgoing particles other than the
photon are changed,
resulting in a modification of the total momentum
$P_{ab} = \sum_j k_j$. We map the momenta $p_a$ and $P_{ab}$ to
\beqar
\tilde p_a^\mu &=& 
\frac{\sqrt{\lambda\Big(P_{ab}^2,m_a^2,m_b^2\Big)}}{\sqrt{\lambda_{ab}}}
\left(p_a^\mu-\frac{p_a p_b}{m_b^2}p_b^\mu\right)
+\frac{P_{ab}^2-m_a^2-m_b^2}{2m_b^2}p_b^\mu,
\nn\\[.5em]
\tilde P_{ab}^\mu &=& \tilde p_a^\mu + p_b^\mu,
\label{eq:tpatP}
\eeqar
leading to the mass-shell relations $\tilde p_a^2=m_a^2$ and 
$\tilde P_{ab}^2=P_{ab}^2$. These relations and the validity of the
required asymptotics in the IR and collinear limits can be checked easily. 
The individual momenta $k_j$ are modified by a Lorentz transformation in
the same way as for light fermions, i.e.\ we have 
$\tilde k_j^\mu = \Lambda^\mu_{\phantom{\mu}\nu} k_j^\nu$ as defined in
\refeq{eq:tkj}. The actual form \refeq{eq:LT} of the transformation
$\Lambda^\mu_{\phantom{\mu}\nu}$ remains valid, but the momenta $P_{ab}$
and $\tilde P_{ab}$ of this section have to be inserted.
This completes the necessary input for the construction of
the differential subtraction contribution $|\M_{\sub,ab}|^2$.

The separation of the photon phase space is again written in terms of a
convolution over an auxiliary parameter $x$,
\beq
\int\rd\phi(k,P_{ab};p_a+p_b) \, \theta(x-x_0) = 
\int_{x_0}^{x_1}\rd x
\int\rd\phi\Big(\tilde P_{ab}(x);\tilde p_a(x)+p_b\Big) 
\int [\rd k(s,x,y_{ab})].
\label{eq:PSabsplit}
\eeq
The $x$-dependent momenta 
\beqar
\tilde p_a^\mu(x) &=& 
R_{ab}(x)\left(p_a^\mu-\frac{\bar s}{2m_b^2}p_b^\mu\right)
+\frac{\bar s x+m_\gamma^2}{2m_b^2}p_b^\mu,
\nn\\[.5em]
\tilde P_{ab}^\mu(x) &=& \tilde p_a^\mu(x) + p_b^\mu
\label{eq:tpatPab}
\eeqar
are obtained from $\tilde p_a$ and $\tilde P_{ab}$ upon replacing 
$P_{ab}^2$ by $(\bar s x+m_a^2+m_b^2+m_\gamma^2)$. Thus, they 
coincide with $\tilde p_a$ and $\tilde P_{ab}$ at $x=x_{ab}$,
\beq
\tilde p_a^\mu(x_{ab}) = \tilde p_a^\mu, \qquad
\tilde P_{ab}^\mu(x_{ab}) = \tilde P_{ab}^\mu.
\eeq
Note that $\tilde p_a^2(x)=m_a^2$ even for $x\neq x_{ab}$.
The measure $[\rd k(s,x,y_{ab})]$ for the photon phase space is
derived in \refapp{app:PS}. The result is
\beq
\int [\rd k(s,x,y_{ab})] =
\frac{1}{4(2\pi)^3}
\frac{\bar s^2}{\sqrt{\lambda_{ab}}}
\int_{y_1(x)}^{y_2(x)}\rd y_{ab} \int_0^{2\pi}\rd\varphi_\gamma,
\label{eq:dkab}
\eeq
where $\varphi_\gamma$ is the azimuthal angle of the photon in the CM
frame. The integration boundary for $y_{ab}$ is given by
\beqar
y_{1,2}(x) &=& \frac{\bar s+2m_a^2}{2s}(1-x) \mp
\frac{\sqrt{\lambda_{ab}}}{2s} 
\sqrt{(1-x)^2-\frac{4m_\gamma^2 s}{\bar s^2}}.
\label{eq:y12}
\eeqar
This relation provides us also with the maximal value $x_1$ of $x$, for
which $x=x_{ab}$ is possible,
\beq
x_1 = 1-\frac{2m_\gamma \sqrt{s}}{\bar s}.
\eeq
Integrating $\gsub_{ab,\tau}(p_a,p_b,k)
\left|\M_0\left(\tilde\Phi_{0,ab};\tau\kappa_a\right)\right|^2$ 
over $\varphi_\gamma$ results in a trivial factor of $2\pi$. 
The integration of $\gsub_{ab,\tau}$ over $y_{ab}$ 
is also performed easily. Defining
\beq
\cGsub_{ab,\tau}(s,x) 
= \frac{x\bar s^2}{2\sqrt{\lambda_{ab}}}
\int_{y_1(x)}^{y_2(x)}\rd y_{ab} \, \gsub_{ab,\tau}(p_a,p_b,k),
\eeq
we obtain
\beqar
\cGsub_{ab,+}(s,x) &=&
\frac{\bar s}{\sqrt{\lambda_{ab}}} \, \left\{
\frac{1+x^2}{1-x} \ln\left[\frac{y_2(x)}{y_1(x)}\right]
+\frac{2m_a^2 x}{\bar s} \left[
\frac{1}{y_2(x)}-\frac{1}{y_1(x)}\right] \right\}
-\cGsub_{ab,-}(s,x),
\nn\\[.5em]
\cGsub_{ab,-}(s,x) &=& 1-x.
\label{eq:cGsubab}
\eeqar
The singularity at $x\to 1$, which appears for $m_\gamma=0$, is split
off by introducing the $[\dots]_+$ distribution.
Integration over $x$ yields the endpoint contributions
\beqar
\Gsub_{ab,+}(s,x_0) &=&
\ln\left[\frac{m_\gamma^2 s}{\bar s^2(1-x_0)^2}\right] + 2(1-x_0)
+ \frac{\bar s}{\sqrt{\lambda_{ab}}} \left\{ 
\ln\left[\frac{m_\gamma^2\lambda_{ab}}
{m_a^2 \bar s^2(1-x_0)^2}\right] \ln(d_1)
\right.
\nn\\ && \left. \hspace*{1em} {}
+\left[ \frac{1}{2}(1-2x_0-x_0^2)-\frac{2m_a^2}{\bar s} \right]\ln(d_1)
+2\Li_2(d_1)+\frac{1}{2}\ln^2(d_1)-\frac{\pi^2}{3} \right\}
\nn\\ && {}
-\Gsub_{ab,-}(s,x_0),
\nn\\[.5em]
\Gsub_{ab,-}(s,x_0) &=& \frac{1}{2}(1-x_0)^2,
\eeqar
where
\beq
d_1 = \frac{\bar s+2m_a^2-\sqrt{\lambda_{ab}}}
{\bar s+2m_a^2+\sqrt{\lambda_{ab}}}.
\eeq

The integral of the $ab$ part of the subtraction function finally reads
\beqar
\lefteqn{\int\rd\Phi_1\,|\M_{\sub,ab}(\Phi_1)|^2 =
-\frac{\alpha}{2\pi} Q_a\sigma_a Q_b\sigma_b } &&
\nn\\*
&& {} \times \, \Biggl\{ \int_{x_0}^1\rd x\, \cGsub_{ab,\tau}(s,x) 
\Biggl[ \frac{1}{x} \, \int\rd\tilde\Phi_{0,ab}(s,x)\, 
\left|\M_0\Big(\tilde p_a(x),\tilde k_j(x);\tau\kappa_a\Big)\right|^2
\nn\\
&& \phantom{ {} \times \,
	    \Biggl\{ \int_{x_0}^1\rd x\, \cGsub_{ab,\tau}(s,x) \Biggl[} {}
-\int\rd\tilde\Phi_{0,ab}(s,1)\,
\left|\M_0\Big(\tilde p_a(1),\tilde k_j(1);\tau\kappa_a\Big)\right|^2 \Biggr]
\nn\\
&& \phantom{ {} \times \, \Biggl\{ } {}
+\Gsub_{ab,\tau}(s,x_0) \int\rd\tilde\Phi_{0,ab}(s,1) \, 
\left|\M_0\Big(\tilde p_a(1),\tilde k_j(1);\tau\kappa_a\Big)\right|^2
\Biggr\},
\label{eq:intab}
\eeqar
where the IR singularity is contained in the
function $\Gsub_{ab,+}(s,x_0)$, and the mass singularity, which appears
in $\cGsub_{ab,+}(s,x_0)$ as well, is factorized 
(see \refse{se:subfunc0}). The convolution over $x$ itself can be carried 
out with $m_\gamma=0$, and with $m_f=0$ for light fermions.
The relation between the momenta $\tilde\Phi_{0,ab}(x)$ and the variable
$x$ is much simpler than in the case with an emitter or a spectator in
the final state. Contracting the first equation of \refeq{eq:tpatPab}
with $p_{b,\mu}$ and cancelling some terms, yields
\beq
\tilde s = P_{ab}^2 =\bar s x+m_a^2+m_b^2+{\cal O}(m_\gamma^2),
\eeq
i.e.\ the CM energy $\sqrt{\tilde s}$ of $\tilde\Phi_{0,ab}(s,x)$ is
completely determined by the original CM energy $\sqrt{s}$ and $x$.
Knowing $\tilde s$ from $s$ and $x$, it is straightforward to
parametrize $\rd\tilde\Phi_{0,ab}(s,x)$.

\section{Applications}
\label{se:appl}

In this section, we compare some numerical results on QED corrections
obtained by the phase-space slicing method with the ones of the
subtraction formalism described in this paper. 
In this context, we mention that we have
adjusted all phase-space parametrizations to the peaking structure of
the integrand in the application of the slicing method. For instance,
$\ln(E_\gamma)$ is used as integration variable, in order to flatten the
IR pole $1/E_\gamma$ for small values of the energy $E_\gamma$ of the
outgoing photon. Photon emission angles are treated in a similar way if 
collinear photon emission from light fermions can take place.
These reparametrizations have improved the efficiency of the slicing method
considerably, whereas such improvements are not necessary for the 
subtraction method. 

We consider the sample processes $\gamma\gamma\to f\bar f(\gamma)$, 
$\Pem\gamma\to\Pem\gamma(\gamma)$, and 
$\mu^+\mu^-\to\nu_\Pe\bar\nu_\Pe(\gamma)$. This choice provides separate
applications for the cases $ij$, $ia+ai$, and $ab$ of 
emitter/spectator pairs $ff'$.

For the numerical evaluations we take the following set of parameters
\cite{ca98}:
\beq
\begin{array}[b]{rlrlrl}
\alpha &= 1/137.0359895, & & & & \\
\MW &= 80.41\GeV, \qquad &
\MZ &= 91.187\GeV, \qquad & \Gamma_\PZ &= 2.49\GeV,
\\
\Me &= 0.51099907\MeV, \qquad &
m_\mu &= 105.658389\MeV, \qquad &
\Mt &= 173.8\GeV.
\end{array}
\eeq
The weak mixing angle $\theta_\Pw$ is fixed by 
\beq
\cos\theta_\Pw = \cw = \frac{\MW}{\MZ}, \qquad 
\sw = \sqrt{1-\cw^2}.
\eeq
The fermionic couplings to the Z~boson are expressed in terms of the
vector and axial-vector factors
\beq
v_f = \frac{I^3_{\Pw,f}}{2\cw\sw}-\frac{\sw}{\cw}Q_f, \qquad
a_f = \frac{I^3_{\Pw,f}}{2\cw\sw},
\eeq
where $I^3_{\Pw,f}=\pm 1/2$ is the third component of the weak
isospin of the fermion $f$.

\subsection{\boldmath{The processes $\gamma\gamma\to f\bar f(\gamma)$}}

The QED and weak corrections to the production of light
fermion--anti-fermion pairs have been discussed recently in
\citere{de98}. Details about different variants of phase-space slicing 
and about the dipole formalism presented here can also be found there. 
In particular, the treatment of angular cuts in the phase-space integral
is described for the dipole formalism.
Actually, the subtraction functions of \citere{de98} and the
ones given in this paper differ by a non-singular factor, leading to a
different constant contribution in the integrated counterparts. 
We have repeated the numerics of \citere{de98} for the functions defined
in this paper and found results of the same quality. Using the same number of
phase-space points in the Monte Carlo integration, which is performed by
{\sl Vegas} \cite{vegas}, the integration error of the results obtained
by phase-space slicing are larger by factors of 10--20.

Here we focus on the case of massive fermions; specifically, we consider
the process
\beq
\gamma(k_1,\lambda_1) + \gamma(k_2,\lambda_2) \; \longrightarrow \;
\Pt(p,\sigma) + \bar\Pt(\bar p,\bar\sigma) 
\; [+ \gamma(k,\lambda)],
\label{eq:aatt}
\eeq
where $k_{1,2}$, $p$, $\bar p$ are the particle momenta, and
$\lambda_{1,2}$, $\sigma$, $\bar\sigma$ are the corresponding
helicities.
There are two emitter/spectator pairs, $\Pt\bar\Pt$ and $\bar\Pt\Pt$,
both of type $ij$, and the subtraction function \refeq{eq:m2sub} is
given by
\beqar
|\M_\sub(p,\bar p,k;\sigma,\bar\sigma)|^2 &=& \phantom{{}+{}}
Q_\Pt^2 e^2 \gsub_{\Pt\bar\Pt,\tau}(p,\bar p,k) 
\left|\M_0(\tilde p_1,\tilde{\bar p}_1;\tau\sigma,\bar\sigma)\right|^2
\nn\\ && {}
+ Q_\Pt^2 e^2 \gsub_{\bar\Pt\Pt,\tau}(\bar p,p,k) 
\left|\M_0(\tilde p_2,\tilde{\bar p}_2;\sigma,\tau\bar\sigma)\right|^2.
\label{eq:aattsub}
\eeqar
The construction of the auxiliary functions $\gsub_{ij,\tau}$ with 
$ij=\Pt\bar\Pt,\bar\Pt\Pt$ proceeds as described in \refse{se:ij}. In
particular, the invariant masses $P_{ij}$ are given by the square of
the CM energy $\sqrt{s}$,
\beq
P_{\Pt\bar\Pt}^2=P_{\bar\Pt\Pt}^2=(p+\bar p+k)^2=(k_1+k_2)^2=s.
\eeq
The pairs of auxiliary momenta $(\tilde p_l,\tilde{\bar p}_l)$ with
$l=1,2$ are obtained from \refeq{eq:tmomij} upon setting 
$p_i=p$, $p_j=\bar p$ and $p_i=\bar p$, $p_j=p$, respectively.
We recall that the spatial parts of the spectator momenta and their
corresponding auxiliary momenta have the same direction in the CM frame,
i.e.\ $\bar{\bf p} \| \tilde{\bar {\bf p}}_1$ and 
${\bf p} \| \tilde{\bf p}_2$. This fact is useful for the implementation
of angular cuts (see \citere{de98}).
The integrated counterpart to the differential subtraction function
\refeq{eq:aattsub} reads
\beqar
\lefteqn{ \hspace*{-1em}
\int\rd\phi(p,\bar p,k;k_1+k_2)\,
|\M_{\sub}(p,\bar p,k;\sigma,\bar\sigma)|^2} &&
\nn\\
&=& \frac{Q_\Pt^2\alpha}{2\pi} \, 
\Gsub_{\Pt\bar\Pt,\tau}(s) \biggl[ \phantom{{}+{}}
\int\rd\phi(\tilde p_1,\tilde{\bar p}_1;k_1+k_2)\, 
|\M_0(\tilde p_1,\tilde{\bar p}_1;\tau\sigma,\bar\sigma)|^2
\nn\\ && 
\phantom{\frac{Q_\Pt^2\alpha}{2\pi} \,
	 \Gsub_{\Pt\bar\Pt,\tau}(s) \biggl[ } {} 
+ \int\rd\phi(\tilde p_2,\tilde{\bar p}_2;k_1+k_2)\,
|\M_0(\tilde p_2,\tilde{\bar p}_2;\sigma,\tau\bar\sigma)|^2 \biggr],
\eeqar
where we have exploited that the auxiliary functions
$\Gsub_{\Pt\bar\Pt,\tau}(s)$ and $\Gsub_{\bar\Pt\Pt,\tau}(s)$ 
are already fixed by the initial state and coincide.

For the numerical evaluation of the matrix element $\M_1$ of the
radiative process $\gamma\gamma\to\Pt\bar\Pt\gamma$, we apply crossing
relations to the result on the related reaction 
$f\bar f\to\gamma\gamma\gamma$, which is listed in \citere{di99}.
The phase-space integration is performed by {\sl Vegas} \cite{vegas}.
Finally, we combine the real-photonic corrections with the virtual
photonic corrections, the evaluation of which is described in
\citere{de95}. 
\begin{table}
\begin{center}
\begin{tabular}{|c||c|c||c|l@{$\,\pm\,$}l|}
\hline
$\sqrt{s}/\GeV$ & $\sigma_0/\pb$ & Method & $\Delta E/E$ & 
\multicolumn{2}{c|}{$\delta_\QED/\%$} 
\\ \hline\hline
360 & 0.351422 & Phase-space slicing 
            & $10^{-2}$ & 1.73343 & 0.00002
\\ \cline{4-6}
          &&& $10^{-4}$ & 1.72936 & 0.00011
\\ \cline{4-6}
          &&& $10^{-6}$ & 1.72932 & 0.00020
\\ \cline{3-6}
&& Dipole formalism &-- & 1.72931 & 0.00001
\\ \hline\hline
500 & 0.869434 & Phase-space slicing 
            & $10^{-2}$ & 0.33592 & 0.00066
\\ \cline{4-6}
          &&& $10^{-4}$ & 0.3309 & 0.0017
\\ \cline{4-6}
          &&& $10^{-6}$ & 0.3307 & 0.0026
\\ \cline{3-6}
&& Dipole formalism &-- & 0.33043 & 0.00013
\\ \hline\hline
1000 & 0.428565 & Phase-space slicing 
            & $10^{-2}$ & 0.1881  & 0.0059
\\ \cline{4-6}
          &&& $10^{-4}$ & 0.184   & 0.013
\\ \cline{4-6}
          &&& $10^{-6}$ & 0.191   & 0.021
\\ \cline{3-6}
&& Dipole formalism &-- & 0.17431 & 0.00059
\\ \hline\hline
2000 & 0.154450 & Phase-space slicing 
            & $10^{-2}$ & 0.366 & 0.020 
\\ \cline{4-6}
          &&& $10^{-4}$ & 0.346 & 0.045
\\ \cline{4-6}
          &&& $10^{-6}$ & 0.362 & 0.071
\\ \cline{3-6}
&& Dipole formalism &-- & 0.3498 & 0.0021
\\ \hline
\end{tabular}
\end{center}
\caption{Results on the QED correction $\delta_{\QED}$
to the unpolarized total cross section of
$\gamma\gamma\to\Pt\bar\Pt(\gamma)$.}
\label{tab:aatt}
\end{table}
The resulting QED correction $\delta_\QED$ to the total
unpolarized cross section is given in \refta{tab:aatt} for some CM
energies $\sqrt{s}$.
As expected, the statistical error of the result of phase-space slicing
grows roughly proportional to $\ln(\Delta E/E)$, where $E=\sqrt{s}/2$ is
the photon beam energy in the CM frame. It is obvious that the value
$\Delta E/E=10^{-2}$ is still not small enough to guarantee reliable
results. For smaller values of $\Delta E$ the integration error is
again larger by a factor of 10--20 than the corresponding error obtained by
the application of the dipole subtraction method.

\subsection{\boldmath{The process $\Pem\gamma\to\Pem\gamma(\gamma)$}}
\label{se:eaea}

\paragraph{(i) Moderate scattering energies}

We consider the Compton process
\beq
\Pem(p,\sigma) + \gamma(k_\gamma,\lambda) \; \longrightarrow \;
\Pem(p',\sigma') + \gamma(k'_1,\lambda'_1) \; [+ \gamma(k'_2,\lambda'_2)],
\label{eq:eaeaproc}
\eeq
where the momenta and helicities are given in parentheses. Owing to the
strong polarization dependence of its polarized cross sections, this
process is well-suited to determine the degrees of beam polarization of
$\Pe^\pm$ beams. For incoming laser photons and $\Pe^\pm$ beams 
in the energy region of $1\GeV$ to $1\TeV$, the CM energy is in the MeV
range, i.e.\ the CM energy is not large with respect to the electron mass.
Details of precision calculations, which include the photonic corrections of
$\O(\alpha)$, for such Compton polarimeters can be found in 
\citeres{ve89,de99}. In the following we make use of the
analytical results on the virtual corrections and on the amplitudes for
real-photonic bremsstrahlung given in \citere{de99} and evaluate the
real corrections with the dipole formalism. 

The subtraction function receives contributions of the mixed emitter/spectator 
types $ia$ and $ai$. Denoting the incoming and outgoing electrons in 
\refeq{eq:eaeaproc} by $\Pe$ and $\Pe'$, respectively, these
contributions are labelled by $\Pe'\Pe$ and $\Pe\Pe'$. Since both
outgoing photons can become soft, we have to introduce subtraction
functions for each individual final-state photon. Note the IR regions
of the two photons are separated in phase space so that the two subtraction
functions can simply be added. Thus, the full subtraction function reads
\beq
|\M_\sub|^2 =
\sum_{l=1,2} |\M^{(l)}_\sub(p,p',k'_l;\sigma,\sigma')|^2
\label{eq:eaeasub1}
\eeq
with
\beqar
|\M^{(l)}_\sub(p,p',k'_l;\sigma,\sigma')|^2 &=& \phantom{{}+{}}
e^2 \gsub_{\Pe'\Pe,\tau}(p',p,k'_l) 
\left|\M_0(\tilde p_l,\tilde p'_l;\sigma,\tau\sigma')\right|^2
\nn\\ && {}
+ e^2 \gsub_{\Pe\Pe',\tau}(p,p',k'_l) 
\left|\M_0(\tilde p_l,\tilde p'_l;\tau\sigma,\sigma')\right|^2.
\label{eq:eaeasub2}
\eeqar
The functions $\gsub_{\Pe'\Pe,\tau}$ and $\gsub_{\Pe\Pe',\tau}$ are
defined in \refses{se:ia} and \ref{se:ai}, where we have to identify
$p_a=p$, $p_i=p'$, and $k=k'_l$. The auxiliary electron 
momenta $\tilde p_l$ and $\tilde p'_l$ play the roles of $\tilde p_a$
and $\tilde p_i$ in \refeq{eq:tpitpa}, respectively, where the index $l$
refers to the inserted photon momentum $k=k'_l$. The subtraction
function is completely fixed by the above identifications. Because of
$m_a=m_i=\Me$, we can take $x_0=0$ as the lower limit on $x_{ia}$ 
[see \refeq{eq:x0lim}].

The integrated counterpart to the subtraction function receives 
contributions from convolutions of the form \refeq{eq:intia}. Owing to
Bose symmetry with respect to the interchange of the outgoing photons,
the two contributions corresponding to the two photons are equal. Therefore,
we calculate only the integrated subtraction contribution for the photon
with momentum $k'_1$ and weight this contribution with a factor of 2. 
Let us first consider the phase-space integration in the convolution.
The squares of the momenta $P_{ia}$ and $K_{ia}$ are given by
\beq
P_{ia}^2 = (p'+k'_1-p)^2 = (k_\gamma-k'_2)^2=\tilde t, \qquad
K_{ia}^2 = (k_\gamma-P_{ia})^2 = k^{\prime 2}_2 = 0.
\eeq
Inserting these quantities and $m_b=0$ into \refeq{eq:tildesiaai},
we obtain 
\beq
\tilde s = \Me^2-\frac{\tilde t}{2}
+\frac{2x s +\tilde t-2\Me^2}{2R_{\Pe'\Pe}(x)}
\label{eq:tildeseaea}
\eeq
for the new squared CM energy used in the convolution over $x$.
The phase-space measure $\rd\tilde\Phi_{0,ia}(K_{ia}^2,P_{ia}^2,x)$ 
reads
\beq
\int\rd\tilde\Phi_{0,\Pe'\Pe}(0,\tilde t,x) =
\frac{1}{4(2\pi)^2} \int_{\tilde t_{\mathrm{min}}(x)}^0\rd\tilde t \,
\int_0^{2\pi}\rd\tilde\varphi'_2 \,
\frac{1}{\tilde s-\Me^2}.
\eeq
The lower limit $\tilde t_{\mathrm{min}}(x)$ on $\tilde t$ is determined
by two kinematical conditions. Firstly, $\tilde t$ cannot be lower than
$-4E_\gamma^2$, where $E_\gamma$ is the energy of the incoming photon in
the original CM frame. This condition corresponds to the ``edge'' of 
phase space where $k'_1\to 0$. Secondly, the requirement $\tilde s>\Me^2$ 
sets another lower limit on $\tilde t$ in the calulation of 
$\tilde s$ from \refeq{eq:tildeseaea} for fixed $x$. Hence, 
$\tilde t_{\mathrm{min}}(x)$ is the maximum of these two limits.
The integration over the
azimuthal angle $\tilde\varphi'_2$ in the CM frame of 
$\tilde p_1(x)+k_\gamma$ yields a factor of $2\pi$ owing to the
rotational invariance of the integrand. In the integrand of the
convolution we insert the distributions $\cGsub_{\Pe'\Pe,\tau}$ and 
$\cGsub_{\Pe\Pe',\tau}$ defined in \refses{se:ia} and \ref{se:ai}, 
respectively. The endpoint contributions $\Gsub_{\Pe'\Pe,\tau}$ and 
$\Gsub_{\Pe\Pe',\tau}$ for $x_0=0$ are taken from \refapp{app:x0}.
The auxiliary function $\rho_{ia}(\tilde s)$ is given by
\beq
\rho_{ia}(\tilde s) = \frac{\tilde s-\Me^2}{s-\Me^2}.
\eeq
For the squared Born amplitudes
$\left|\M_0\Big(\tilde p_a(x),\tilde p_i(x)\Big)\right|^2$
the invariants $\tilde s$ and $\tilde t$ correspond to the Mandelstam 
variables $s$ and $t$ as defined in \citere{de99}, respectively.

\begin{table}
\begin{center}
\begin{tabular}{|c||c|c||c|l@{$\,\pm\,$}l|}
\hline
$P_\Pe$, $P_\gamma$ & $\sigma_0/\mb$ & Method & $\Delta E/E_\gamma$ & 
\multicolumn{2}{c|}{$\delta_\QED/\%$} 
\\ \hline\hline
$+$, $+$ & 110.946 & Phase-space slicing 
            & $10^{-2}$ & 0.4094 & 0.0014
\\ \cline{4-6}
          &&& $10^{-4}$ & 0.4016 & 0.0030
\\ \cline{4-6}
          &&& $10^{-6}$ & 0.4014 & 0.0047
\\ \cline{3-6}
&& Dipole formalism &-- & 0.40131 & 0.00033
\\ \hline\hline
$+$, $-$ & 65.2608 & Phase-space slicing 
            & $10^{-2}$ & 0.4996 & 0.0016
\\ \cline{4-6}
          &&& $10^{-4}$ & 0.4921 & 0.0035
\\ \cline{4-6}
          &&& $10^{-6}$ & 0.4898 & 0.0053
\\ \cline{3-6}
&& Dipole formalism &-- & 0.49699 & 0.00092
\\ \hline
\end{tabular}
\end{center}
\caption{Results on the ${\cal O}(\alpha)$ QED correction $\delta_{\QED}$
to the total Born cross section $\sigma_0$ 
of $\Pem\gamma\to\Pem\gamma(\gamma)$ for
$\sqrt{s}=2.21836\MeV$ and different degrees of beam
polarization $P_\Pe$ and~$P_\gamma$.}
\label{tab:eaea}
\vspace*{1em}
\begin{center}
\begin{tabular}{|c||c|c|c|c||l@{$\,\pm\,$}l|}
\hline
$P_\Pe$, $P_\gamma$ & $\sigma_0/\pb$ & Method & $\Delta E/E$ & 
$\Delta\theta/\mathrm{rad}$ &
\multicolumn{2}{c|}{$\delta_\QED/\%$} 
\\ \hline\hline
$+$, $+$ & 90.4372 & IR slicing and
            & $10^{-2}$ & -- & 5.441 & 0.016
\\ \cline{4-7}
         && effective mass factor & $10^{-4}$ & -- & 5.416 & 0.031
\\ \cline{4-7}
          &&& $10^{-6}$ & -- & 5.468 & 0.047
\\ \cline{3-7}
&& Phase-space slicing 
            & $10^{-2}$ & $10^{-2}$ & 5.3783 & 0.0074
\\ \cline{4-7}
          &&& $10^{-4}$ & $10^{-4}$ & 5.385 & 0.026
\\ \cline{4-7}
          &&& $10^{-6}$ & $10^{-6}$ & 5.454 & 0.055
\\ \cline{3-7}
&& Dipole formalism &-- & -- & 5.3588 & 0.0041
\\ \hline\hline
$+$, $-$ & 12.2425 & IR slicing and
            & $10^{-2}$ & -- & 15.686 & 0.015
\\ \cline{4-7}
         && effective mass factor & $10^{-4}$ & -- & 15.685 & 0.026
\\ \cline{4-7}
          &&& $10^{-6}$ & -- & 15.679 & 0.037
\\ \cline{3-7}
&& Phase-space slicing 
            & $10^{-2}$ & $10^{-2}$ & 15.655 & 0.0056
\\ \cline{4-7}
          &&& $10^{-4}$ & $10^{-4}$ & 15.656 & 0.019
\\ \cline{4-7}
          &&& $10^{-6}$ & $10^{-6}$ & 15.687 & 0.045
\\ \cline{3-7}
&& Dipole formalism &-- & -- & 15.649 & 0.011
\\ \hline
\end{tabular}
\end{center}
\caption{Results on the ${\cal O}(\alpha)$ QED correction $\delta_{\QED}$
to the total Born cross section $\sigma_0$ 
of $\Pem\gamma\to\Pem\gamma(\gamma)$ for
$\sqrt{s}=100\GeV$, $20^\circ<\theta'_\Pe<160^\circ$, and
different degrees of beam polarization $P_\Pe$ and $P_\gamma$.}
\label{tab:eaea2}
\end{table}
In \refta{tab:eaea} we give the total cross sections for polarized 
incoming particles and unpolarized outgoing particles for a
CM energy that is typical for a Compton polarimeter of a future 
$\Pep\Pem$ collider; the beam energies are $\bar E_\Pe=500\GeV$ and
$\bar E_\gamma=2.33\eV$.
The statistical error of the result of phase-space slicing
grows with decreasing $\Delta E/E_\gamma$. For 
$\Delta E/E_\gamma=10^{-2}$ the influence of the finite value of $\Delta E$ 
is still visible.
For the smaller values of $\Delta E$ the integration error of the results
obtained by the dipole subtraction formalism is smaller than the one of
the slicing method by at least an order of magnitude.

\paragraph{(ii) High scattering energies}

Compton scattering represents an important reference process in possible
future elec\-tron--pho\-ton colliders with CM energies in the GeV to TeV
range. In this case the electron mass $\Me$ is small with respect to the
CM energy and can be neglected in predictions whenever mass
singularities are avoided. Detailed discussions of the corresponding
lowest-order cross sections and the electroweak ${\cal O}(\alpha)$
corrections can be found in \citeres{de93,di94}. In the following we
take over the results on the virtual QED corrections given there and
supplement the calculation of the real-photonic corrections of
\citere{di94} by the application of the dipole formalism.

The construction of the subtraction function and its integrated
counterpart proceeds analogously to the case of finite $\Me$ above. One
can either expand the above results for $\Me\to 0$ or make direct use of
the general results presented in \refse{se:subfunc0} for light fermions.
There is, however, a difference to the massive case as far as the
kinematics is concerned. For $\Me\to 0$, exact backward Compton scattering 
has to be excluded by appropriate cuts because of a kinematical
$u$-channel pole in the lowest-order cross section, which is only
regularized by a finite electron mass. We avoid this singular region by
requiring a finite angle $\theta'_\Pe$ of the outgoing electron with the 
beam axis in the CM frame. To this end, we introduce the step function
\beq
g_\cut(\theta) = \Theta(\theta-\theta_\cut)
\Theta(180^\circ-\theta_\cut-\theta)
\label{eq:step}
\eeq
and set $\theta_\cut=20^\circ$ in the numerical evaluation.
While the original squared matrix element $|\M_1|^2$ is simply
multiplied by $g_\cut(\theta'_\Pe)$ in the phase-space integration, the
cuts on the subtraction function have to be chosen in such a way that
the same cuts can be applied in the integrated counterpart to the
subtraction function. At the same time, one has to ensure that the
subtraction function still compensates all singularities of
$|\M_1|^2 g_\cut(\theta'_\Pe)$. Applying the cuts to the polar angles 
$\tilde\theta'_{\Pe,l}$ of the two momenta $\tilde p'_l$ in the original 
CM frame fulfills these requirements. Thus, $|\M^{(l)}_\sub|^2$ in
\refeq{eq:eaeasub1} is replaced by 
$|\M^{(l)}_\sub|^2 g_\cut(\tilde\theta'_{\Pe,l})$.

In the limit $\Me\to 0$ the integrated counterpart to the subtraction
function simplifies drastically. The boost that relates the CM frames of
$p+k_\gamma$ and $\tilde p_1+k_\gamma=x p+k_\gamma$ goes along the beam
axis, and the squared CM energies are related by $\tilde s=x s$.
Therefore, the phase-space measure and the auxiliary function given
above reduce to
\beq
\int\rd\tilde\Phi_{0,\Pe'\Pe}(0,\tilde t,x) =
\frac{1}{8\pi x s} \, \int_{-x s}^0\rd\tilde t,
\qquad
\rho_{ia}(\tilde s) = x,
\eeq
where rotational invariance is already exploited to perform the
integration over $\tilde\varphi'_2$. According to the cutting procedure
described above, we have to apply the angular cut on the angle of
$\tilde p'_1$ in the original CM frame, i.e.\ we have to transform the
polar angle $\tilde\vartheta'_{\Pe,1}(x)$ of $\tilde p'_1$ defined in 
the CM frame of $x p+k_\gamma$ back to the CM frame of $p+k_\gamma$. 
Denoting the angle in the latter
frame by $\tilde\theta'_{\Pe,1}(x)$, the two angles are related by
\beq
\cos\tilde\theta'_{\Pe,1}(x) = 
\frac{x-1+(1+x)\cos\tilde\vartheta'_{\Pe,1}(x)}
{1+x+(x-1)\cos\tilde\vartheta'_{\Pe,1}(x)}.
\eeq
The cuts are consistently introduced in the convolutions over $x$ if 
all squared matrix elements 
$\left|\M_0\Big(\tilde p_a(\xi),\tilde p_i(\xi)\Big)\right|^2$
get the factor
$g_\cut\Big(\tilde\theta'_{\Pe,1}(\xi)\Big)$, where $\xi$ is equal to 
$x$ or 1.

\begin{sloppypar}
Let us inspect the IR and mass singularities explicitly. From the results
of \citere{di94} we deduce that the factor $\delta_\QED^\virt$ for the
virtual corrections can be decomposed into a polarization-independent
singular part and a polarization-dependent regular part
$\delta_{\rem}(\sigma,\lambda,\sigma',\lambda'_1)$,
\beq
\delta_\QED^\virt = -\frac{\alpha}{\pi}{\cal L}(-t,\Me^2) +
\delta_{\rem}(\sigma,\lambda,\sigma',\lambda'_1),
\eeq
with the auxiliary function ${\cal L}$ of \refeq{eq:L}. Since the
Mandelstam variable $t=(p-p')^2$ of \citere{di94} corresponds to 
$\tilde t$ in the convolution over $x$ described above, the IR and mass
singularities of the virtual correction exactly cancel against the ones
contained in the endpoint parts $\Gsub_{\Pe'\Pe,+}$ and 
$\Gsub_{\Pe\Pe',+}$. Therefore, the only uncancelled mass-singular 
corrections are the ones contained in the distribution
$\cGsub_{\Pe\Pe',+}$, where they are weighted with 
the splitting function $P_{ff}(x)$ in the convolution over $x$.
\end{sloppypar}

Table~\ref{tab:eaea2} shows our results on the ${\cal O}(\alpha)$ QED
corrections to the integrated cross section for $\sqrt{s}=100\GeV$ and
different beam polarizations. The table does not only contain the
results from the slicing and subtraction methods, but also includes the
results obtained by a formalism called ``IR slicing and effective mass 
factor''. In this approach only the IR regions are removed from phase 
space by cuts, and the collinear poles are regularized by applying
appropriate factors that replace the poles by the correct mass-dependent
behaviour. More details about the application of this procedure and of 
the slicing method can be found in \citere{di94}. For both slicing
variants, the statistical integration errors increase with decreasing
cut parameters $\De E/E$ and $\De\theta$. Here $E$ is the beam energy in
the CM frame, and the cut angle $\De\theta$ defines cones around the 
electron directions that are excluded from phase space. For a cut size
of $10^{-2}$, the integration errors of the different methods are of the
same order of magnitude, but at least for the approach with effective
mass factors the finiteness of the cut is still visible. Therefore,
smaller cuts are advisable. In this case the superiority of the
subtraction formalism becomes obvious. For the inspected cuts, there is
an improvement of a factor of 2 or more in the integration error.%
\footnote{We expect that the superiority of the subtraction method is
more enhanced if more realistic cuts are applied. Cutting the electron
angle directly, without taking into account a recombination with soft
or collinear photons in the detector, is a strong idealization.
Technically this leads to regions in phase space where 
$g_\cut(\theta'_\Pe)=1$ and $g_\cut(\tilde\theta'_{\Pe,l})=0$ or vice
versa. For collinear photons these regions shrink to zero, but
nevertheless induce strong peaks in the integrand. Realistic cuts should
avoid such pathologies, leading to an improvement in the numerical
integration.}

\subsection{\boldmath{The process $\mu^+\mu^-\to\nu_\Pe\bar\nu_\Pe(\gamma)$}}

\paragraph{(i) Moderate scattering energies}

As a final example, we consider the process
\beq
\mu^-(p,\sigma) + \mu^+(p',\sigma') \; \longrightarrow \;
\nu_\Pe(q,-) + \bar\nu_\Pe(q',+) \; [+ \gamma(k,\lambda)],
\label{eq:mmnnproc}
\eeq
which is phenomenologically less important, but---owing to its
simplicity---it is well suited for demonstrating the application of the 
dipole formalism in situations with two charged fermions in the initial state.
At lowest order, there is only an $s$-channel diagram with Z-boson
exchange, and the Born amplitude reads
\beqar
\M_0(p,p',q,q';\sigma,\sigma') &=&
\frac{e^2(v_{\nu_\Pe}+a_{\nu_\Pe})}{2(s-\MZ^2+\ri\MZ\Gamma_\PZ)} \,
\Big[ \bar v_{\mu^+}(p';\sigma') \gamma^\rho (v_\mu-a_\mu\gamma_5)
u_{\mu^-}(p;\sigma) \Big] \, 
\nn\\[.5em] && {} \times
\Big[ \bar u_{\nu_\Pe}(q;-) \gamma_\rho (1-\gamma_5) 
v_{\bar\nu_\Pe}(q';+) \Big].
\label{eq:mmnnborn}
\eeqar
For our purposes, it is sufficient to describe the Z~resonance with the
constant experimental width $\Gamma_\PZ$ given above.
The virtual photonic corrections consist of a correction to
the $Z\mu\mu$ vertex and the muon wave-function correction. The
derivation of the Born cross section and the virtual correction is
standard and has been performed using the techniques described in
\citere{de93a}. The results are listed in \refapp{app:mmnn} for a finite
muon mass $m_\mu$. 
The bremsstrahlung corrections involve photon emission 
from the muons in the initial state only. The amplitudes for these real
corrections have been obtained from the general results for
$\mu^+\mu^-\to f\bar f\gamma$ given in \citere{di99}.

The subtraction function receives the two contributions
$\gsub_{\mu^-\mu^+,\tau}$ and $\gsub_{\mu^+\mu^-,\tau}$, which are both
of type $ab$, and reads
\beqar
|\M_\sub(p,p',k,q,q';\sigma,\sigma')|^2 &=& \phantom{{}+{}}
e^2 \gsub_{\mu^-\mu^+,\tau}(p,p',k) 
\left|\M_0(\tilde p,p',\tilde q_1,\tilde q'_1;\tau\sigma,\sigma')\right|^2
\nn\\ && {}
+ e^2 \gsub_{\mu^+\mu^-,\tau}(p',p,k) 
\left|\M_0(p,\tilde p',\tilde q_2,\tilde q'_2;\sigma,\tau\sigma')\right|^2.
\label{eq:mmnnsub}
\eeqar
In the following, we only describe the construction of the auxiliary
momenta $\tilde p$, $\tilde q_1$, and $\tilde q'_1$ for the contribution of
$\gsub_{\mu^-\mu^+,\tau}$. The case of $\gsub_{\mu^+\mu^-,\tau}$ can be
treated analogously. The auxiliary variables $x=x_{ab}$ and $y=y_{ab}$ for the
$\gsub_{\mu^-\mu^+,\tau}$ contribution read
\beq
x = \frac{pp'-pk-p'k}{pp'} = \frac{\tilde s-2m_\mu^2}{s-2m_\mu^2},
\qquad y = \frac{pk}{pp'},
\eeq
where we have included the relation between $x$ and the two squared CM
energies $s=(p+p')^2$ and $\tilde s=(p+p'-k)^2=(\tilde p+p')^2$. 
The subtraction function is consistently set to zero for $x<x_0$ with
\beq
x_0 = \frac{2m_\mu^2}{s-2m_\mu^2}.
\eeq
Inserting $p_a=p$, $p_b=p'$, and
$P=P_{ab}=p+p'-k$ into \refeq{eq:tpatP}, we get the new momenta
\beq
\tilde p^\rho = \sqrt{\frac{\lambda_{\tilde s}}{\lambda_s}} \, p^\rho
+\frac{4m_\mu^2(pp')(1-x^2)}
{\sqrt{\lambda_s}(x\sqrt{\lambda_s}+\sqrt{\lambda_{\tilde s}})} \, 
p^{\prime\rho}, \qquad
\tilde P^\rho = \tilde p^\rho + p^{\prime\rho},
\eeq
where we have rearranged some terms in order to reveal the behaviour of
$\tilde p$ in the limit $m_\mu\to 0$. The abbreviations $\lambda_s$ and
$\lambda_{\tilde s}$ are given by
\beq
\lambda_s = \lambda(s,m_\mu^2,m_\mu^2) = \sqrt{s(s-4m_\mu^2)}, \qquad
\lambda_{\tilde s} = \lambda(\tilde s,m_\mu^2,m_\mu^2) = 
\sqrt{\tilde s(\tilde s-4m_\mu^2)}.
\eeq
For the evaluation of the Born matrix element 
$\M_0(\tilde p,p',\tilde q_1,\tilde q'_1)$ in \refeq{eq:mmnnsub}
we need a scalar product between an initial-state and a final-state
momentum, such as 
$p'\tilde q'_1 = p'_\mu\Lambda^\mu_{\phantom{\mu}\nu}q^{\prime\nu}_1$, 
in addition to the already known 
quantity $\tilde s$. At this point, the explicit form \refeq{eq:LT} of 
the Lorentz transformation $\Lambda^\mu_{\phantom{\mu}\nu}$ enters. 
The calculation of the desired scalar product simply requires some
contractions among the original momenta and $\tilde p$.

The evaluation of the integrated counterpart to the subtraction function
leads to convolutions of the form given in \refeq{eq:intab}. Since the
variable $x$ enters the phase space $\tilde\Phi_{0,ab}(s,x)$ only by the
CM energy $\sqrt{\tilde s}$, the phase-space integration over the
squared amplitudes $|\M_0|^2$ is the same as for the lowest-order cross
section $\sigma_0(\tilde s)$ at the CM energy $\sqrt{\tilde s}$.
Consequently, the convolution can be formulated in terms of lowest-order
cross sections, and the integrated counterpart to the 
$\gsub_{\mu^-\mu^+,\tau}$ contribution in \refeq{eq:mmnnsub} reads
\beqar
\Delta\sigma^{(\sub)}_{\mu^-\mu^+}(s,P_-,P_+) &=& 
\frac{\alpha}{2\pi} \int_{x_0}^1\rd x\, 
\cGsub_{\mu^-\mu^+,\tau}(s,x) \left[ 
\frac{\sqrt{\lambda_{\tilde s}}}{x\sqrt{\lambda_s}} \,
\sigma_0(\tilde s,\tau P_-,P_+)-\sigma_0(s,\tau P_-,P_+) \right]
\nn\\ && {}
+ \frac{\alpha}{2\pi} \, \Gsub_{\mu^-\mu^+,\tau}(s,x_0) \,
\sigma_0(s,\tau P_-,P_+),
\label{eq:intmmnn}
\eeqar
where $P_\mp$ are the degrees of beam polarization of the $\mu^\mp$
beams. The factor $\sqrt{\lambda_{\tilde s}}/\sqrt{\lambda_s}$ stems
from the flux factors in the transition from squared matrix elements to
cross sections.

\begin{table}
\begin{center}
\begin{tabular}{|c||c|c||c|l@{$\,\pm\,$}l|}
\hline
$P_-$ & $\sigma_0/10^{-5}\pb$ & Method & $\Delta E/E$ & 
\multicolumn{2}{c|}{$\delta_\QED/\%$} 
\\ \hline\hline
$+$ & 5.74003 & Phase-space slicing 
            & $10^{-2}$ & 0.3827 & 0.0048
\\ \cline{4-6}
          &&& $10^{-4}$ & 0.352  & 0.011 
\\ \cline{4-6}
          &&& $10^{-6}$ & 0.341  & 0.018
\\ \cline{3-6}
&& Dipole formalism &-- & 0.36637 & 0.00022
\\ \hline\hline
$-$ & 9.35981 & Phase-space slicing 
            & $10^{-2}$ & 0.3275 & 0.0045
\\ \cline{4-6}
          &&& $10^{-4}$ & 0.299  & 0.011
\\ \cline{4-6}
          &&& $10^{-6}$ & 0.292  & 0.017
\\ \cline{3-6}
&& Dipole formalism &-- & 0.31238 & 0.00020
\\ \hline
\end{tabular}
\end{center}
\caption{Results on the ${\cal O}(\alpha)$ QED correction $\delta_{\QED}$
to the total Born cross section $\sigma_0$ of 
$\mu^+\mu^-\to\nu_\Pe\bar\nu_\Pe(\gamma)$ for
$\sqrt{s}=500\MeV$ and different degrees of $\mu^-$-beam
polarization $P_-$ and unpolarized $\mu^+$.}
\label{tab:mmnn}
\vspace*{1em}
\begin{center}
\begin{tabular}{|c||c|c|c|c||l@{$\,\pm\,$}l|}
\hline
$P_-$ & $\sigma_0/\pb$ & Method & $\Delta E/E$ & 
$\Delta\theta/\mathrm{rad}$ &
\multicolumn{2}{c|}{$\delta_\QED/\%$} 
\\ \hline\hline
$+$ & 1.32547 & IR slicing and
            & $10^{-2}$ & -- & $-4.157$ & 0.021
\\ \cline{4-7}
         && effective mass factor & $10^{-4}$ & -- & $-4.331$ & 0.055
\\ \cline{4-7}
          &&& $10^{-6}$ & -- & $-4.353$ & 0.089
\\ \cline{3-7}
&& Phase-space slicing 
            & $10^{-2}$ & $10^{-2}$ & $-4.162$ & 0.018
\\ \cline{4-7}
          &&& $10^{-4}$ & $10^{-4}$ & $-4.321$ & 0.090
\\ \cline{4-7}
          &&& $10^{-6}$ & $10^{-6}$ & $-4.36$  & 0.22
\\ \cline{3-7}
&& Dipole formalism &-- & -- & $-4.29135$ & 0.00022
\\ \hline\hline
$-$ & 2.06497 & IR slicing and
            & $10^{-2}$ & -- & $-4.168$ & 0.021
\\ \cline{4-7}
         && effective mass factor & $10^{-4}$ & -- & $-4.335$ & 0.054
\\ \cline{4-7}
          &&& $10^{-6}$ & -- & $-4.356$ & 0.087
\\ \cline{3-7}
&& Phase-space slicing 
            & $10^{-2}$ & $10^{-2}$ & $-4.151$ & 0.018
\\ \cline{4-7}
          &&& $10^{-4}$ & $10^{-4}$ & $-4.257$ & 0.091
\\ \cline{4-7}
          &&& $10^{-6}$ & $10^{-6}$ & $-4.24$  & 0.22
\\ \cline{3-7}
&& Dipole formalism &-- & -- & $-4.30390$ & 0.00020
\\ \hline
\end{tabular}
\end{center}
\caption{Results on the ${\cal O}(\alpha)$ QED correction $\delta_{\QED}$
to the total Born cross section $\sigma_0$ of 
$\mu^+\mu^-\to\nu_\Pe\bar\nu_\Pe(\gamma)$ for
$\sqrt{s}=50\GeV$ and different degrees of $\mu^-$-beam
polarization $P_-$ and unpolarized $\mu^+$.}
\label{tab:mmnn2}
\end{table}
Table~\ref{tab:mmnn} shows some results on the photonic ${\cal O}(\alpha)$
corrections to the lowest-order cross sections $\sigma_0$ for different
$\mu^-$-beam polarizations. The $\mu^+$ beam is assumed to be
unpolarized. The considered CM energy of $\sqrt{s}=500\MeV$ is too small
for a neglect of the muon mass $m_\mu$ in the non-singular
contributions. Therefore, the $m_\mu$ dependence is treated exactly.
The results of the dipole subtraction formalism are compared to the ones
obtained by phase-space slicing, where $\Delta E$ is the cut energy on
the outgoing photon, and $E=\sqrt{s}/2$ denotes the beam energy in the
CM frame. Similar to the examples inspected previously, for 
$\Delta E/E=10^{-2}$ the influence of the finiteness of $\Delta E$ is
still visible at the chosen level of accuracy. On the other hand, using
the same integration parameters for the subtraction method, the
improvement in the integration error is between one and two orders of
magnitude.

\paragraph{(ii) High scattering energies}

Now we turn to high scattering energies and neglect the muon mass
whenever possible, i.e.\ we apply the results of \refse{se:ab0}. In this
limit, the virtual correction reduces to the simple polarization-independent
factor $\delta_\QED^\virt$ to the Born cross section $\sigma_0$,
\beq
\delta_\QED^\virt = -\frac{\alpha}{\pi} \left[
{\cal L}(s,m_\mu^2) - \frac{2\pi^2}{3} + 2 \right],
\eeq
in agreement with the result given in \citere{be82a} on initial-state
radiation in
$\Pep\Pem\to\PZ^*\to f\bar f$. Concerning the real correction, 
the subtraction procedure described above becomes technically simpler, 
since $m_\mu$ can be neglected in the kinematics everywhere. 
Owing to the simplicity of the total Born cross section for $m_\mu=0$,
\beq
\sigma_0(s) = (1-\tau P_+)(1+\tau P_-) \,
\frac{\pi\alpha^2}{3}(v_{\nu_\Pe}+a_{\nu_\Pe})^2(v_\mu-\tau a_\mu)^2
\frac{s}{|s-\MZ^2+\ri\MZ\Gamma_\PZ|^2},
\eeq
the convolution \refeq{eq:intmmnn} over $x$ can be easily performed
analytically. Note that the IR and mass-singular part of the virtual
correction $\delta_\QED^\virt$ is again exactly cancelled by the
singular terms in the endpoint contributions $\Gsub_{\mu^-\mu^+,+}$ and
$\Gsub_{\mu^+\mu^-,+}$. The remaining mass-singular contributions are
contained in $\cGsub_{\mu^-\mu^+,+}$ and $\cGsub_{\mu^+\mu^-,+}$ and
enter the convolution \refeq{eq:intmmnn} over $x$ weighted by the
splitting function $P_{ff}(x)$.

The application of
the slicing method additionally requires the analytic treatment of
photons that are emitted nearly collinearly from the muon beams, i.e.\
which have emission angles $\theta$ within the ranges
$0^\circ<\theta<\Delta\theta$ or 
$0^\circ<180^\circ-\theta<\Delta\theta$ with $\Delta\theta\ll 1$.
These effects are calculated in the same way as described in
\citere{di94} for initial-state radiation in Compton scattering. 
Details about the variant with effective mass factors can
also be found there.

In \refta{tab:mmnn2} we show the photonic corrections for
$\sqrt{s}=50\GeV$, $P_-=\pm 1$, and $P_+=0$, obtained in the small-mass
limit $m_\mu\to 0$. The numbers
again underline the superiority of the subtraction formalism.
The integration error is reduced by one to two orders of magnitude, 
without the necessity to look for a plateau in auxiliary parameters,
such as $\Delta E$ and $\Delta\theta$.

\section{Discussion and outlook}
\label{se:disc}

\subsection{Features of subtraction methods and the dipole formalism}

As already explained in the introduction, the basic motivation for the
development of subtraction methods is to avoid singular numerical
integrations in the calculation of real-photonic (or real-gluonic) 
corrections. In the previous section, we have compared the results
for various radiative processes obtained by applying the dipole subtraction
formalism of this paper with the ones obtained by phase-space slicing.
We have found that the application of the subtraction formalism typically
reduces the integration error by an order of magnitude with respect to
the results of phase-space slicing, when all integrations are performed
with the same statistics. 
As mentioned at the beginning of \refse{se:appl}, the efficiency
of the slicing method has been improved by introducing appropriate 
parametrizations of phase space, whereas such improvements are not
needed for the subtraction method.

Moreover, a successful application of
the slicing method requires a careful investigation of the dependence on
the soft-photon cut $\Delta E$ and, if relevant, on the angular cut
$\Delta\theta$. It is necessary to optimize the choice of
the cut parameters for all considered observables. The integration error
roughly grows proportional to the logarithm of a cut parameter if the
cut becomes small. The optimal choice of cut parameters loosens the cuts
as much as possible, but still suppresses remnant effects of their 
finiteness. The optimal set of cuts varies with the desired accuracy 
and, in most cases, also with input parameters, such as the scattering energy.
In practice, one often tends to choose rather small cuts at the cost of
accuracy, in order to be on the safe side. Subtraction methods are not
plagued by the need of such an optimization procedure.
Of course, checking the cutoff independence of observables represents a
good consistency test of calculations based on slicing methods that is
not possible for subtraction procedures, but the dipole 
subtraction formalism allows for various other checks, some of which are
described in \refse{se:advice}.

One of the great advantages of the dipole formalism is certainly its
process independence, which distinguishes this approach from most of the
other subtraction procedures. In this paper, the dipole formalism is
worked out for photon radiation in processes involving charged fermions
and any other neutral particles. We stress that all different
configurations of particle masses and helicities are supported. The
subtraction function, which removes IR and possible collinear 
singularities from the differential cross section, is constructed in 
such a way that the transition to the region of small fermion masses 
proceeds smoothly. In other words, there is one subtraction function that
interpolates the regions of large and small masses. 

Finally, one has to admit that the actual application of subtraction
methods, in general, is more involved than the use of phase-space slicing 
for complicated electroweak processes.
The presentation in this paper certainly shows that the 
application of a subtraction procedure can be quite involved for
processes with massive particles. The implementation of phase-space cuts
within subtraction methods is straightforward, but nevertheless can be
laborious (see also next section). On the other hand, once the procedure 
is applied to a process, such complications are completely overcome, and
the advantages described above become apparent.

\subsection{Phase-space cuts and distributions}

In the above formulation of the dipole formalism, we mainly concentrated 
on the calculation of total cross sections, but we did not pay particular
attention to phase-space cuts or to the calculation of distributions.
We recall that the difference of the differential cross section and the
subtraction function is integrated over the full phase space $\Phi_1$ of
$n+1$ particles
numerically, but the integrated counterpart to the subtraction function
implicitly contains the integration over the photonic part of 
phase space, which is carried out analytically. 
The cuts that are applied to the subtraction function have to be
identical with the ones that are applied and to the integrated counterpart
of the subtraction function.
Otherwise these two contributions will not compensate each
other, leading to wrong results. In practice, this means that we have to
distinguish two types of cuts. Firstly, we have the original cuts 
that are applied to the original differential cross section; these cuts
concern the full phase space $\Phi_1$ of $n+1$ particles.
Secondly, we have auxiliary cuts that are applied to the subtraction
function and to its integrated counterpart; they are defined in the
reduced phase spaces $\tilde\Phi_{0,ff'}$ of $n$ particles. Simple
examples for the implementation of angular cuts have been 
described in \refse{se:eaea} for Compton scattering at high energies and
in \citere{de98} for the production of light fermion--anti-fermion pairs
in photon--photon collisions. 

The calculation of distributions is
similar to the application of cuts, since a histogram of a distribution
is nothing but a series of cuts. Hence, the histogram routine that
generates the desired distribution during the Monte Carlo integration
has to handle each column of the histogram in the same way as a cutted
contribution to the integrated cross section. Note that this procedure
implies that the original differential cross section and the subtraction
function may contribute to different columns of the histogram for one
and the same event. The final result for each column is nevertheless
finite, because such events are in general far away from the singular
regions.%
\footnote{At the edges of the histogram columns this can also occur for
``singular events''. The finiteness of such contributions is guaranteed
by the suppression of phase space for those events.}

\subsection{Practical advice}
\label{se:advice}

Subtraction methods offer a number of checks, which are very
useful in practice. The basic principle of subtraction methods is
that all contributions originating from the subtraction function add up
to zero in the final result. In the following we describe some possible
checks for the dipole formalism 
that are mainly based on this principle. The described checks 
have been successfully carried out in the applications discussed in
\refse{se:appl}.

The auxiliary functions $\gsub_{ff',\tau}$, $\cGsub_{ff',\tau}$, and
$\Gsub_{ff',\tau}$ can be checked for consistency without application to
a specific process. To this end, one should carry out all integrations
numerically that have been performed analytically for the derivation of
$\cGsub_{ff',\tau}$ and $\Gsub_{ff',\tau}$. Since some of these
integrations involve IR singularities, a small photon mass $m_\gamma$
has to be consistently used in the numerics. 

For the treatment of specific processes, the construction of the phase
spaces $\tilde\Phi_{0,ff'}$ deserves particular care. It can be very
useful to compare the corresponding phase-space volumes entering the
integrations over the phase spaces of $n+1$ and $n$ particles. 
The two volumes are obtained as follows:
\begin{itemize}
\item[(a)] in the original integration over $\rd\Phi_1$ we set 
$\M_1\to 0$, $\M_0\to 1$, and $\gsub_{ff',\tau}\to 1$;
\item[(b)] in the integrations of the counterparts over 
$\rd\tilde\Phi_{0,ff'}$ we set $\M_0\to 1$ and use the expressions for
$\cGsub_{ff',\tau}$ and $\Gsub_{ff',\tau}$ that correspond to
$\gsub_{ff',\tau}\to 1$. Those expressions can be derived easily, using 
$m_\gamma=0$.
\end{itemize}
Note that this phase-space comparison, in particular, represents a
non-trivial check on the convolutions \refeq{eq:intia} 
in the $ia$ and $ai$ cases, which can be quite
complicated for massive initial-state fermions.

In many cases, the phase-space check can be extended by including the 
full form of the functions $\gsub_{ff',\tau}$, $\cGsub_{ff',\tau}$, and
$\Gsub_{ff',\tau}$, i.e.\ the only substitutions are $\M_1\to 0$ and 
$\M_0\to 1$ in the phase-space integrations. Owing to the IR and
collinear singularities in $\gsub_{ff',\tau}$, this kind of check is not
always possible in a simple way. 
The check is, for instance, useful in the $ia$
and $ai$ cases with $m_a\neq 0$. In these cases, all singularities
appear for $x\to 1$ and can be removed by applying the additional cut
$x_{ia}<1-\Delta x$ with any small $\Delta x>0$ in the integration 
over $\rd\Phi_1$. This additional cut has to be incorporated in the
convolution of $\cGsub_{ff',\tau}$ over $x$, too. The simplest
possibility to achieve this is to omit the introduction of the
$[\dots]_+$ prescription and to perform the convolution in the range
$x_0<x<1-\Delta x$.

Of course, many other variants of such consistency checks may be useful
in actual applications.

\subsection{Generalization to QCD}

In this paper, we have focussed on photon radiation off fermions only.
The presented formalism can, however, be carried over to gluon radiation
for a certain class of processes. Consider, for instance, a process that
involves a heavy quark--anti-quark pair $q\bar q$, but no other QCD 
partons. In this case, the gluonic corrections can be obtained from the 
photonic corrections by the replacement 
$Q_q^2\alpha\to 4\alpha_{\mathrm{s}}/3$, and the infinitesimal photon 
mass $m_\gamma$ turns into an infinitesimal gluon mass $m_{\mathrm{g}}$.
Since the IR singularity is abelian, the transition to dimensional 
regularization is performed by the well-known substitution
\beq
\ln(m_{\mathrm{g}}^2) \;\to\;
\frac{(4\pi\mu^2)^\epsilon\Gamma(1+\epsilon)}{\epsilon} 
+ {\cal O}(\epsilon),
\eeq
where $D=4-2\epsilon$ is the dimension and $\mu$ the reference mass of
dimensional regularization.

The results of this paper can also be used to deal with processes
involving more than one heavy quark--anti-quark pair if the colour flow is
treated properly. The colour algebra is identical to the corresponding
process with massless quarks and can be taken over from \citere{ca96}.

The presented results do not cover the cases of gluon radiation in which
collinear singularities are treated within dimensional regularization.
This includes real-gluonic corrections to all processes involving
massless partons in the initial state. However, the presented results
can serve as a starting point for a full generalization of the dipole
formalism to QCD with heavy quarks.

\section{Summary}
\label{se:sum}

Following the guideline of \citere{ca96}, where the dipole subtraction
formalism is presented for NLO QCD corrections involving massless
unpolarized partons, we have formulated this method for photon
radiation off massive fermions. 
The dipole formalism represents a process-independent subtraction procedure
that removes all IR and collinear singularities from differential cross
sections of bremsstrahlung processes. The subtracted singular structures
are calculated separately, where the integration over the singular regions 
is performed analytically. Consequently, no singular numerical
integrations are needed for the final result. This advantage
distinguishes subtraction formalisms from methods that employ
phase-space slicing. Slicing methods require a careful optimization of
small cuts that exclude the singular regions from the numerical
phase-space integration. 

Since the consistent inclusion of finite fermion masses turned out to 
be highly non-trivial, we have presented the derivation of the method in
a rather detailed way. Our formulation, which allows for 
fermions with definite helicity eigenstates, is applicable to 
processes involving charged fermions and any type of neutral particles.
The generalization to charged bosons is straightforward.
In the limit of small fermion masses, which is of particular importance
phenomenologically, the dipole formalism simplifies considerably and is
easy to use. 

In order to illustrate the use of the method in practice, we have
applied the dipole subtraction method to 
the processes $\gamma\gamma\to\Pt\bar\Pt(\gamma)$, 
$\Pem\gamma\to\Pem\gamma(\gamma)$, and
$\mu^+\mu^-\to \nu_\Pe\bar\nu_\Pe(\gamma)$. Comparing the corresponding
results to the ones obtained by slicing methods, we find improvements in
the integration errors of typically an order of magnitude, where Monte
Carlo integrations are performed with the same statistics in both
approaches.

Finally, we conclude that the dipole subtraction formalism is superior to 
methods that are based on phase-space slicing. Moreover, the presented
procedure for photon radiation off massive fermions is a first step 
towards the full generalization of the dipole formalism to QCD with
heavy quarks.

\section*{Acknowledgement}

I am grateful to M.~Roth for a careful comparison 
of his results on the special case of light fermions with the
corresponding ones of this work. Moreover, I thank S.~Catani, A.~Denner, 
M.~Kr\"amer, M.~Roth, and M.~Spira for helpful discussions and comments 
on the manuscript.

\appendix
\section*{Appendix}

\section{Special cases}

\subsection{Light incoming particles}
\label{app:ma0}

The case of light fermions in the inital state is of particular 
importance. For instance, it is relevant for $\Pep\Pem$ collisions
at high energies, as observed at LEP or the SLC. Since the $ab$ case
with $m_{a,b}\to 0$ is already covered by \refse{se:ab0}, here we 
concentrate on the mixed cases $ia$ and $ai$ with $m_a\to 0$.

Using the auxiliary parameters $x_{ia}$ and $z_{ia}$ of \refeq{eq:xiazia},
the subtraction functions are given by
\beqar
\gsub_{ia,+}(p_i,p_a,k) &=&
\frac{1}{(p_i k)x_{ia}} \biggl[ \frac{2}{2-x_{ia}-z_{ia}}-1-z_{ia} 
-\frac{m_i^2}{p_i k} \biggr]
-\gsub_{ia,-}(p_i,p_a,k),
\nn\\[.5em]
\gsub_{ia,-}(p_i,p_a,k) &=&
\frac{m_i^2}{2(p_i k)^2 x_{ia}^2}\, 
\frac{(1-z_{ia})^2}{z_{ia}}, 
\nn\\[.5em]
\gsub_{ai,+}(p_a,p_i,k) &=&
\frac{1}{(p_a k)x_{ia}} \biggl[ \frac{2}{2-x_{ia}-z_{ia}}
-1-x_{ia} \biggr],
\nn\\[.5em]
\gsub_{ai,-}(p_a,p_i,k) &=& 0.
\eeqar
The lower limit $x_0$ on $x_{ia}$ can be set to zero consistently. 
The construction of the auxiliary momenta $\tilde p_a$ and $\tilde p_i$,
which is given in \refeq{eq:tpitpa} for finite fermion masses,
becomes particularly simple for $m_a\to 0$. 
The result is formally identical with \refeq{eq:tpitpa0}
for the fully massless case, but one should note that $\tilde p_i^2=m_i^2$
still holds. Using the above relations, the $ai$ and $ia$ contributions
to the subtraction function $|\M_\sub|^2$ of \refeq{eq:m2sub}
can be constructed easily.

The calculation of the integrated counterparts to $|\M_\sub|^2$ also
considerably simplifies in the limit $m_a\to 0$. Since the construction
of the auxiliary momenta $\tilde p_i$ and $\tilde p_a$ proceeds as in
the fully massless case described in \refse{se:subfunc0}, the
convolution over $x$ takes the simple form \refeq{eq:intiaai02} even for
finite $m_i$. The distributions for these convolutions read
\beqar
\cGsub_{ia,+}(P_{ia}^2,x) &=&
\frac{1}{(1-x)} \, \left\{
2\ln\left[\frac{2-x-z_1(x)}{1-x}\right]
\right.
\nn\\*
&& \left. \hspace*{1em} {}
+\frac{1}{2}[z_1(x)-1] \left[
3+z_1(x)
-\frac{4m_i^2 x}{(P_{ia}^2-m_i^2)(1-x)} \right] \right\}
-\cGsub_{ia,-}(P_{ia}^2,x),
\nn\\[.5em]
\cGsub_{ia,-}(P_{ia}^2,x) &=&
\frac{m_i^2}{P_{ia}^2-m_i^2}
\frac{1}{(1-x)^2} \left\{ \ln\left[z_1(x)\right]
+\frac{1}{2}[1-z_1(x)][3-z_1(x)] \right\},
\nn\\[.5em]
\cGsub_{ai,+}(P_{ia}^2,x) &=&
P_{ff}(x)\left\{
\ln\left(\frac{m_i^2-P_{ia}^2}{m_a^2 x}\right)
+\ln[1-z_1(x)] -1 \right\}
\nn\\ && {}
-\frac{2}{1-x}\ln[2-x-z_1(x)]+(1+x)\ln(1-x),
\nn\\[.5em]
\cGsub_{ai,-}(P_{ia}^2,x) &=& 1-x,
\eeqar
where 
\beq
z_1(x)= \frac{m_i^2 x}{m_i^2-P_{ia}^2(1-x)}
\eeq
is the lower limit on $z_{ia}$ for $m_\gamma=m_a=0$. The endpoint
contributions are given by
\beqar
\Gsub_{ia,+}(P_{ia}^2) &=&
\ln\left(\frac{m_i^2}{m_\gamma^2}\right)
\ln\left(2-\frac{P_{ia}^2}{m_i^2}\right)
+2\ln\left(\frac{m_\gamma m_i}{m_i^2-P_{ia}^2}\right)
-2\Li_2\left(\frac{P_{ia}^2}{P_{ia}^2-2m_i^2}\right)
\nn\\ && {}
+\frac{1}{2}\ln^2\left(2-\frac{P_{ia}^2}{m_i^2}\right)
+\frac{(P_{ia}^2-m_i^2)^2}{2P_{ia}^4}\ln\left(1-\frac{P_{ia}^2}{m_i^2}\right)
-\frac{\pi^2}{6}+1+\frac{m_i^2}{2P_{ia}^2},
\nn\\[.5em]
\Gsub_{ai,+}(P_{ia}^2) &=&
\ln\left(\frac{m_\gamma^2}{m_a^2}\right)
\ln\left[\frac{m_a^2(2m_i^2-P_{ia}^2)}{(m_i^2-P_{ia}^2)^2}\right]
+\ln\left(\frac{m_\gamma^2}{m_a^2}\right)
+2\Li_2\left(\frac{P_{ia}^2}{2m_i^2-P_{ia}^2}\right)
\nn\\ && {}
-2\Li_2\left(\frac{m_i^2}{2m_i^2-P_{ia}^2}\right)
+2\ln\left[\frac{m_a^2 m_i^2}{(m_i^2-P_{ia}^2)(2m_i^2-P_{ia}^2)}\right]
\ln\left(\frac{2m_i^2-P_{ia}^2}{m_i^2-P_{ia}^2}\right)
\nn\\ && {}
+\frac{1}{2}\ln^2\left(\frac{m_a^2}{2m_i^2-P_{ia}^2}\right)
+\frac{3}{2}\ln\left(\frac{m_a^2}{m_i^2-P_{ia}^2}\right)
+\frac{m_i^2(m_i^2-4P_{ia}^2)}{2P_{ia}^4}
\ln\left(1-\frac{P_{ia}^2}{m_i^2}\right)
\nn\\ && {}
+\frac{\pi^2}{3}-\frac{3}{2}+\frac{m_i^2}{2P_{ia}^2},
\nn\\[.5em]
\Gsub_{ia,-}(P_{ia}^2) &=&
\Gsub_{ai,-}(P_{ia}^2) = \frac{1}{2}. 
\eeqar
The fully massless limit, in which we additionally have $m_i\to 0$, can
be read off from the above results easily, and we get back the corresponding
results of \refse{se:subfunc0}.

\subsection{\boldmath{Endpoint contributions for $x_0=0$}}
\label{app:x0}

In \refse{se:subfunc} we have given the endpoint contributions 
$\Gsub_{ff',\tau}(P_{ff'}^2,x_0)$ with $ff'=ia,ai,ab$ for an arbitrary 
lower limit $x_0\ge 0$ and finite fermion masses. In many applications 
it is possible to set $x_0$ to zero, which simplifies the formulas for 
these contributions. For convenience, we list the results on
$\Gsub_{ff',\tau}(P_{ff'}^2)=\Gsub_{ff',\tau}(P_{ff'}^2,0)$
explicitly,
\beqar
\Gsub_{ia,+}(P_{ia}^2) &=&
2\ln\left(\frac{m_\gamma m_i}{|\bar P_{ia}^2|}\right)
+\frac{\bar P_{ia}^4}{2(\bar P_{ia}^2+m_i^2)^2}
\ln\left(\frac{|\bar P_{ia}^2|}{m_i^2}\right)
-\frac{\bar P_{ia}^2}{2(\bar P_{ia}^2+m_i^2)} + 2
\nn\\ &&{}
+\frac{\bar P_{ia}^2}{\sqrt{\lambda_{ia}}} \left\{
2\ln(b_1)\ln\left[\frac{b_0\sqrt{\lambda_{ia}(m_a^2+m_i^2-\bar P_{ia}^2)}}
{m_\gamma m_a^2}\right]
-\frac{1}{2}\ln^2(b_1)+\frac{\pi^2}{3}
\right.
\nn\\ && \hspace*{4em} \left. {}
+2\sum_{k=1}^5 (-1)^k \Li_2(b_k) 
    \vphantom{\left[\frac{\sqrt{\bar\lambda_{ia}}}{m_a^2}\right]}
\right\}
-\Gsub_{ia,-}(P_{ia}^2),
\nn\\[.5em]
\Gsub_{ia,-}(P_{ia}^2) &=&
\frac{\bar P_{ia}^4}{2\lambda_{ia}}
+\frac{2m_a m_i(\bar P_{ia}^2+2m_i^2)}{\lambda_{ia}}
{\mathrm{arctan}}\left(\frac{m_a}{m_i}\right)
+\frac{m_a^2 m_i^2}{\lambda_{ia}} \left(
\frac{\bar P_{ia}^2}{\bar P_{ia}^2+m_i^2}-4 \right)
\nn\\ &&{}
+\frac{m_i^2 \bar P_{ia}^2}{\lambda_{ia}} \left\{
\frac{2m_a^2-\bar P_{ia}^2}{\bar P_{ia}^2}
\ln\left(\frac{m_a^2+m_i^2}{m_i^2}\right)
-\frac{m_a^2 \bar P_{ia}^2}{(\bar P_{ia}^2+m_i^2)^2}
\ln\left(\frac{|\bar P_{ia}^2|}{m_i^2}\right)
\right\},
\nn\\[.5em]
\Gsub_{ai,+}(P_{ia}^2) &=&
2\ln\left(\frac{m_\gamma m_i}{|\bar P_{ia}^2|}\right)
+\frac{\bar P_{ia}^4(\bar P_{ia}^2+\sqrt{\lambda_{ia}})}
{4m_a^2(\bar P_{ia}^2+m_i^2)\sqrt{\lambda_{ia}}} + \frac{3}{2}
+\frac{\bar P_{ia}^4(3\bar P_{ia}^2+2m_i^2)}
{2\sqrt{\lambda_{ia}}(\bar P_{ia}^2+m_i^2)^2} 
\nn\\ && {}
\times \left\{ \frac{1}{\gamma}\ln\left[
\frac{\bar P_{ia}^2 \gamma^2+2m_a^2+\gamma\sqrt{\lambda_{ia}}}
{\bar P_{ia}^2\gamma(\gamma-1)} \right] 
+\ln\left[\frac{\bar P_{ia}^2(1-2\gamma^2)-2m_a^2+\sqrt{\lambda_{ia}}}
{2\gamma^2 m_i^2}\right]
\right\}
\nn\\ && {}
+\frac{\bar P_{ia}^2}{\sqrt{\lambda_{ia}}} \left\{
2\ln\left(\frac{m_\gamma m_a}{b_0\sqrt{\lambda_{ia}}}\right)
\ln\left(\frac{c_1}{c_0}\right)
-\ln\left(\frac{m_a^2+m_i^2-\bar P_{ia}^2}{m_a^2}\right)\ln(c_1)
\right.
\nn\\ && \hspace*{4em}\left. {}
+\frac{1}{2}\ln(c_0 c_1)\ln\left(\frac{c_1}{c_0}\right)
-\frac{1}{2}\ln(c_0)
-2\sum_{k=0}^5 (-1)^k \Li_2(c_k)
\right\},
\nn\\[.5em]
\Gsub_{ai,-}(P_{ia}^2) &=& \frac{1}{2}, 
\nn\\[.5em]
\Gsub_{ab,+}(s) &=&
\ln\left(\frac{m_\gamma^2 s}{\bar s^2}\right) + \frac{3}{2}
+ \frac{\bar s}{\sqrt{\lambda_{ab}}} \left\{ 
\ln\left(\frac{m_\gamma^2\lambda_{ab}}{m_a^2 \bar s^2}\right) \ln(d_1)
+\left( \frac{1}{2}-\frac{2m_a^2}{\bar s} \right)\ln(d_1)
\right.
\nn\\ && \left. \hspace*{4em} {}
+2\Li_2(d_1)+\frac{1}{2}\ln^2(d_1)-\frac{\pi^2}{3} \right\},
\nn\\[.5em]
\Gsub_{ab,-}(s) &=& \frac{1}{2},
\eeqar
where the abbreviations $b_i$, $c_i$, $d_i$, $\gamma$ are defined as in
\refse{se:subfunc}. In particular, we have
\beq
b_0 = -\frac{2m_a^2}{2m_a^2+\bar P_{ia}^2-\sqrt{\lambda_{ia}}}.
\eeq

\section{Derivation of phase-space splittings}
\label{app:PS}

In this appendix, we outline the derivation of the photonic parts of phase 
space needed for the analytical integration of the subtraction function. 
The emitter/spectator cases $ij$, $ia+ai$, and $ab$ are kinematically
different and are treated separately.

\subsection{Final-state emitter and final-state spectator}

\begin{figure}
\centerline{
\setlength{\unitlength}{1cm}
\begin{picture}(14,8.4)
\put(-5.5,-9.3){\includegraphics{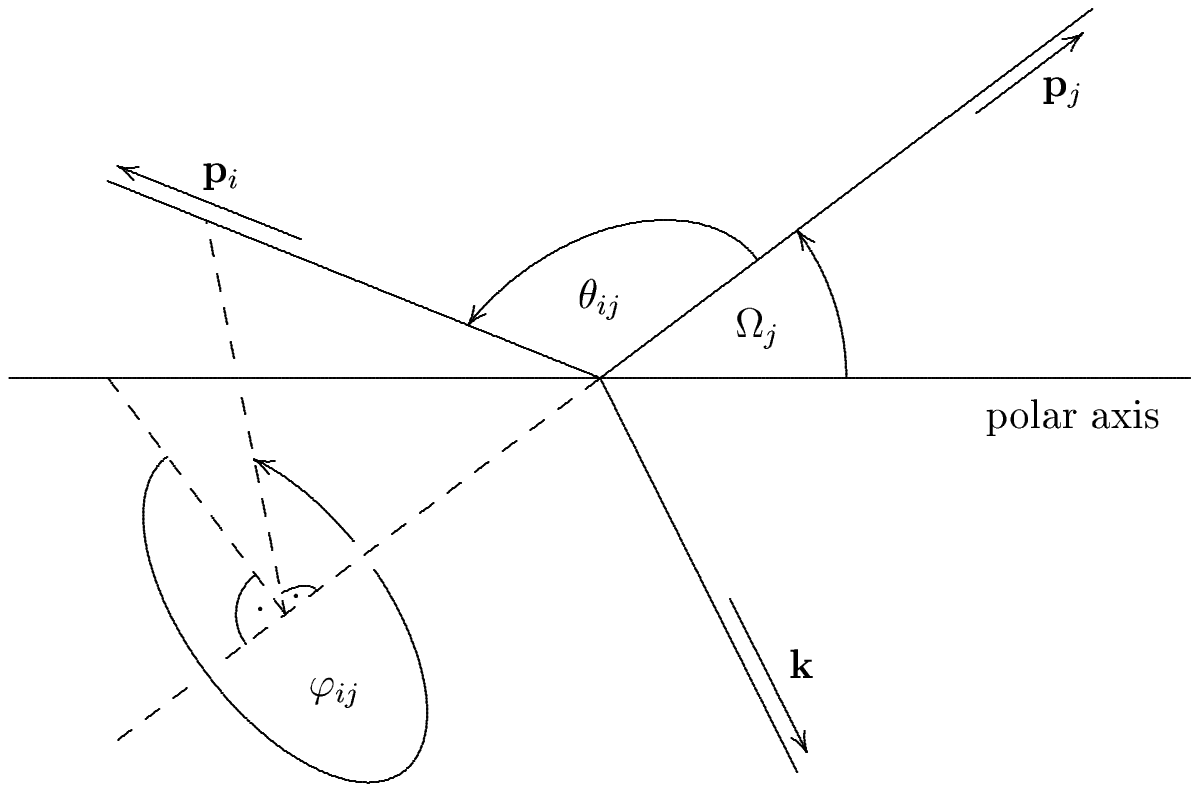}}
\end{picture} }
\caption{Definition of kinematical variables for
$\rd\phi(p_i,p_j,k;P_{ij})$ in the CM frame of $P_{ij}$.}
\label{fig:PSijk}
\end{figure}

The photonic phase-space measure $[\rd k(P_{ij}^2,y_{ij},z_{ij})]$
of the $ij$ case is defined in \refeq{eq:PSijsplit}. In order
to derive this measure, we inspect the explicit parametrizations 
\beqar
\int\rd\phi(p_i,p_j,k;P_{ij}) &=& \frac{1}{8(2\pi)^5} 
\int\rd\Omega_j \int\rd\varphi_{ij} \int\rd p_i^0 \int\rd p_j^0,
\nn\\[.5em]
\int\rd\phi(\tilde p_i,\tilde p_j;P_{ij}) &=& \frac{1}{8(2\pi)^2} 
\frac{\sqrt{\lambda_{ij}}}{P_{ij}^2} \int\rd\Omega_j 
\label{eq:PSij}
\eeqar
in the CM frame of $P_{ij}$. In \refeq{eq:PSij} we have
exploited the fact that the momenta $p_j$ and $\tilde p_j$ possess the
same solid angle $\Omega_j$ in this frame, as a consequence of definition 
\refeq{eq:tmomij}. The phase-space variables of
$\rd\phi(p_i,p_j,k;P_{ij})$ are illustrated in \reffi{fig:PSijk}.
The angles $\theta_{ij}$ and $\varphi_{ij}$ assign the polar
and azimuthal angles of ${\bf p}_i$ with respect to the ${\bf p}_j$ axis,
respectively. The particle energies 
$p_i^0$ and $p_j^0$ can be expressed in terms of the variables $y_{ij}$ and
$z_{ij}$, 
\beq
p_i^0 = \frac{2m_i^2+\bar P_{ij}^2(y_{ij}+z_{ij}-y_{ij}z_{ij})}
{2\sqrt{P_{ij}^2}}, \qquad
p_j^0 = \frac{2m_j^2+\bar P_{ij}^2(1-y_{ij})}
{2\sqrt{P_{ij}^2}}, 
\eeq
so that \refeq{eq:PSijsplit} and \refeq{eq:PSij} directly lead to
$[\rd k(P_{ij}^2,y_{ij},z_{ij})]$ as given in \refeq{eq:dkij}.
The integration boundary for the particle energies, which is determined by 
\beq
1 \ge |\cos\theta_{ij}| =
\frac{\left|P_{ij}^2-2\sqrt{P_{ij}^2}(p^0_i+p^0_j)+2p^0_i
p^0_j+m_i^2+m_j^2-m_\gamma^2\right|}{2|{\bf p}_i||{\bf p}_j|}
\eeq
translates into the boundary \refeq{eq:yijzijbound} of the variables
$y_{ij}$ and $z_{ij}$.

\subsection{Final-state emitter and initial-state spectator, and vice
versa}

The photonic phase-space measure $[\rd k(P_{ia}^2,x,z_{ia})]$ of the
$ia$ and $ai$ cases is defined in \refeq{eq:PSiasplit}. We derive the
form of this measure by comparing appropriate parametrizations of both
sides of \refeq{eq:PSiasplit}. On the l.h.s.\ we factorize the phase
space into two two-particle phase spaces,
\beq
\int\rd\phi(p_i,k,K_{ia};p_a+p_b) 
=\int\frac{\rd (p_i+k)^2}{2\pi} \,
\int\rd\phi(p_i+k,K_{ia};p_a+p_b) \,
\int\rd\phi(p_i,k;p_i+k), 
\eeq
and insert the parametrizations
\beqar
\int\rd\phi(p_i+k,K_{ia};p_a+p_b) &=&
\frac{1}{4(2\pi)^2} 
\frac{1}{\sqrt{\lambda(s,m_a^2,m_b^2)}} \int\rd P_{ia}^2 \int\rd\varphi_K,
\nn\\[.5em]
\int\rd\phi(p_i,k;p_i+k) &=&
\frac{1}{4(2\pi)^2} 
\frac{1}{\sqrt{\lambda((p_i+k)^2,P_{ia}^2,m_a^2)}} 
\int\rd (p_i-p_a)^2 \int\rd\varphi_\gamma,
\hspace*{1em}
\eeqar
where $\varphi_K$ is the azimuthal angle of $K_{ia}$ in the CM frame of
$p_a+p_b$, and $\varphi_\gamma$ is the one of the photon in the CM frame of
$p_i+k$. Using the relations
\beq
x_{ia} = \frac{-\bar P_{ia}^2}{(p_i+k)^2-P_{ia}^2+m_a^2}, \qquad
z_{ia} = \frac{m_a^2+m_i^2-(p_a-p_i)^2}{(p_i+k)^2-P_{ia}^2+m_a^2},
\eeq
the integrations over the invariants $(p_i+k)^2$ and $(p_i-p_a)^2$ can
be replaced by integrations over $x_{ia}$ and $z_{ia}$. The integration
limits $z_{1,2}(x)$, which are given in \refeq{eq:z12}, follow from
the limits on $(p_a-p_i)^2$ for fixed 
$(p_i+k)^2=(P_{ia}+p_a)^2=P_{ia}^2-m_a^2-\bar P_{ia}^2/x$. They can be
easily derived in the CM frame of $p_i+k$. Finally, we use
\beq
\sqrt{\lambda((p_i+k)^2,P_{ia}^2,m_a^2)} =
\frac{\sqrt{\lambda_{ia}}R_{ia}(x_{ia})}{x_{ia}}
\eeq
on the l.h.s.\ of \refeq{eq:PSiasplit}. 
On the r.h.s.\ we make use of the parametrization
\beq
\int\rd\phi\Big(\tilde p_i(x),K_{ia};\tilde p_a(x)+p_b\Big) =
\frac{1}{4(2\pi)^2} 
\int\rd P_{ia}^2 \,
\frac{1}{\sqrt{\lambda(\tilde s,m_a^2,m_b^2)}} 
\int\rd\tilde\varphi_K,
\eeq
where $\tilde\varphi_K$ denotes the azimuthal angle of $K_{ia}$ in the
CM frame of $\tilde p_a(x)+p_b$.
Note that $\tilde s$, which is the squared CM energy of 
$\tilde p_a(x)+p_b$, depends on $s$, $x$, $P_{ia}^2$, and $K_{ia}^2$.

Since we are not dealing with transverse polarizations, but with
helicity eigenstates or unpolarized configurations, we can assume
rotational invariance of 
$\left|\M_0\left(\tilde\Phi_{0,ia}\right)\right|^2$ 
around the beam axis in the corresponding CM frame of $\tilde p_a(x)+p_b$. 
This implies that the integration over the azimuthal angle 
$\tilde\varphi_K$ yields a trivial factor of $2\pi$. 
The integration of 
$\left|\M_0\left(\tilde\Phi_{0,ia}\right)\right|^2$ 
over $\varphi_K$ yields a factor of $2\pi$ as well because of the
one-to-one correspondence of $\varphi_K$ and $\tilde\varphi_K$.%
\footnote{The angles are related by 
$\tan\varphi_K=f(s,x,P_{ia}^2,K_{ia}^2)\tan\tilde\varphi_K$, if
$\varphi_K=\tilde\varphi_K=0$ is defined in the plane spanned by 
${\bf p}_a$, ${\bf p}_b$, and $\tilde{\bf p}_a$. This follows from the
fact that components of the direction orthogonal to this plane are not
affected by the Lorentz transformation that relates the CM frames of
$p_a+p_b$ and $\tilde p_a+p_b$. The integrals over $\varphi_K$ and
$\tilde\varphi_K$ remain unchanged by the transformation:
$\int_0^{2\pi}\rd\varphi_K = 
\int_0^{2\pi}\rd\tilde\varphi_K 
|f|/(\cos^2\tilde\varphi_K+f^2\sin^2\tilde\varphi_K)=2\pi$.}
Inserting the above relations into
\refeq{eq:PSiasplit}, the photonic phase space
$[\rd k(P_{ia}^2,x,z_{ia})]$ can be identified for fixed values of
$x=x_{ia}$ and $P_{ia}^2$. The result is given in \refeq{eq:dkia}.

\subsection{Initial-state emitter and initial-state spectator}

The measure $[\rd k(s,x,y_{ab})]$ for the photon phase space is
derived by considering explicit representations of the two phase-space
volumes $\rd\phi(\dots)$ in \refeq{eq:PSabsplit}. The full phase
space of $P_{ab}$ and $k$ is parametrized in the CM frame of $p_a+p_b$
by
\beq
\int\rd\phi(k,P_{ab};p_a+p_b) = \frac{1}{8(2\pi)^2} 
\frac{\sqrt{\lambda(s,P_{ab}^2,m_\gamma^2)}}{s} \int\rd\Omega_\gamma, 
\label{eq:PSab}
\eeq
where $\Omega_\gamma$ is the solid angle of the photon.
The one-particle phase space of $\tilde P_{ab}(x)$ reads
\beq
\int\rd\phi\Big(\tilde P_{ab}(x);\tilde p_a(x)+p_b\Big) =
(2\pi)\, \delta\Big(\tilde P_{ab}^2(x)-P_{ab}^2\Big) =
\frac{2\pi}{\bar s} \, \delta(x-x_{ab}).
\eeq
Putting everything together and expressing the
polar angle $\theta_\gamma$ of the photon in terms of $y_{ab}$, we get
the result \refeq{eq:dkab} for $[\rd k(s,x,y_{ab})]$.
The integration boundary \refeq{eq:y12} on $y_{ab}$ is determined by
$|\cos\theta_\gamma|<1$.

\section{Sketch of the calculation of the non-trivial integrals}
\label{app:int}

In \refse{se:subfunc} we have seen that the analytical integration of
the subtraction function $|\M_\sub|^2$ over the photonic parts 
$[\rd k(\dots)]$ of phase space leads to integrals of a non-trivial
structure. Therefore, we sketch the calculation of those integrals in
this appendix.

\subsection{Final-state emitter and final-state spectator}

We first consider the integral for $\Gsub_{ij,+}$, as defined in
\refeq{eq:Gijdef}, for an emitter $i$ and a spectator $j$ in the final state.
The integration over the variable $z_{ij}$ is simple and yields
\beqar
\Gsub_{ij,+}(P_{ij}^2) &=& \int_{y_1}^{y_2} \rd y\, 
\frac{\bar P_{ij}^2}{\sqrt{\lambda_{ij}}R_{ij}(y)} \left\{
\frac{2}{y}\ln\left[\frac{1-(1-y)z_1(y)}{1-(1-y)z_2(y)}\right] \right.
\nn\\ && \left. {}
-\frac{1-y}{2y}[2+z_1(y)+z_2(y)][z_2(y)-z_1(y)]
-\frac{2m_i^2}{\bar P_{ij}^2} \frac{1-y}{y^2} [z_2(y)-z_1(y)] \right\}
\nn\\ && {}
-\Gsub_{ij,-}(P_{ij}^2),
\eeqar
where we have renamed $y_{ij}$ to $y$. The contribution of
$\Gsub_{ij,-}(P_{ij}^2)$ will be calculated below.
The explicit integral over $y$ involves
two types of square roots of quadratic forms in $y$, entering via the
limits $z_{1,2}(y)$ given in \refeq{eq:yijzijbound}. The limits
$y_{1,2}$ are also defined there. Either of those roots
can be removed by splitting the $y$ range into two pieces:
\begin{itemize}
\item[(a)] $y_1 < y < \Delta y \ll 1$,
\item[(b)] $\Delta y < y < y_2$.
\end{itemize}
The IR singularity is contained in part (a) so that part (b) can be
evaluated with $m_\gamma=0$, replacing the root $\sqrt{y^2-y_1^2}$ by
$y$. The integration over part (a) is simplified by choosing the
auxiliary parameter $\Delta y$ small so that in ${\cal O}(\Delta y)$
the parameter $y$ can be set to zero in
the non-singular factors of the integrand. Thus, for $m_\gamma\to 0$ we
can replace $R_{ij}(y)$ by 1 in this part.
Explicitly we get
\beqar
\Gsub_{ij,+}(P_{ij}^2)\Big|_{\mathrm{(a)}} &=&
\int_{y_1}^{\Delta y} \rd y\, 
\frac{2\bar P_{ij}^2}{y\sqrt{\lambda_{ij}}}
\ln\left[\frac{(\bar P_{ij}^2+2m_i^2)y+\sqrt{\lambda_{ij}}\sqrt{y^2-y_1^2}}
	{(\bar P_{ij}^2+2m_i^2)y-\sqrt{\lambda_{ij}}\sqrt{y^2-y_1^2}}\right]
\nn\\ && {}
- \int_{y_1}^{\Delta y} \rd y\, \frac{2}{y^2} \sqrt{y^2-y_1^2},
\nn\\[.5em] 
\Gsub_{ij,+}(P_{ij}^2)\Big|_{\mathrm{(b)}} &=& 
\int_{\Delta y}^{y_2} \rd y\, 
\frac{2\bar P_{ij}^2 \disp
  \ln\left[\frac{2m_i^2+\bar P_{ij}^2(1+y)
	+\sqrt{(2m_j^2+\bar P_{ij}^2-\bar P_{ij}^2 y)^2-4P_{ij}^2 m_j^2}}
 		{2m_i^2+\bar P_{ij}^2(1+y)
	-\sqrt{(2m_j^2+\bar P_{ij}^2-\bar P_{ij}^2 y)^2-4P_{ij}^2 m_j^2}}
     \right]}
  {y \sqrt{(2m_j^2+\bar P_{ij}^2-\bar P_{ij}^2 y)^2-4P_{ij}^2 m_j^2}}
\nn\\ && {}
-\int_{\Delta y}^{y_2} \rd y\, \left\{
\frac{2}{y} - \frac{\bar P_{ij}^4 y}{2(m_i^2+y\bar P_{ij}^2)^2} 
\right\}
-\Gsub_{ij,-}(P_{ij}^2).
\eeqar
The second integrals in both parts are elementary. In the first integrals 
we remove the square roots by the substitutions
\begin{itemize}
\item[(a)] $y_1\sqrt{1+x} = y + \sqrt{y^2-y_1^2}$,
\item[(b)] $x = y_2 - y + \disp \sqrt{(y_2-y)
	\left(2-y-y_2+\frac{4m_j^2}{\bar P_{ij}^2}\right)}$.
\end{itemize}
The resulting integrals are of the form 
\beq
\int_{x_1}^{x_2}\rd x\, f(x) \ln[g(x)],
\label{eq:genint}
\eeq
where $f(x)$ and $g(x)$ are algebraic functions. Upon decomposing
$f(x)$ into partial fractions and factorizing $g(x)$, such integrals
yield subintegrals that can be expressed in terms of logarithms and 
dilogarithms. A convenient way to obtain compact results is to transform
the limits $x_{1,2}$ into 0 and $\infty$ in a first step. This is
achieved by the substitution $\xi=(x-x_1)/(x_2-x)$. The
subintegrals that lead to dilogarithms can then be calculated by using
the standard integral%
\footnote{The contributions of the function 
$\eta(x,y)=\ln(xy)-\ln(x)-\ln(y)$ 
are necessary to put the arguments of the dilogarithms onto the first
Riemann sheet for complex constants $\alpha_{0,1}$ and $\beta$.}
\beq
\int_0^\infty\rd\xi\, 
\left(\frac{1}{\xi-\alpha_0}-\frac{1}{\xi-\alpha_1}\right) 
\ln(1+\beta\xi) = \sum_{l=0,1}(-1)^l
[ \Li_2(1+\beta\alpha_l)+\eta(-\alpha_l,\beta)\ln(1+\beta\alpha_l) ].
\eeq
Although these steps are straightforward, they nevertheless involve a
lot of algebra. Therefore, we omit the details. Instead we comment on
the IR singularity and the role of the parameter $\Delta y$. 
In part (a) the upper
limit of the integration over $x$ tends to infinity like
$\Delta y^2 \bar P_{ij}^4/(m_i^2 m_\gamma^2)$ for fixed $\Delta y$,
since the photon mass $m_\gamma$ is infinitesimal. This induces terms
proportional to $\ln(\Delta y/m_\gamma)$ in part (a). On the other hand, part
(b) is logarithmically divergent for $\Delta y\to 0$. The artifical
$\ln(\Delta y)$ terms, of course, cancel in the sum of parts (a) and
(b). Finally, we note that we had to exploit some identities for
the dilogarithms in order to obtain the compact form of the final result 
\refeq{eq:Gij} for $\Gsub_{ij,+}$. 

The calculation of $\Gsub_{ij,-}$ proceeds in a different way.
Since this function is IR-finite, we can set $m_\gamma$ to zero from the
beginning. The defining integral \refeq{eq:Gijdef} explicitly reads
\beq
\Gsub_{ij,-}(P_{ij}^2) = 
\frac{m_i^2}{\sqrt{\lambda_{ij}}} \, 
\int_0^{y_2} \rd y\, 
\frac{1-y}{y^2} \, \frac{r_{ij}(y)}{R_{ij}(y)} \,
\int_{z_1(y)}^{z_2(y)} \rd z \, \frac{(1-z)^2}{z}.
\label{eq:Gijm}
\eeq
Note that only the behaviour of $g_{ij,-}$ at $y\to 0$ is relevant in
the IR and collinear limits [see \refeq{eq:yzijlim}]. Therefore, we have
chosen a form of the auxiliary function $r_{ij}(y)$ that simplifies the
integration. We have defined $r_{ij}(y)$ in such a way that
\beq
\frac{1-y}{y^2} \frac{r_{ij}(y)}{R_{ij}(y)} = 
-\frac{\rd}{\rd y}\left[\frac{R_{ij}(y)}{y}\right] + {\cal O}(m_\gamma^2).
\eeq
This choice allows us to perform the integration in \refeq{eq:Gijm} over
$z$ implicitly upon applying integration by parts in the integration
over $y$. The boundary terms of this integration by parts vanish, and
the result is
\beq
\Gsub_{ij,-}(P_{ij}^2) = 
\frac{m_i^2}{\sqrt{\lambda_{ij}}} \, 
\int_0^{y_2} \rd y\, 
\frac{R_{ij}(y)}{y} \,
\left\{ \frac{[1-z_2(y)]^2}{z_2(y)}z'_2(y) 
-\frac{[1-z_1(y)]^2}{z_1(y)}z'_1(y) \right\},
\eeq
where $z'_{1,2}(y)=\rd z_{1,2}(y)/\rd y$. 
The final integration over $y$ is elementary.
Note that the above trick avoids terms such as $\ln[z_{1,2}(y)]$ after
the integration over $z$; such terms would lead to dilogarithms in the
final result.

\subsection{Final-state emitter and initial-state spectator, and vice
versa}

The integrals for the endpoint contributions defined in 
\refeq{eq:Gsubiaaidef} for the mixed cases $ff'=ia,ai$ are calculated in 
a similar way. Therefore, we outline only the basic steps. 

\begin{sloppypar}
Inspecting the explicit form of the distributions $\cGsub_{ff',+}$, we
find that the integrals \refeq{eq:Gsubiaaidef} for $\Gsub_{ff',+}$ again
contain two different square roots of quadratic forms in $x$. Analogously
to the $ij$ case, we first separate these roots by splitting the range of 
the $x$ integration as follows:
\begin{itemize}
\item[(a)] $x_1 > x > 1-\Delta x$, \quad with $\Delta x \ll 1$,
\item[(b)] $1-\Delta x > x > x_0$.
\end{itemize}
Part (a) contains the IR singularity and involves only values of $x$ in
the vicinity of 1. Thus, in ${\cal O}(\Delta x)$ we can set $x$ to 1 in
all non-singular terms of the integral. In particular, this replaces the
function $R_{ia}(x)$ by $1+{\cal O}(m_\gamma^2)$ and removes the root 
implicitly contained in $R_{ia}(x)$. In ${\cal O}(m_\gamma)$ we can
replace the explicit root 
$\sqrt{\bar P_{ia}^4(1-x)^2-4m_i^2 m_\gamma^2 x^2}$, which appears in
$z_{1,2}(x)$ given in \refeq{eq:z12}, by
$\sqrt{\bar P_{ia}^4(1-x)^2-4m_i^2 m_\gamma^2}$.
This root is removed by the substitution
\begin{itemize}
\item[(a)] $\disp\frac{2m_i m_\gamma}{-\bar P_{ia}^2}\sqrt{1+y}= 
1-x+\sqrt{(1-x)^2-\frac{4m_i^2 m_\gamma^2}{\bar P_{ia}^4}}$.
\end{itemize}
The resulting integral is of the form \refeq{eq:genint} and can be
reduced to logarithms and dilogarithms, as described above. The IR
singularity appears in contributions proportional to $\ln(\Delta x/m_\gamma)$.
In part (b) we can set $m_\gamma$ to zero,
since the IR singularity is avoided by the finite value of $\Delta x$.
This eliminates the explicit root in $z_{1,2}(x)$ given in \refeq{eq:z12}.
The root in $R_{ia}(x)$ is removed by the substitution
\begin{itemize}
\item[(b)] $\disp y=-\frac{2m_a A}{\bar P_{ia}^2} x
-\frac{\sqrt{\lambda_{ia}}}{\bar P_{ia}^2} R_{ia}(x)$, \quad
with $A=\sqrt{-\bar P_{ia}^2-m_i^2} > 0.$
\end{itemize}
Note that this substitution is only allowed for $A>0$. This is, e.g.,
fulfilled if $P_{ia}^2<0$, but in general not for all $P_{ia}^2$. We
evaluate the integrals for the allowed range with $A>0$ and cover the
full parameter space in the final result by analytical continuation in
$P_{ia}^2$. The reduction of the obtained integral, which is again of
the general form \refeq{eq:genint}, to logarithms and dilogarithms
proceeds as above. However, particular care is needed in the arguments
of those multivalued functions. As required, the singular $\ln(\Delta x)$
terms cancel in the sum of parts (a) and (b).
\end{sloppypar}

The calculation of $\Gsub_{ia,-}$ is simplified by an appropriate choice
of the auxiliary function $r_{ia}(x)$. Similar to the $ij$ case, we have
defined this function in such a way that
\beq
\frac{x}{(1-x)^2} \frac{r_{ia}(x)}{R_{ia}(x)} = 
\frac{\rd}{\rd x}\left[\frac{R_{ia}(x)}{1-x}\right] + {\cal O}(m_\gamma^2).
\eeq
Hence, integration by parts can be applied as above, and the resulting
integral over $x$ is elementary. 

Finally, the calculation of $\Gsub_{ai,-}$ is trivial.

\subsection{Initial-state emitter and initial-state spectator}

In view of the analytical integrations, the $ab$ case turns out to be
the simplest one. The integrals of the endpoint parts are given by
\beq
\Gsub_{ab,\tau}(s,x_0) = \int_{x_0}^{x_1}\rd x\, \cGsub_{ab,\tau}(s,x)
\eeq
with the distributions $\cGsub_{ab,\tau}$ of \refeq{eq:cGsubab}.
The calculation of $\Gsub_{ab,-}$ is trivial.

The integral for $\Gsub_{ab,+}$ involves only the square root of the 
quadratic form in $x$ that is contained in the limits $y_{1,2}(x)$ given 
in \refeq{eq:y12}. Note that the auxiliary function $R_{ab}(x)$ does not 
occur in the integral. As above, we first split the range of the 
integration over $x$ as follows:
\begin{itemize}
\item[(a)] $x_1 > x > 1-\Delta x$, \quad with $\Delta x \ll 1$,
\item[(b)] $1-\Delta x > x > x_0$.
\end{itemize}
In part (a) we can set $x$ to 1 in all non-singular terms, and the root is
removed by the substitution
\begin{itemize}
\item[(a)] $\disp\frac{2m_\gamma\sqrt{s}}{\bar s}\sqrt{1+y}= 
1-x+\sqrt{(1-x)^2-\frac{4s m_\gamma^2}{\bar s^2}}$.
\end{itemize}
This leads to an integral of the form \refeq{eq:genint}, which is evaluated
as described above. In part (b) we can set $m_\gamma$ to zero, directly
resulting in an elementary integral, which is expressed in terms of
logarithms.

\section{\boldmath{Photonic corrections to 
$\mu^+\mu^-\to\nu_\Pe\bar\nu_\Pe(\gamma)$}}
\label{app:mmnn}

Using the methods described in \citere{de93a}, we have calculated the
virtual photonic corrections of ${\cal O}(\alpha)$ for arbitrary muon mass. 
The one-loop amplitude reads
\beqar
\M_\virt &=& \frac{\alpha}{\pi} \left\{
\frac{s-2m_\mu^2}{4\beta s}\left[
\pi^2-4\ln\left(\frac{m_\gamma}{m_\mu}\right)\ln(x_s)-\ln^2(x_s)
-4\Li_2(1+x_s) + 4\pi\ri\ln(1+x_s) \right] \right.
\nn\\ && \left. \hspace*{2em} {}
+\ln\left(\frac{m_\mu}{m_\gamma}\right)-\frac{3\beta}{4}\ln(x_s)-1 
\right\} \M_0
\nn\\ && {}
+\frac{e(v_{\nu_\Pe}+a_{\nu_\Pe})}{2(s-\MZ^2+\ri\MZ\Gamma_\PZ)} \,
\frac{\alpha m_\mu}{\pi\beta s} \, \ln(x_s) \,
\left\{
2m_\mu a_\mu \Big[ \bar v_{\mu^+} \gamma^\rho \gamma_5 u_{\mu^-} \Big] \, 
\Big[ \bar u_{\nu_\Pe} \gamma_\rho (1-\gamma_5) v_{\bar\nu_\Pe} \Big]
\right.
\nn\\ && \left. \hspace*{2em} {}
-v_\mu \Big[ \bar v_{\mu^+} u_{\mu^-} \Big] \, 
\Big[ \bar u_{\nu_\Pe} \slash{p} (1-\gamma_5) v_{\bar\nu_\Pe} \Big]
\right\},
\label{eq:mmnnvirt}
\eeqar
where $\beta$ denotes the muon velocity in the CM frame, and $x_s$ is an
auxiliary variable,
\beq
\beta = \sqrt{1-\frac{4m_\mu^2}{s}}, \qquad
x_s = \frac{\beta-1}{\beta+1}+\ri\epsilon.
\eeq
The fermion spinors in \refeq{eq:mmnnvirt} carry the same arguments as
indicated in the Born amplitude $\M_0$ given in \refeq{eq:mmnnborn}. The
spinor chains have been evaluated by applying the Weyl--van der Waerden
spinor technique, following the formulation of \citere{di99}.
The amplitudes $\M_1$ for the radiative process
$\mu^+\mu^-\to\nu_\Pe\bar\nu_\Pe\gamma$ are contained in
\citere{di99} explicitly. The application of the slicing method to
the real corrections requires the separate calculation of the
soft-photonic corrections. They are contained in the factor correction
$\delta_\soft$ to the Born cross section $\sigma_0$, 
\beqar
\delta_\soft &=& -\frac{\alpha}{\pi} \left\{
\frac{s-2m_\mu^2}{2\beta s} \left[
4\ln\left(\frac{2\Delta E}{m_\gamma}\right)\ln(-x_s)
+4\Li_2(1+x_s)+\ln^2(-x_s)
\right] \right.
\nn\\ && \left. \hspace*{2em} {}
+2\ln\left(\frac{2\Delta E}{m_\gamma}\right)
+\frac{\ln(-x_s)}{\beta}
\right\},
\eeqar
which has been deduced from the general results given in \citere{de93a}.

The above results can be easily expanded in the limit $m_\mu\to 0$,
which can be used for high energies. 

\def\vol#1{{\bf #1}}
\def\mag#1{{\sl #1}}

\end{document}